\documentclass[reprint,aps,amsmath,amssymb,nofootinbib,superscriptaddress]{revtex4-1}

\usepackage{graphicx}
\graphicspath{{images/}}
 
\usepackage{mathrsfs}
\usepackage{dcolumn} 	% Align table columns on decimal point
\usepackage{bm} 		% bold math
\usepackage{hyperref}	
\usepackage{siunitx}
\usepackage{color, xcolor, colortbl}
\usepackage{subfigure}
\usepackage{siunitx}
\usepackage{tabularx}
\usepackage{enumerate}
\newcommand{\Ar}{\text{Ar}}
\newcommand{\Efield}{\mathscr{E}}
\newcommand{\Wion}{W_{\text{ion}}}
\newcommand{\Wph}{W_{\text{ph}}}
\newcommand{\Ni}{N_{\text{i}}}
\newcommand{\Nex}{N_{\text{ex}}}
\newcommand{\vis}{\text{vis}}

\begin{document}
%\preprint{APS/123-QED}

\title{Calorimetry for low-energy electrons using charge and light in liquid argon}

%#################################################
% Authors
%#################################################
\newcommand{\ufederalABC}{Universidade Federal do ABC, Santo Andr\'{e}, SP 09210-580, Brasil}
\newcommand{\ucampinas}{Universidade Estadual de Campinas, Campinas, SP 13083-859, Brasil}
\newcommand{\uchicago}{University of Chicago, Chicago, IL 60637, USA}
\newcommand{\cincinnati}{University of Cincinnati, Cincinnati, OH 45221, USA}
\newcommand{\fnal}{Fermi National Accelerator Laboratory (FNAL), Batavia, IL 60510, USA}
\newcommand{\ufedgoi}{Universidade Federal de Goi\'{a}s, Goi\'{a}s, CEP 74690-900, Brasil}
\newcommand{\utarlington}{University of Texas at Arlington, Arlington, TX 76019, USA} 
\newcommand{\boston}{Boston University, Boston, MA 02215, USA}
\newcommand{\michigan}{Michigan State University, East Lansing, MI 48824, USA}
\newcommand{\duluth}{University of Minnesota, Duluth, Duluth, MN 55812, USA}
\newcommand{\infn}{Istituto Nazionale di Fisica Nucleare (INFN), Rome 00186, Italy}
\newcommand{\louisiana}{Louisiana State University, Baton Rouge, LA 70803, USA}
\newcommand{\manchester}{University of Manchester, Manchester M13 9PL, UK} 
\newcommand{\syracuse}{Syracuse University, Syracuse, NY 13244, USA}
\newcommand{\utaustin}{University of Texas at Austin, Austin, TX 78712, USA}
\newcommand{\ucollegelondon}{University College London, London WC1E 6BT, UK}
\newcommand{\williamandmary}{College of William \& Mary, Williamsburg, VA 23187, USA}
\newcommand{\yale}{Yale University, New Haven, CT 06520, USA}
\newcommand{\kek}{High Energy Accelerator Research Organization (KEK), Tsukuba 305-0801, Japan}

\author{W.~Foreman}\thanks{wforeman@iit.edu}
\altaffiliation{Current address: Illinois Institute of Technology, Chicago, IL 60616, USA}
\affiliation{\uchicago}

\author{R.~Acciarri}
\affiliation{\fnal}

\author{J.~A.~Asaadi}
\affiliation{\utarlington}

\author{W.~Badgett}
\affiliation{\fnal}

\author{F.~d.~M.~Blaszczyk}
\affiliation{\boston}

\author{R.~Bouabid}
\affiliation{\uchicago}

\author{C.~Bromberg}
\affiliation{\michigan}

\author{R.~Carey}
\affiliation{\boston}

\author{F.~Cavanna}
\affiliation{\fnal}
\affiliation{\yale}

\author{J.~I.~Cevallos~Aleman}
\affiliation{\uchicago}

\author{A.~Chatterjee}
\affiliation{\utarlington}

\author{J.~Evans}
\affiliation{\manchester}

\author{A.~Falcone}
\affiliation{\utarlington}

\author{W.~Flanagan}
\altaffiliation{Current address: Univ. of Dallas, Irving, TX 75062, USA}
\affiliation{\utaustin}

\author{B.~T.~Fleming}
\affiliation{\yale}

\author{D.~Garcia-Gamez}
\altaffiliation{Current address: Univ. of Granada, 18010 Granada, Spain}
\affiliation{\manchester}

\author{B.~Gelli}
\affiliation{\ucampinas}

\author{T.~Ghosh}
\author{R.~A.~Gomes}
\affiliation{\ufedgoi}

\author{E.~Gramellini}
\altaffiliation{Current address: FNAL, Batavia, IL 60510, USA}
\affiliation{\yale}

\author{R.~Gran}
\affiliation{\duluth}

\author{P.~Hamilton}
\affiliation{\syracuse}

\author{C.~Hill}
\affiliation{\manchester}

\author{J.~Ho}
\affiliation{\uchicago}

\author{J.~Hugon}
\affiliation{\louisiana}

\author{E.~Iwai}
\affiliation{\kek}

\author{E.~Kearns}
\affiliation{\boston}

\author{E.~Kemp}
\affiliation{\ucampinas}

\author{T.~Kobilarcik}
\affiliation{\fnal}

\author{M.~Kordosky}
\affiliation{\williamandmary}

\author{P.~Kryczy\'nski}
\affiliation{\fnal}
\affiliation{Institute of Nuclear Physics PAN, 31-342 Krak\'{o}w, Poland}

\author{K.~Lang}
\affiliation{\utaustin}

\author{R.~Linehan}
\affiliation{\boston}

\author{A.~A.~B.~Machado}
\affiliation{\ucampinas}

\author{T.~Maruyama}
\affiliation{\kek}

\author{W.~Metcalf}
\affiliation{\louisiana}

\author{C.~A.~Moura}
\affiliation{\ufederalABC}

\author{R.~Nichol}
\affiliation{\ucollegelondon}

\author{M.~Nunes}
\affiliation{\ucampinas}

\author{I.~Nutini}
\affiliation{\fnal}
\affiliation{\infn}

\author{A.~Olivier}
\altaffiliation{Current address: Univ. of Rochester, Rochester, NY 14627, USA}
\affiliation{\louisiana}

\author{O.~Palamara}
\affiliation{\fnal}
\affiliation{\yale}

\author{J.~Paley}
\affiliation{\fnal}

\author{L.~Paulucci}
\affiliation{\ufederalABC}

\author{G.~Pulliam}
\affiliation{\syracuse}

\author{J.~L.~Raaf}
\affiliation{\fnal}

\author{B.~Rebel}
\affiliation{\fnal}
\affiliation{University of Wisconsin-Madison, Madison, Wisconsin 53706, USA}

\author{O.~Rodrigues}
\altaffiliation{Current address: Syracuse Univ., Syracuse, NY 13244, USA}
\affiliation{\ufedgoi}

\author{L.~Mendes~Santos}
\affiliation{\ucampinas}

\author{D.~W.~Schmitz}
\affiliation{\uchicago}

\author{E.~Segreto}
\affiliation{\ucampinas}

\author{D.~Smith}
\affiliation{\boston}

\author{M.~Soderberg}
\affiliation{\syracuse}

\author{F.~Spagliardi}
\altaffiliation{Current address: Univ. of Oxford, Oxford OX1 3PJ, UK}
\affiliation{\manchester}

\author{J.~M.~St.~John}
\altaffiliation{Current address: FNAL, Batavia, IL 60510, USA}
\affiliation{\cincinnati}

\author{M.~Stancari}
\affiliation{\fnal}

\author{A.~M.~Szelc}
\affiliation{\manchester}

\author{M.~Tzanov}
\affiliation{\louisiana}

\author{D.~Walker}
\affiliation{\louisiana}

\author{Z.~Williams}
\affiliation{\utarlington}

\author{T.~Yang}
\affiliation{\fnal}

\author{J.~Yu}
\affiliation{\utarlington}

\author{S.~Zhang}
\affiliation{\boston}

\collaboration{The LArIAT Collaboration}
\date{\today}

\begin{abstract}

Precise calorimetric reconstruction of 5-\SI{50}{MeV} electrons in liquid argon time projection chambers (LArTPCs) will enable the study of astrophysical neutrinos in DUNE and could enhance the physics reach of oscillation analyses. Liquid argon scintillation light has the potential to improve energy reconstruction for low-energy electrons over charge-based measurements alone. Here we demonstrate light-augmented calorimetry for low-energy electrons in a single-phase LArTPC using a sample of Michel electrons from decays of stopping cosmic muons in the LArIAT experiment at Fermilab.  Michel electron energy spectra are reconstructed using both a traditional charge-based approach as well as a more holistic approach that incorporates both charge and light. A maximum-likelihood fitter, using LArIAT's well-tuned simulation, is developed for combining these quantities to achieve optimal energy resolution. A sample of isolated electrons is simulated to better determine the energy resolution expected for astrophysical electron-neutrino charged-current interaction final states.  In LArIAT, which has very low wire noise and an average light yield of \SI{18}{pe/MeV}, an energy resolution of $\sigma/E \simeq 9.3\%/\sqrt{E} \oplus 1.3\%$ is achieved. Samples are then generated with varying wire noise levels and light yields to gauge the impact of light-augmented calorimetry in larger LArTPCs. At a charge-readout signal-to-noise of S/N~$\simeq$~30, for example, the energy resolution for electrons below \SI{40}{MeV} is improved by $\approx$~10\%, $\approx$~20\%, and $\approx$~40\% over charge-only calorimetry for average light yields of \SI{10}{pe}/MeV, \SI{20}{pe}/MeV, and \SI{100}{pe}/MeV, respectively.

\end{abstract}

\pacs{Valid PACS appear here}
\maketitle

%#################################################
% Section 1: Introduction
%#################################################
\section{\label{sec:intro}Introduction}

Open questions in neutrino physics are inspiring a new generation of experiments which utilize liquid argon time projection chamber (LArTPC) technology~\cite{sbn,dune}. The largest of these efforts is the Deep Underground Neutrino Experiment (DUNE), which will employ four 10-kiloton LArTPCs at a depth of nearly \SI{1.5}{km}. By detecting neutrinos produced \SI{1300}{km} away at Fermilab, DUNE aims to determine the neutrino mass ordering, measure the \emph{CP}-violating phase, and carry out precision tests of the three-flavor oscillation framework. DUNE's large underground active volume also enables new searches for nucleon decay, solar neutrino studies, and the observation of an electron neutrinos ($\nu_e$) from a galactic supernova.

The $\nu_e$ signal from the neutronization burst of a core-collapse supernova, as well as $\nu_e$ produced by the sun, can be used to probe a rich variety of topics related to neutrinos, stellar astrophysics, and theories beyond the Standard Model~\cite{dune_solar,dune_supernova}. Liquid argon is particularly sensitive to the $\nu_e$-induced charged-current (CC) interaction, $\nu_e + \Ar \rightarrow e^- + K^*$, which produces an outgoing electron with energy ranging from from zero to several tens of MeV. %from a few MeV to several tens of MeV. 
Excellent calorimetric resolution for low-energy electrons in LAr is therefore needed to maximize the physics potential for studies of astrophysical neutrinos in LArTPCs.
For supernova $\nu_e$ in particular, it is estimated that a resolution of $\sigma/E \lesssim 15$ \%/$\sqrt{E\text{[MeV]}}$ is required in the DUNE detector~\cite{dune_supernova}.

Low-energy capabilities also enhance DUNE's oscillation physics program through more efficient tagging of \SI{4.1}{MeV} muons from pion decay-at-rest and Michel electrons from muon decay-at-rest. Better identification of final state particles from neutrino interactions facilitates a more accurate determination of the neutrino's energy and interaction type~\cite{dune_supernova}.

\begin{figure}
\includegraphics[width=\columnwidth]{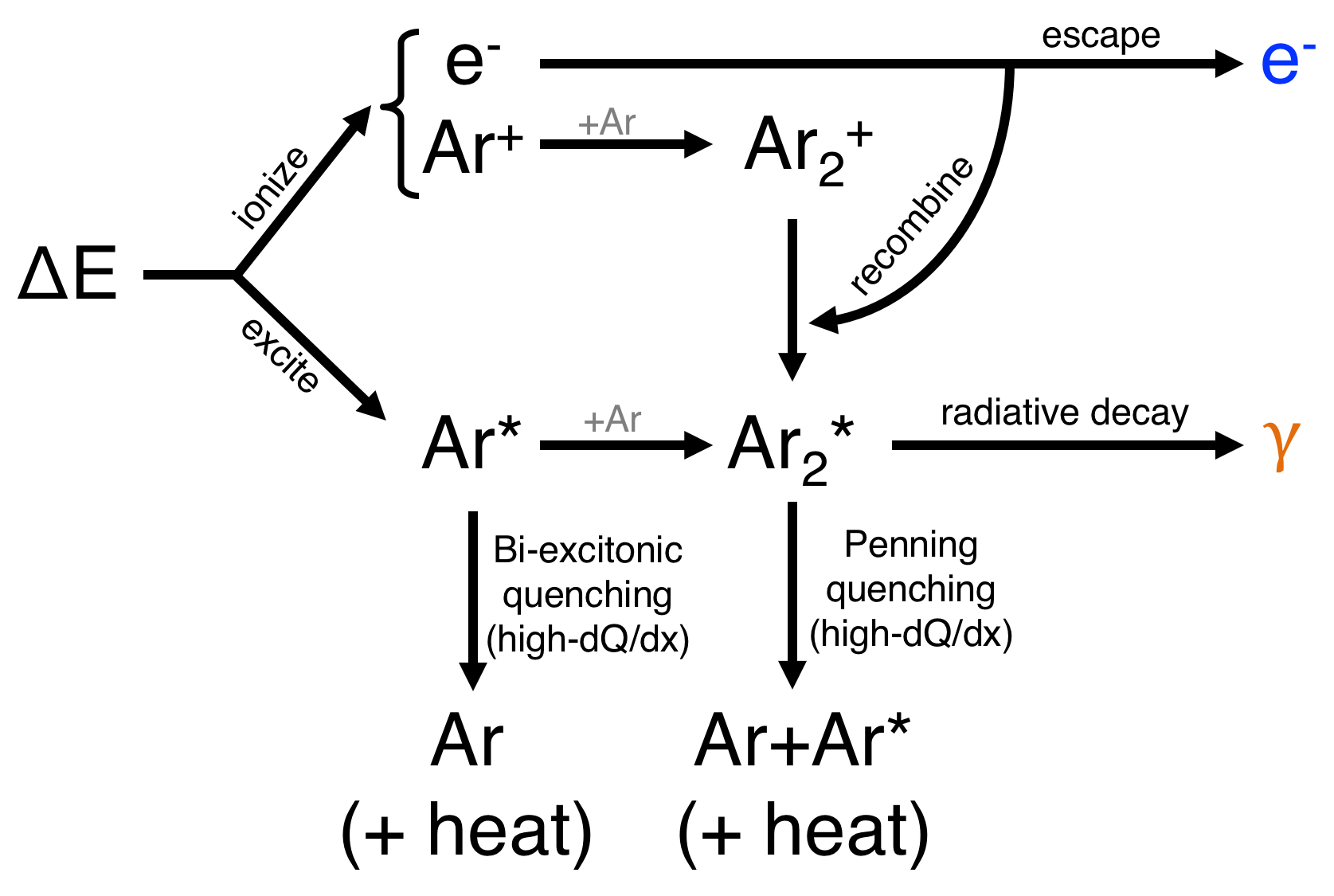}
\caption{Schematic diagram illustrating the production of free ionization electrons ($e^-$) and scintillation photons ($\gamma$) from energy deposited in liquid argon.}
\label{fig:recomb_diag}
\end{figure}

The two observables generated from energy loss by charged particles in LAr are charge ($Q$) and light ($L$).  As illustrated in Fig.~\ref{fig:recomb_diag}, energy $\Delta E$ deposited by a particle first goes into producing $\Nex$ excitons ($\Ar^*$) and $\Ni$ electron-ion pairs ($e^-, \Ar^+$) with a known excitation ratio $\alpha = \Nex / \Ni~=~0.21$~\cite{w_ion,excitons_in_lar, aprile_book}. Short-lived excimers ($\Ar_2^*$) are formed through collisions of excitons with bulk Ar. These excimers undergo dissociative decay to their ground state by emitting a vacuum ultraviolet (VUV) photon with a singlet-state ($^1\Sigma_u^+$) lifetime of $\tau_s = 6\pm\SI{2}{ns}$~\cite{hitachi_timedependence} and a longer triplet-state ($^3\Sigma_u^+$) lifetime of $\tau_t = 1300 \pm \SI{60}{ns}$~\cite{scintoflar}. However, these photon-producing $\Ar_2^*$ are also formed by free electrons recombining with surrounding molecular argon ions. In heavily-ionizing nuclear recoils, several quenching effects~\cite{nuclear_recoil_scint, lindhard, biexcitonic_quenching, penning_quenching} are believed to cause nonradiative destruction of some fraction of $\Ar_2^*$, resulting in some energy being lost to atomic motion (heat). For MeV-scale electrons, which are minimally-ionizing, we may neglect these effects and assume all deposited energy goes into observable charge and light. With that assumption, $Q$ and $L$ may be expressed as:
%\begin{linenomath*}
\begin{align}
	\label{eq:QL}
	\begin{split}
	Q &= N_e = \Ni R, \\
	L &= N_\gamma = \Nex + \Ni(1-R),
	\end{split}
	\end{align}
%\end{linenomath*}
where $R$ is the electron recombination survival probability, commonly referred to as the \emph{recombination factor}.  Charge and light are therefore anticorrelated for particles like electrons, with their sum directly proportional to the total energy deposited,
\begin{equation}\label{eq:QplusL}
Q+L = \Nex + \Ni = \frac{\Delta E}{\Wph},
\end{equation}
where $\Wph~=~19.5\pm\SI{1.0}{eV}$~\cite{w_ph,doke} is the average amount of energy deposited by a charged particle manifesting in the production of an ion or exciton. The quantity $\Wph$ is related to the commonly used ionization work function, $\Wion = 23.6~\pm~\SI{0.3}{eV}$~\cite{w_ion}, through the excitation ratio~$\alpha$: $\Wion~=~(1~+~\alpha)~\Wph$.

When reconstructing deposited (or ``visible'') energy in a LArTPC using $Q$, only the electrons that escape electron-ion recombination ($N_e = N_iR$) and successfully drift to the wire planes can be used, so corrections must be applied to account for the charge lost to recombination.  The dependence of $R$ on the external electric field $\Efield$, as well as the local ionization charge density $dQ/dx$, has been described phenomenologically~\cite{birks_law,thomas_imel} and modeled using fits to data from the ICARUS and ArgoNeuT experiments~\cite{icarus_recomb,argoneut_recomb}.  For tracklike topologies that can be spatially reconstructed, these recombination models are used to calculate the linear energy deposition density, $dE/dx$, along the track.
%The total deposited energy is then taken as the sum of 
%recombination models are used to find the particle's total deposited energy by summing together the individual energy depositions calculated from the linear energy density, $dE/dx$, for points along the track. 
However, for electromagnetic (EM) showers, %which often include isolated charge deposited by bremsstrahlung photons,
the $dE/dx$ (and thus $R$) cannot easily be determined at all deposition sites. Therefore, to calculate energy deposited by EM showers, the simplest method is to assume an average $R$ and uniformly scale up the total charge:
%\begin{linenomath*}
\begin{align}\label{eq:EQ_intro}
E_Q = ( Q \times  R^{-1} )  \times \Wion.
\end{align}
%\end{linenomath*}

Due to the stochastic nature of bremsstrahlung radiation, energy is deposited over a wider and more variable range of $dE/dx$ in EM showers compared to energy deposited by simple, minimally ionizing particle tracks.  Assuming a uniform $R$ is therefore not realistic. However, by combining Eqs.~(\ref{eq:QL}) and (\ref{eq:QplusL}), the deposited energy can be more holistically determined by exploiting the complementarity of $Q$ and $L$, foregoing the need to correct for recombination:
%\begin{linenomath*}
\begin{equation}\label{eq:EQL_intro}
E_{QL} = (Q + L) \times \Wph.
\end{equation}
%\end{linenomath*}
%This method is particularly useful for measuring the total energy deposited by EM showers which, due to the stochasticity of bremsstrahlung radiation and normal Poisson fluctuations in energy deposition, are likely to include charge deposited over a wider range of $dE/dx$ compared to simple particle tracks.

The prospect of combining charge and light for calorimetric neutrino reconstruction in LArTPCs was explored previously through simulations by Sorel~\cite{sorel}, though that work focused on neutrino interactions at the GeV scale. In this paper, we demonstrate light-augmented calorimetric measurements relevant to astrophysical $\nu_e$ interactions using data from the LArIAT experiment~\cite{lariat_detpaper}.  We then utilize the LArIAT detector simulation, which has been carefully tuned and cross-checked with data, to study a variety of realistic performance scenarios for large LArTPC neutrino detectors.

% *********************************************************
\section{The LArIAT Experiment}\label{sec:lariat} 

\begin{figure} \centering
\includegraphics[width=0.8\columnwidth]{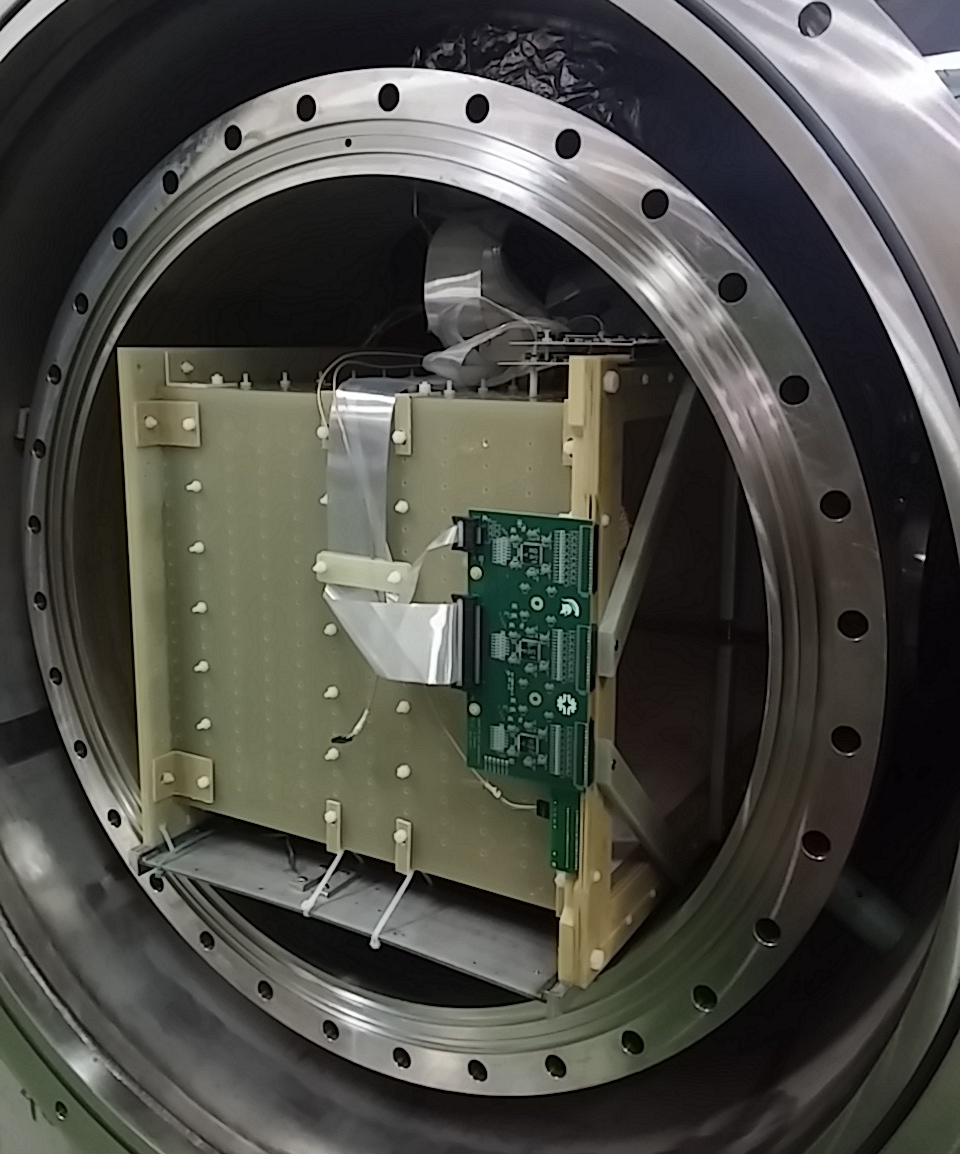}\\
\vspace{2mm}
\includegraphics[width=0.8\columnwidth]{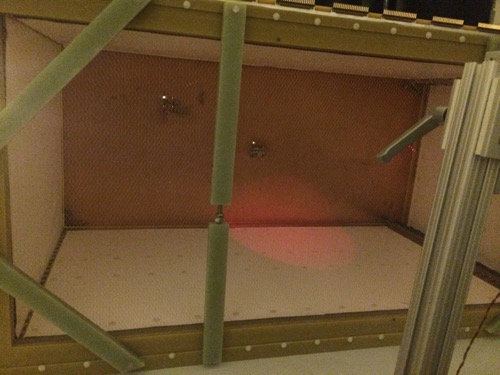}
\caption{The LArIAT TPC sitting inside the inner cryostat (top), and a view of the TPB-coated foils mounted to the field cage walls from behind the anode wire planes (bottom).
}
\label{fig:tpc}
\end{figure}

\begin{figure*}\centering
\includegraphics[height=2in]{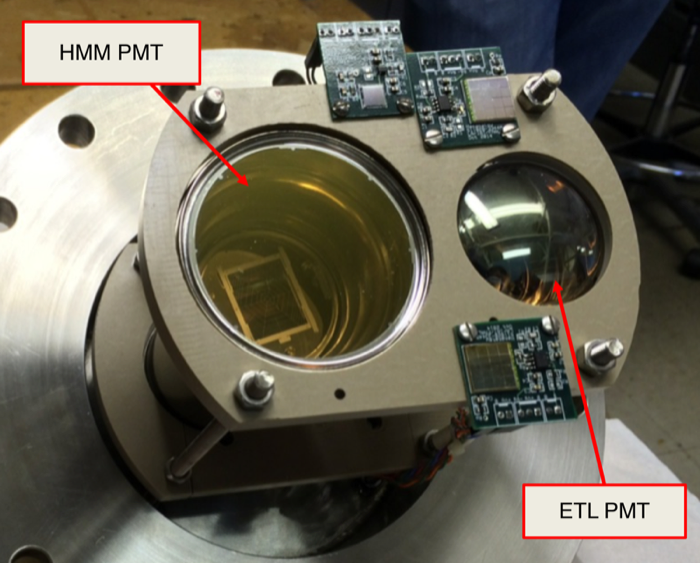}
\hspace{1.5cm}
\includegraphics[height=2in]{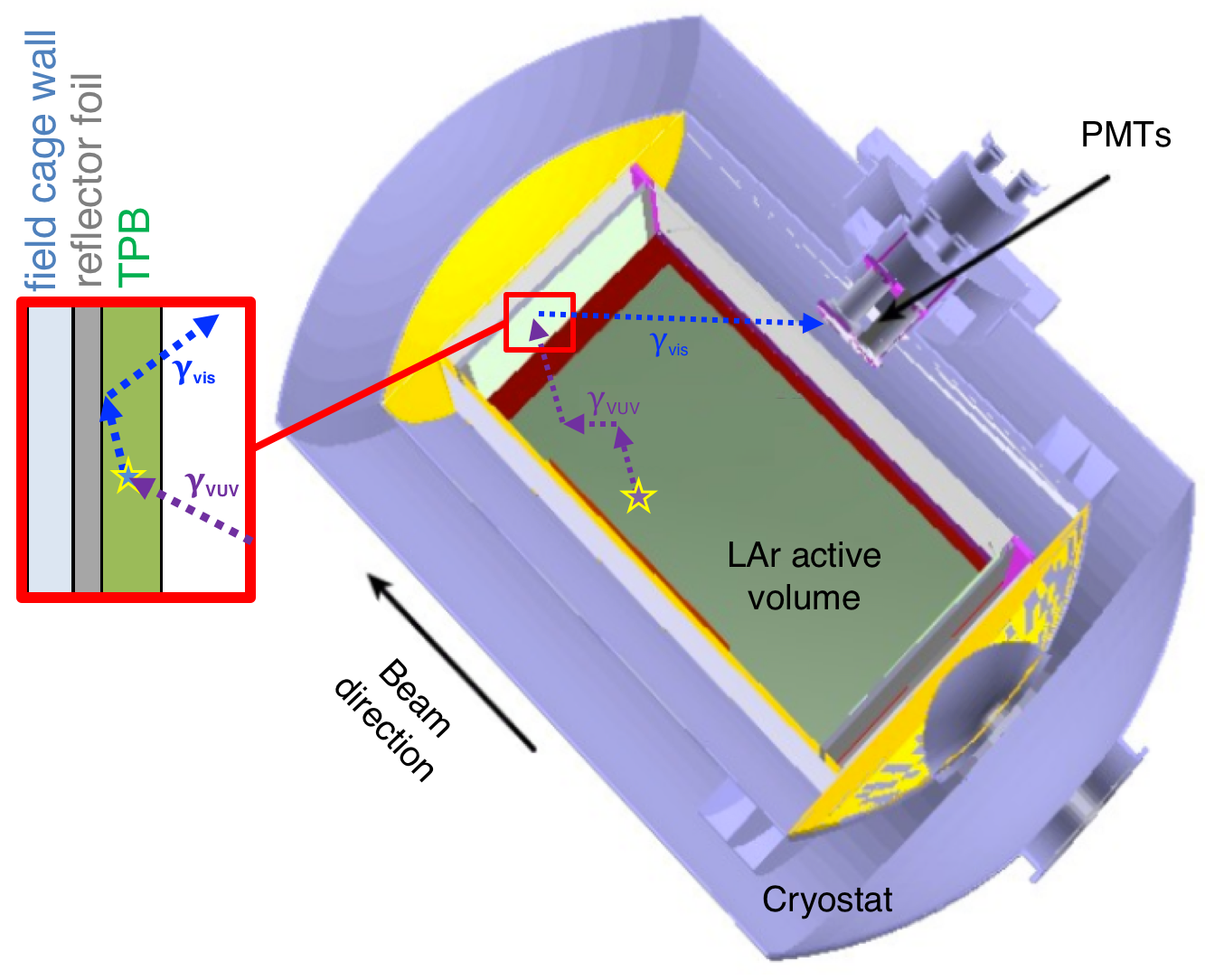}
\caption{LArIAT's photodetector system for collecting light inside the TPC (left), and a schematic which illustrates a VUV scintillation photon propagating from an energy deposition site, undergoing Rayleigh scattering, and subsequently being wavelength-shifted and reflected by TPB-coated foils (right).}
\label{fig:lightsys}
\end{figure*}

LArIAT (Liquid Argon In A Testbeam) is a LArTPC that ran in a charged particle beamline at Fermilab's Test Beam Facility~\cite{ftbf} from 2015 to 2017.  Its cryostat and TPC, shown in Fig.~\ref{fig:tpc}, were inherited from ArgoNeuT~\cite{argoneut}. The TPC's 170-liter active volume is 40~cm tall ($\hat{y}$) and 90~cm long in the beam direction ($\hat{z}$) with a width of 47.5~cm along the electron drift direction ($\hat{x}$). New wire planes and readout electronics were installed on the TPC. The wire planes are comprised of one nonreadout-instrumented 225-wire shield plane, as well as 240-wire induction and collection readout planes, each with an in-plane wire separation of 4~mm.  Wires on the induction and collection planes are oriented at $\pm60^{\circ}$ relative to the beam direction.  At the nominal electric field strength of $\Efield = \SI{484}{V/cm}$, the total electron drift time is approximately $\SI{320}{\mu s}$. Wire signals are digitized with a sampling period of $\SI{128}{ns}$.

LArIAT's photon detection system is unique among existing LArTPCs. To match the spectral sensitivity of most photodetectors, light collection in LAr typically relies on the use of wavelength-shifting tetraphenyl butadiene (TPB) to down-convert the VUV scintillation photons ($\lambda$ = \SI{128}{nm}~\cite{scintoflar}) into the visible regime ($\lambda \approx$~\SI{430}{nm}).  To accomplish this in LArIAT, the four walls of the TPC field cage are lined with highly reflective dielectric foils that have been evaporatively coated with a layer of TPB. Compared to more traditional methods where TPB is coated on or suspended in front of the windows of the photodetectors, LArIAT's use of reflector foils increases the average light yield by a factor of $\approx$~2 and improves LY spatial uniformity within the active volume.

The wavelength-shifted light is then detected primarily by two cryogenic photomultiplier tubes (PMTs): a 3-inch-diameter Hamamatsu (HMM) R-11065 and a 2-inch-diameter Electron Tubes Limited (ETL) D757-KFL.   These PMT models were previously tested at cryogenic temperature for WArP~\cite{pmt_tests}.  Each PMT is suspended behind the wire planes with about 5~cm of clearance using a plastic support structure, shown in Fig.~\ref{fig:lightsys}, attached to the side access flange of the cryostat. Prior to LArIAT's Run~II, a translucent film of a TPB/polystyrene solution was added to the window of the ETL PMT, allowing some of the VUV scintillation light to down-convert directly at the face of the PMT.  Optical signals for each triggered event are digitized with a sampling period of \SI{1}{ns} for a duration of $\SI{28}{\mu s}$ using a CAEN~V1751 digitizer modified to have a dynamic range of \SI{200}{mV}~\cite{warp_fast_dig}.

\begin{figure}
\centering
\includegraphics[width=0.95\columnwidth]{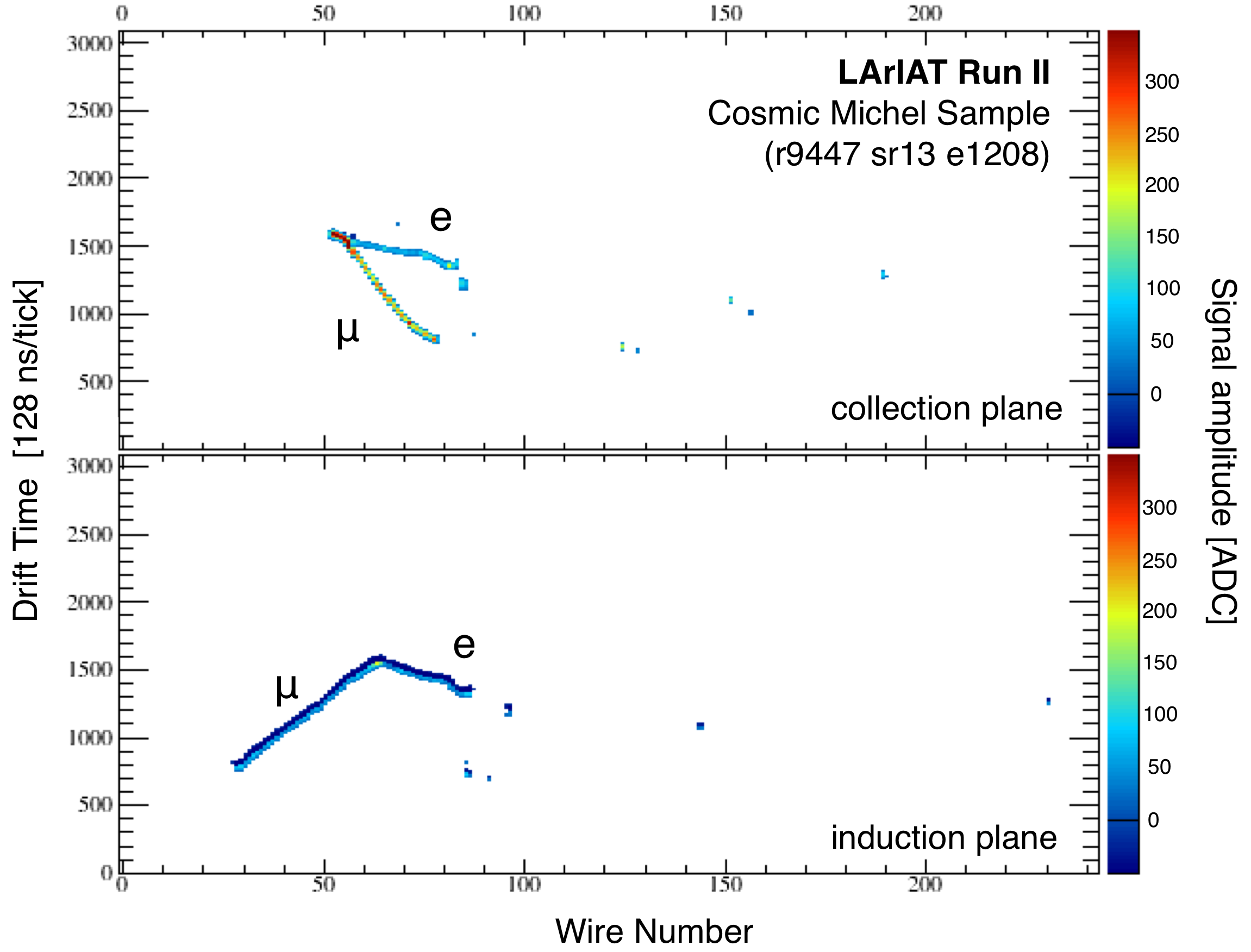}
\caption{
A stopping muon candidate from data, with its decay electron, acquired by the Michel electron trigger. Vertical columns of pixels represents the raw signal collected on each wire over the drift time.  Samples less than 20 ADC in absolute value are uncolored.}
\label{fig:evt_display}
\end{figure}

In LArIAT, decay electrons from at-rest muons, known as \emph{Michel electrons}, serve as a proxy for $\nu_e$-CC final states.  Michel electrons have a well-measured energy spectrum in the range of 0-53~MeV~\cite{nevis}, closely approximating the expected energy range of solar or supernova $\nu_e$. To acquire a sample of Michel electrons in LArIAT, a hardware-level trigger~\cite{lariat_detpaper} was set up to prompt readout of events in which both PMTs observe a delayed double-pulse topology, with a delayed coincidence window set to accept a maximum pulse separation of $\SI{7}{\mu s}$.  This trigger was used during a dedicated period of cosmic readout following each beam spill in order to select events containing cosmic muons that stop and decay within the active volume of the TPC. A Michel electron candidate event from LArIAT is shown in Fig.~\ref{fig:evt_display}.    

The following sections cover the reconstruction and analysis of a sample of Michel electrons collected over a 10-day period during LArIAT's Run~IIB when the light-based trigger was stable and functioning optimally.

%#################################################
% Section 2: Reconstruction of Michels
%#################################################
\section{\label{sec:reco} Michel Electron Reconstruction} 

Here we review the process of identifying and reconstructing Michel electrons in our sample.   Data processing is performed in LArSoft~\cite{larsoft}, a software framework containing algorithms and modules tailored for common LArTPC reconstruction tasks.  

\subsection{Charge clustering and shower reconstruction}

Wire signals from both planes are first de-convolved with the known charge response function of the LAr preamplifiers,
and the resulting unipolar 
%Gaussian-shaped 
pulses corresponding to charge depositions are identified as {\it hits} and fit to Gaussians.  Using LArSoft's general-purpose clustering and tracking algorithms~\cite{trajcluster,pmtrack}, linelike groups of wire hits are formed into 3D tracks.  Events with a track extending from the TPC boundaries to a point within a fiducial volume are tagged as stopping cosmic muon candidates.

Distinctive characteristics of the muon-plus-electron event topology help identify the boundary between charge from the muon and the decay electron. For instance, a muon deposits an increasing amount of energy per unit length as it loses energy, resulting in a visible Bragg peak just prior to its stopping point. In addition, the outgoing decay electron will emanate in a random direction, often creating a visible ``kink'' in the spatial pattern of charge.  A cluster profiling procedure adapted from MicroBooNE~\cite{microboone_michels}, with modifications to account for LArIAT's smaller size, is used to search for these two defining features. 

\begin{figure}
\centering
\includegraphics[width=\columnwidth]{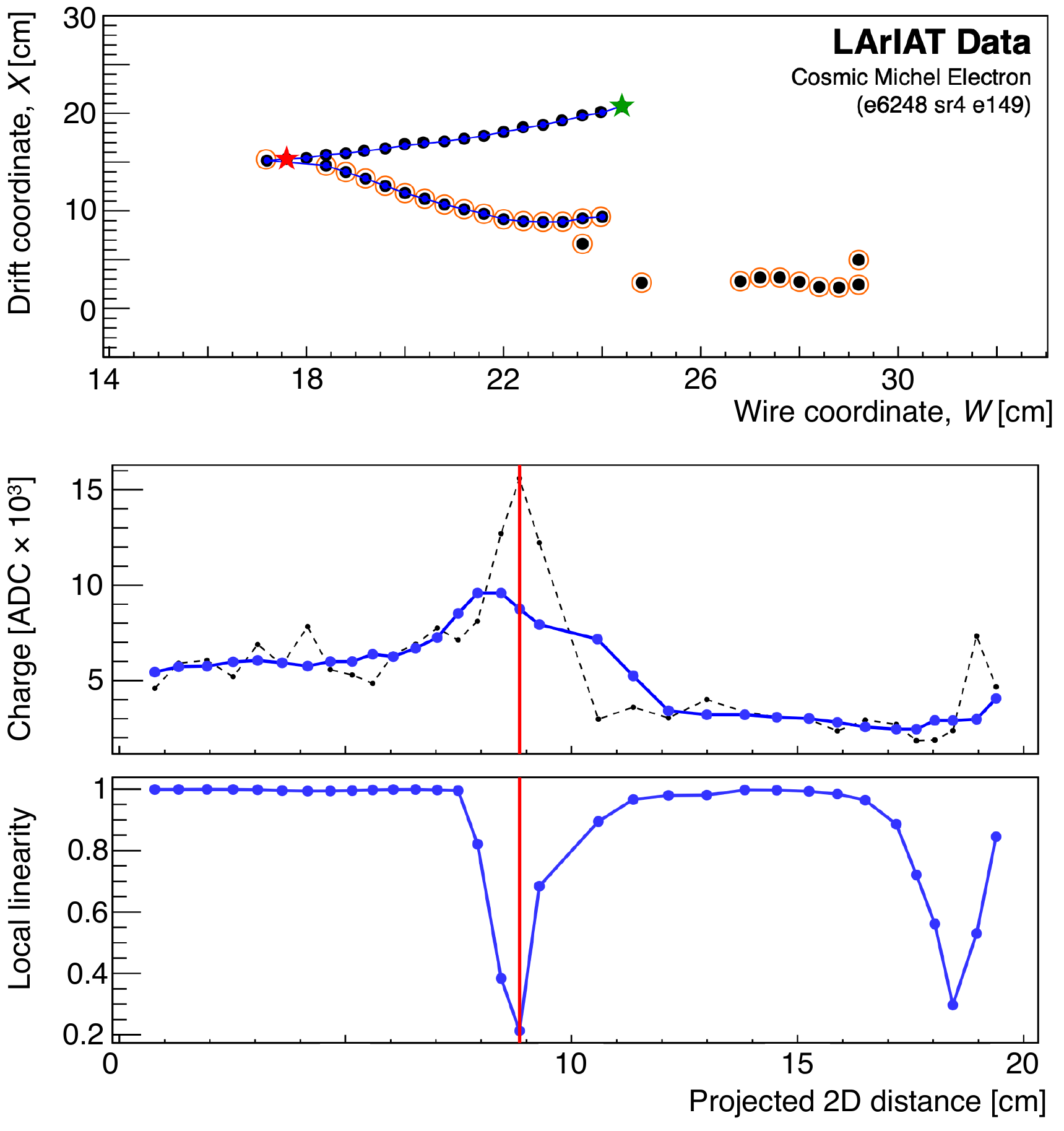}
\caption{An example of the 2D Michel cluster profiling procedure on the collection plane.  In the top plot, wire hits are represented as filled circles. The blue circles connected by lines indicate clustered hits.  The green star marks the hit used as the starting point for the clustering, and the red star shows the identified muon endpoint.  Hits included in the final Michel electron shower are circled in orange. The two lower plots show the charge and linearity profiles as functions of the cumulative distance \emph{measured along the 2D cluster}, with a vertical red line marking the muon endpoint in each. For the charge profile in the middle plot, the truncated mean hit charge is shown in blue while the dotted black line traces the charge of each individual hit.  
%The local linearity along the cluster is shown in the bottom plot.}
}
\label{fig:cluster_demo}
\end{figure}

First, wire hits are mapped into a 2D space of wire coordinate ($W$), defined as the wire number multiplied by the wire separation distance, and drift distance coordinate ($X$) calculated as $X = t\times v_d$ where $t$ is the hit's drift time and $v_d$ the electron drift velocity.   A proximity-based clustering is performed within this 2D W-X space on each wire plane, starting from the hit corresponding to the entrance point of the candidate muon track. This cluster will naturally include hits from both the muon and the decay electron in proper sequential order.  A charge density profile is constructed using the local truncated mean charge in a neighborhood surrounding each hit along the cluster trajectory. A profile of local linearity or spatial covariance along the cluster is also constructed to quantify deviations from a perfectly linear trajectory. The local linearity at hit $i$, $\chi^2_i$, is calculated as
\begin{equation}
\chi_i^2 = \frac{1}{N \sigma_x \sigma_w} \left| \sum\limits_{i-b}^{i+b} \left[ (X_i - \bar{X})\times(W_i - \bar{W}) \right]\right|,
\end{equation}
where $b$ is the number of hits defining the local neighborhood window, $N = 2b+1$ is the total number of hits in the neighborhood, and $\sigma_x$ and $\sigma_w$ are the standard deviations of the $X$ and $W$ coordinates.  The kink of the decay electron emanating from the end of the muon track should produce a sudden deviation in linearity.  An example of these charge and linearity profiles for an event are displayed in Fig.~\ref{fig:cluster_demo}.

The proximity of the stopping muon's Bragg peak and the kink formed by the decay electron is used to identify the muon-electron boundary:
\begin{enumerate}
\item First, the maximum in the truncated mean charge along the cluster is located. Then the hit with the highest individual raw charge within a local neighborhood of this point is identified as the boundary candidate.
\item If the local linearity at the candidate boundary hit is less than some threshold, then this {\it boundary hit} divides the cluster into a muon segment and an electron segment.
\end{enumerate}

A vector is fit to hits in regions of high local linearity at the end of the muon segment of the cluster to determine the terminal direction of the muon, $\vec{d_{\mu}}$. A vector corresponding to the outgoing electron, $\vec{d_e}$, is then drawn from the muon endpoint to the charge-weighted mean (in W-X space) of the electron-tagged hits.  The 2D decay angle ($\theta_{2D}$) is defined as the angle between $\vec{d_\mu}$ and $\vec{d_e}$.  All hits included in the electron-tagged portion of the cluster, in addition to all other previously unclustered hits falling within a 2D ``cone'' with opening angle $\delta\theta = 30^{\circ}$ directed along $\vec{d_e}$, are grouped into the Michel electron \emph{shower}. In Fig.~\ref{fig:cluster_demo}, hits belonging to the electron shower are outlined in orange. In principle, this will include both the direct ionization from the electron as well as any displaced charge deposited by bremsstrahlung photons emitted by the electron.

The 2D clustering and shower reconstruction procedures are repeated on both the collection and induction wire planes. Attempts are then made to form 3D ``space points'' within the shower by combining information from both wire planes.  If a shower is successfully reconstructed on both planes, interplane hits separated in time by less than some threshold, defined by their respective hit widths, are paired up starting with those that are most closely matched in time. For each pair of matched wire hits, a $Y$ and $Z$ coordinate is calculated from the intersection of the two wires. Together with the drift coordinate $X$, a full 3D space point is thus formed.

% **********************************************************************
\subsection{Optical reconstruction}

We now turn to the reconstruction of optical signals for events in LArIAT's Michel electron sample.  The \emph{prompt} singlet-state $\Ar_2^*$ component of a typical optical pulse in LArIAT is characterized by a rise time of
$\sim$\SI{10}{ns} and a width of $\sim$\SI{50}{ns} at half-max, with pulse amplitudes in the range of $\sim$50-500~ADC counts. Each prompt pulse is followed by an exponential tail of \emph{late} light from triplet-state $\Ar_2^*$ decays ($\tau_t~= \SI{1.3}{\mu s}$) and delayed TPB fluorescence~\cite{segreto}.

\begin{figure} 
\centering
\includegraphics[width=0.9\columnwidth]{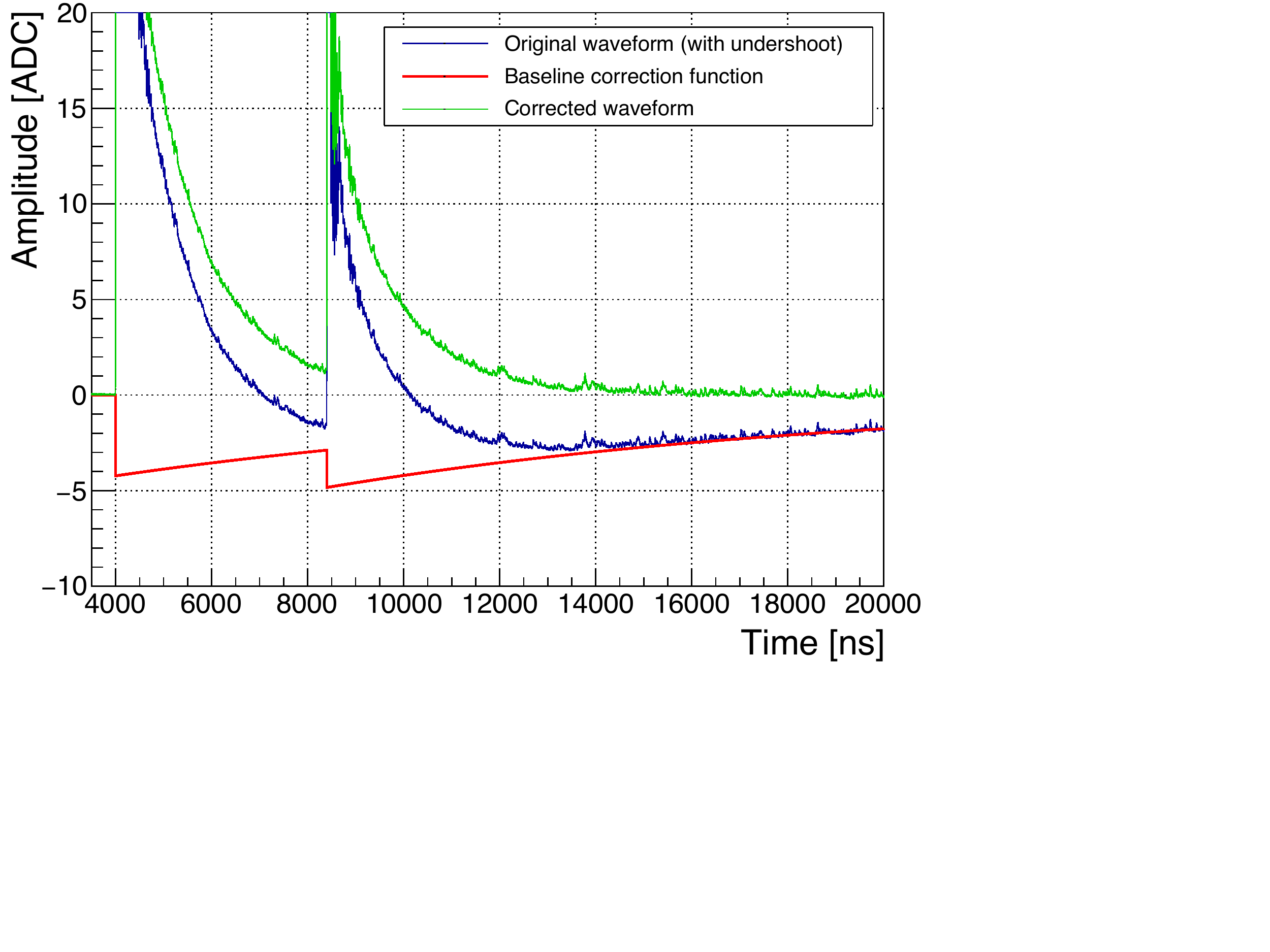}
\caption{A demonstration of the undershoot correction for a representative \emph{inverted} HMM PMT signal from a Michel electron event containing two identified optical pulses. A modified baseline (red) is constructed based on information from each identified pulse in the original PMT waveform (blue). The corrected waveform is shown in green. The amplitude range of the plot is intentionally limited to emphasize the scale of the baseline correction.}
\label{fig:undershoot_corr}
\end{figure}

%Waveforms from the PMTs are first smoothed ($\pm$\SI{1}{ns} averaging), pedestal-subtracted, and inverted to appear positive-going.
To prevent a systematic mismeasurement of pulse areas due to a negative undershoot observed in the HMM PMT signal, a simple correction is crafted by exploiting an observed linear relationship between the integral around the peak of each pulse and the amplitude of the undershoot that follows it.  % when modeled as a negated falling exponential. 
Inverted average waveforms from the HMM PMT were compiled for nine ranges of pulse amplitudes and each waveform was fitted from 9~$\mu$s to 18~$\mu$s to the function $f(t) = -A e^{-t/\tau_r}$. The nine fitted baseline recovery lifetimes, $\tau_r$, were averaged to determine $\tau_{r}^{\text{ave}} = 11.5$ $\mu$s.  The fits were repeated with the recovery lifetime fixed to $\tau_r = \tau_r^{\text{ave}}$ and the normalization parameter $A$ was recorded for each.  A direct linear relationship %shown in Fig.~\ref{fig:overshoot_corr} 
was found between this overshoot normalization and the 50-ns integral of the pulse peak.  To apply the undershoot correction during reconstruction, for each  pulse found on the waveform, a baseline correction of
\begin{equation}
%f^{\text{BS}}_{\text{corr}}(t) = \begin{cases}
%	0  											& t < t_i \\
%	-A \cdot e^{-(t-t_i)/\tau_r^{\text{ave}}} 		& t \ge t_i,
%\end{cases}
f^{\text{BS}}_{\text{corr}}(t) = -A_i \cdot e^{-(t-t_i)/\tau_r^{\text{ave}}}
\end{equation}
is subtracted from all samples with $t > t_i$, where $t_i$ is the reconstructed start time of the rising edge of pulse $i$ and $A_i$ is determined from the pulse's peak integral.  %Since this is repeated for each pulse, the total baseline correction is cumulative. 
This correction procedure is illustrated in Fig.~\ref{fig:undershoot_corr}.

\begin{figure}
\includegraphics[width=0.98\columnwidth]{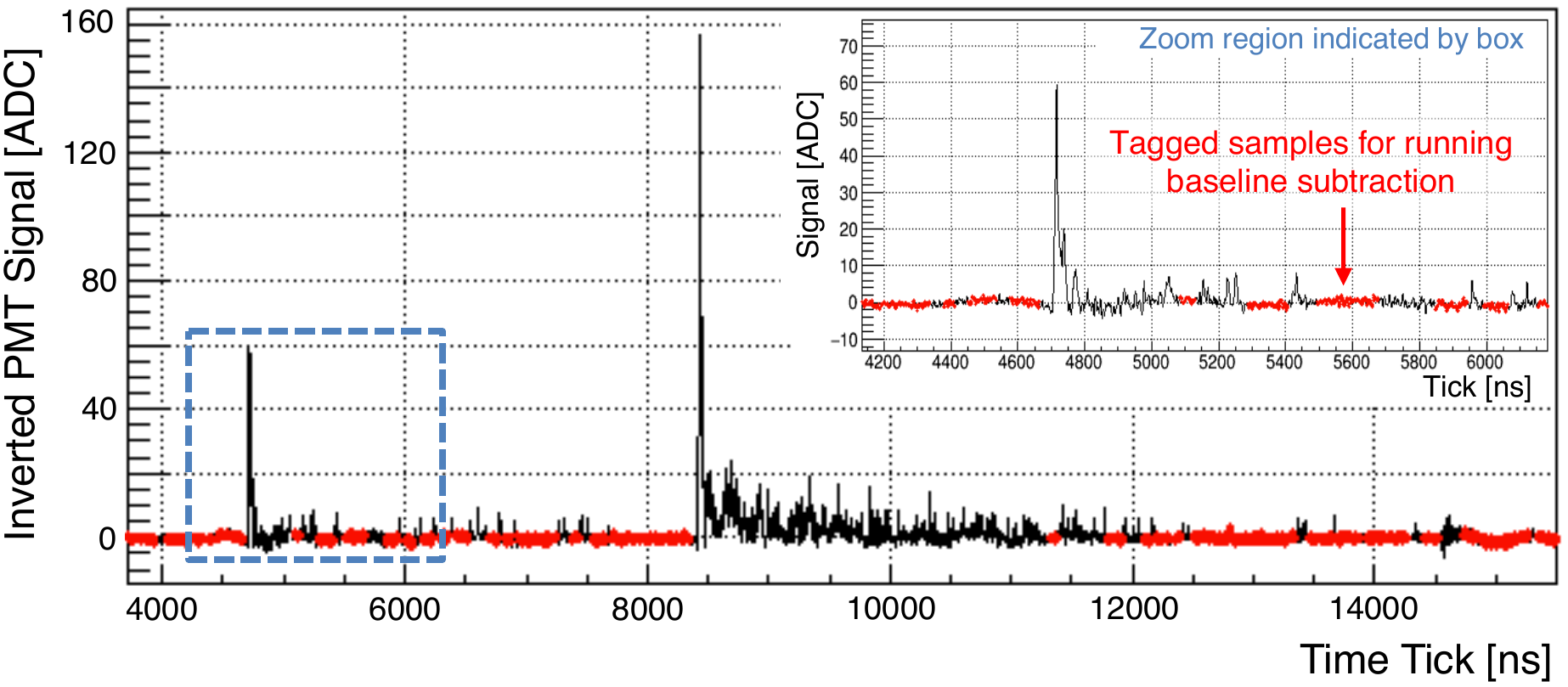}\\
\vspace{0.3cm}
\includegraphics[width=0.98\columnwidth]{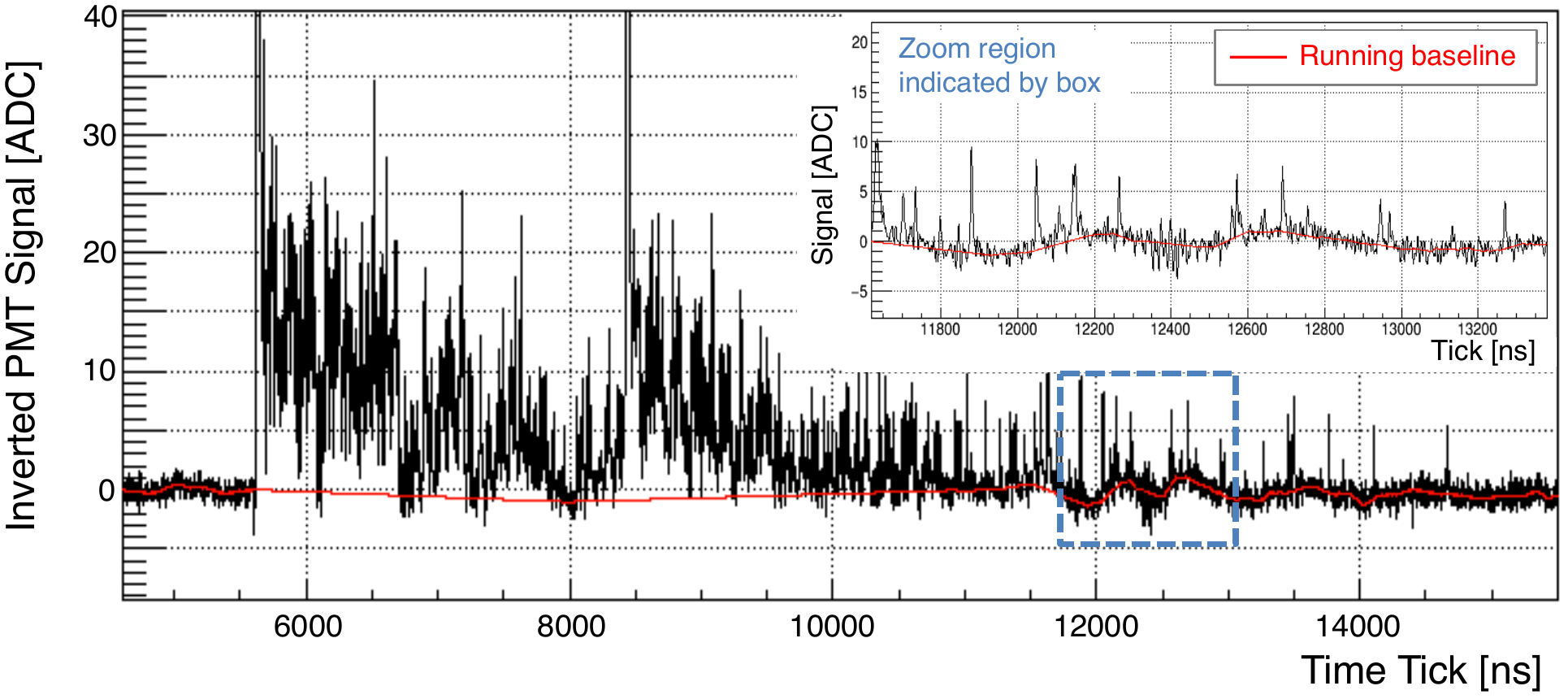}
\caption{Tagged waveform regions used in the calculation of the running baseline (top), as well as the final computed running baseline (bottom), for an example waveform.}
\label{fig:runningbaseline}
\end{figure}   

To remove slow oscillations in the PMT signal, a modified running baseline is constructed using a gradient-based technique that masks out regions where activity is detected.  This process is illustrated in Fig.~\ref{fig:runningbaseline}.  
First, a signal gradient $g(t)$ is computed at each sample in the signal $s(t)$ using a 5th-order central numerical differential,
\begin{equation}
g(t_i) = \frac{1}{12} \Big[ s(t_{i-2}) - 8s(t_{i-1}) + 8s(t_{i+1}) - s(t_{i+2}) \Big].
\label{eq:gradient}
\end{equation}
The RMS of the gradient, $\sigma_g$, is calculated over a 1-$\mu$s baseline region on the waveform preceding any optical activity.  Samples with $s(t) > 5$ ADC or $g(t) > 3 \sigma_g$ are marked as the start of regions containing activity, and the following 50 samples are designated as ``active.''  A standard running baseline is computed over the inactive or ``quiet'' regions using a truncated mean taken from quiet regions that lie within 50 clock ticks of the current sample.  A linear interpolation extends this running baseline across tagged active regions, connecting the first sample of each quiet region to the last sample of the preceding quiet region. This running baseline is then subtracted from the waveform.

\begin{figure}
\includegraphics[width=0.95\columnwidth]{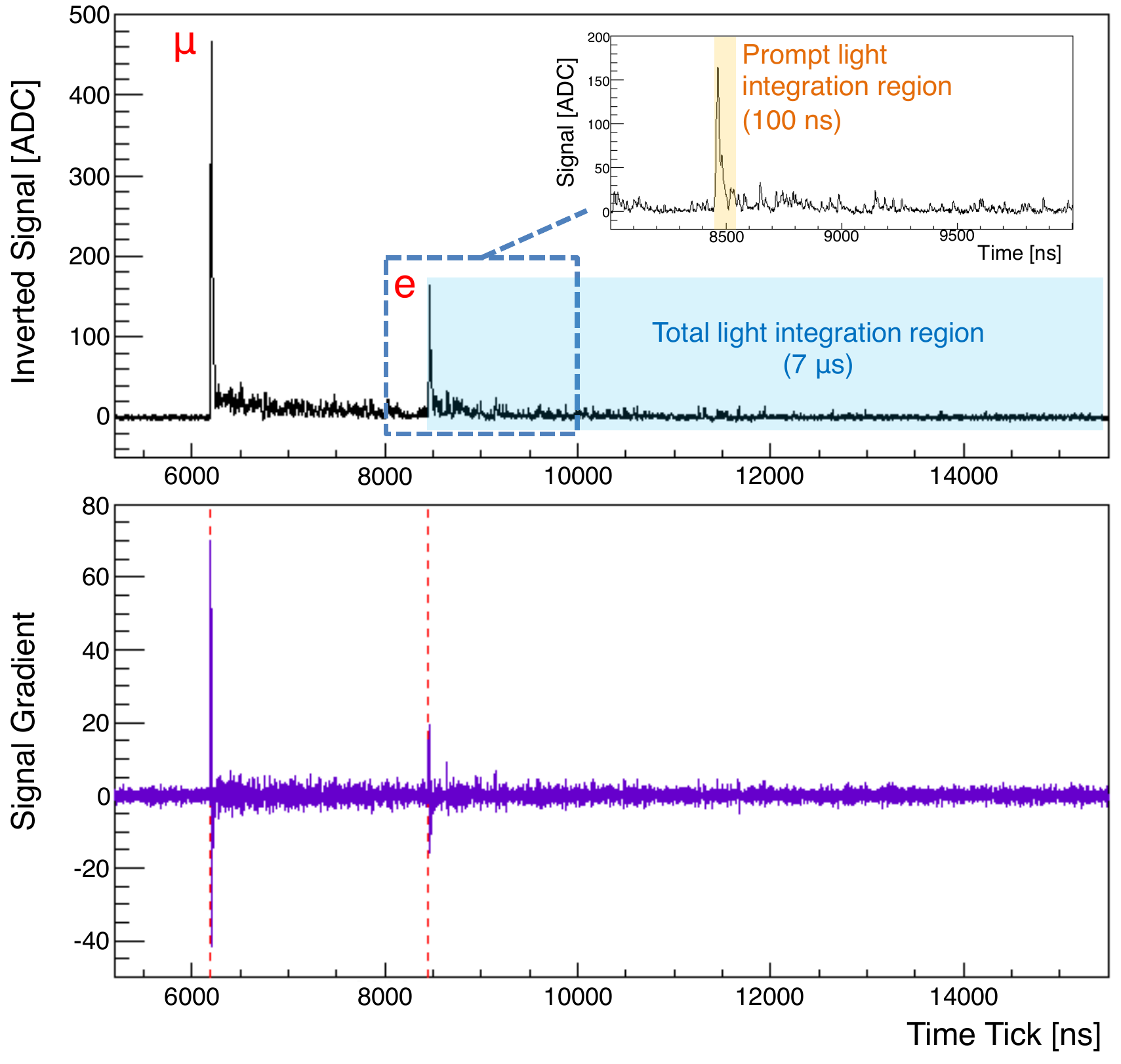}
\caption{An optical signal from the ETL PMT for a Michel electron candidate event. The prompt and total light integration regions are highlighted.}
\label{fig:michelwfm}
\end{figure} 

Following waveform cleanup, $g(t)$ is thresholded to identify \emph{optical hits}.  If there are exactly two optical hits found, the muon decay time $\Delta T$ is measured as the difference between the time of the first (muon candidate) and second (Michel electron candidate) pulses as illustrated in Fig.~\ref{fig:michelwfm}. An integral of the first \SI{100}{ns} of the electron candidate pulse determines the prompt light, which is labeled $S_\textrm{100}$, while an integral over $\SI{7}{\mu s}$ determines the total light, $S_\textrm{total}$. The integration start point precedes the pulse's identified hit time by \SI{5}{ns} in both cases. 

Late light from the muon will naturally overlap with the Michel candidate integration window. This contamination is exaggerated by the time structure of TPB fluourescence. Attempts are made to correct for this by fitting the muon light in the range 0.4~$\mu$s to 1.8~$\mu$s to a function approximating the distribution of late light,
	\begin{equation}\label{eq:s_late}
	S_{\text{late}}(t) = Ae^{-t/\tau_t^{\prime}} + Be^{-t/\tau^*},
	\end{equation}
where $A$ is treated as a free parameter and $\tau_t^{\prime}$ is set to the quenched triplet excimer lifetime inferred from fits to average muon waveforms for different data-taking periods (see Sec.~\ref{sec:sim-ql}). The second term in $S_{\text{late}}$ is meant to account for the component of TPB fluorescence with the longest lifetime, $\tau^*$ = 3.55 $\mu$s, which comprises 8\% of the re-emitted visible light~\cite{segreto}.  %This additional late TPB light is not effectively modeled by a simple single-exponential fit. 
The normalization factor $B$ is fixed to values based on fits of Eq.~\ref{eq:s_late} to average waveforms from independent samples of cosmic muons that cross the TPC. % for different data-taking periods.  
The fitted function $S_{\text{late}}$ is extrapolated through the integration regions of the Michel pulse to get an estimation of the muon late-light contamination, which is subtracted from the $S_\text{100}$ and $S_\text{total}$ integrals accordingly.

To validate the data sample, the measured distribution of muon decay times is fitted to extract the negative muon lifetime and estimate the sample purity. Based on a flat background term, the sample is estimated to have a purity of $>95\%$. The relative normalization of fit components representing the positive and negative muon populations is also used to estimate the cosmic muon charge ratio. These calculations are detailed in Appendix~\ref{sec:mudecaytimes}.

%#################################################
% Section 3: Simulation
%#################################################
\section{\label{sec:sim} Simulation} 

An accurate simulation of the LArIAT detector is critical for assessing our energy resolution. The Geant4~\cite{geant4} framework is used to generate and propagate a sample of cosmic muons through the LArIAT geometry following a $\cos^2\theta$ angular distribution relative to the zenith. The momenta of generated muons is limited to a range of $p=$~50-300~MeV/c in order to maximize the number of muons that stop and decay in the detector. The simulation of charge deposition, recombination, drift, and detector response is then handled by LArSoft~\cite{larsoft}.

\begin{figure}
\centering
\includegraphics[width=0.85\columnwidth]{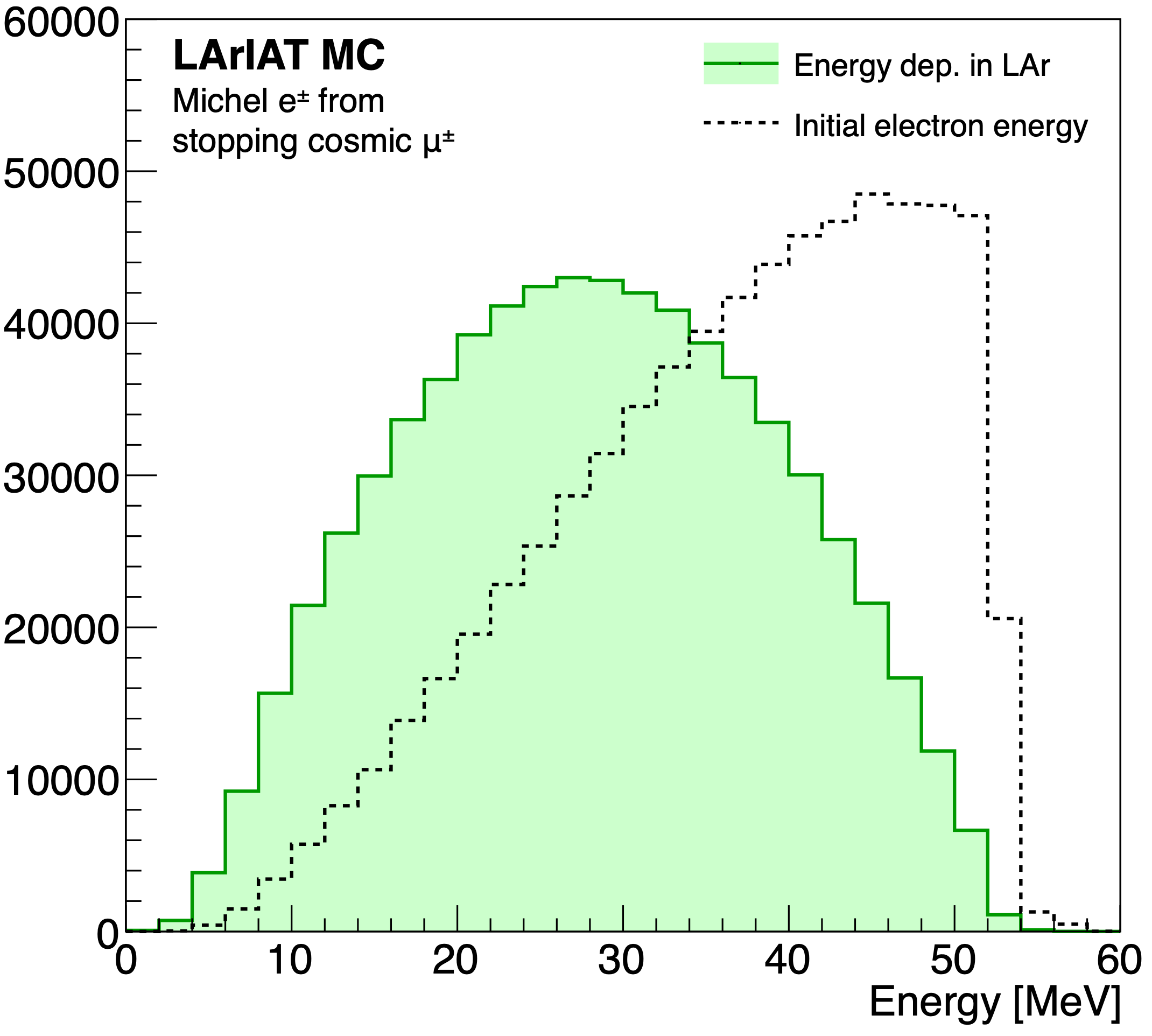}
\caption{Simulated distribution of the initial Michel electron energy ($\langle E \rangle = 36.7$~MeV) compared to the visible energy deposited in the LArIAT active volume ($\langle E_{\text{dep}} \rangle = 28.2$~MeV).}
\label{fig:energydep}
\end{figure}

Due to LArIAT's limited size, many bremsstrahlung photons emitted by the Michel electron escape the LAr active volume before pair-producing or Compton scattering, or the electron itself will sometimes leave the TPC.  Because these effects limit the detectable energy as shown in Fig.~\ref{fig:energydep}, the characteristic energy cutoff in the Michel spectrum is not resolvable without detailed containment corrections, which are not attempted in this analysis. 

\subsection{\label{sec:sim-ql} Charge and light production} 
\begin{figure}\centering
\includegraphics[width=0.95\columnwidth]{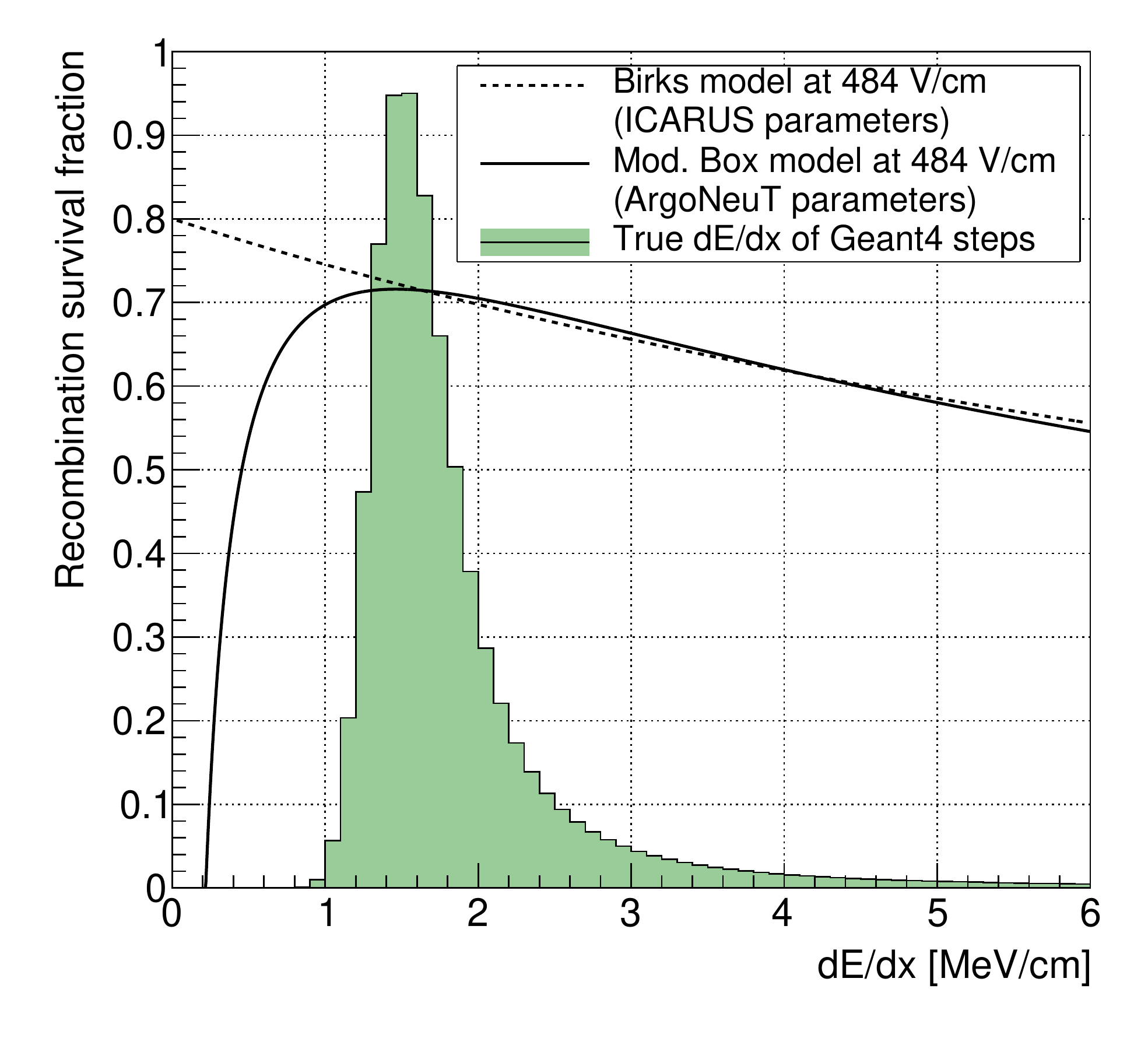}
\caption{The Birks (dashed line) and Modified Box (solid line) recombination models as functions of $dE/dx$ for LArIAT's field of $\Efield$ = 484~V/cm. Parametrizations from fits to data from the ICARUS \cite{icarus_recomb} and ArgoNeuT \cite{argoneut_recomb} experiments are used. An arbitrarily-normalized distribution of $dE/dx$ for Geant4 particle steps is overlaid.}
\label{fig:recombcurves}
\end{figure}

Each energy deposition simulated in Geant4 is first apportioned to $\Ar_2^*$ excimers ($\Nex$) and electron-ion pairs ($\Ni$) according to the excitation ratio in LAr, $\alpha = \Nex/\Ni~=~0.21$~\cite{w_ion, aprile_book}.  
The number of ionization electrons surviving recombination ($N_e$) is determined from the $dE/dx$ of the particle step using one of two parametrized models: the Modified Box Model for $dE/dx \gtrsim$~\SI{1.7}{MeV/cm}, since it more accurately describes data at higher ionization densities~\cite{argoneut_recomb}; and the Birks Model~\cite{icarus_recomb} for smaller $dE/dx$ where the Modified Box model starts to fail as illustrated in Fig.~\ref{fig:recombcurves}.   Electrons are then drifted to the wireplanes with an impurity attenuation lifetime, $\tau_e$, set to match average values measured from independent samples of cosmic muons in data that cross the long diagonal of the TPC~\cite{lariat_detpaper}.

The final number of photon-producing $\Ar_2^*$ generated in the ionization ($N_{\gamma}$) is determined from Eq.~\ref{eq:QL} such that the anticorrelation expected between $N_{\gamma}$ and $N_e$ is preserved after recombination.  Excimers are then divided into fast (singlet) and slow (triplet) populations using ratios from literature.  A singlet-to-triplet ratio of $I_s/I_t = 0.51$ is applied for muon and electron-induced ionization while $I_s/I_t = 0.3$ is used for ionization induced by bremsstrahlung photons that pair-produce or Compton scatter~\cite{hofmann,singlet-to-triplet}. 

Nonradiative destruction or \emph{quenching} of $\Ar_2^*$ excimers can occur through collisions with residual impurities like O$_2$ and N$_2$. Quenching competes with radiative decay, reducing the effective scintillation lifetime to $\tau_i^\prime = (1/\tau_q + 1/\tau_i)^{-1}$, where $\tau_q$ is the quenching timescale and $i$ denotes either the triplet or singlet excimer state. This quenching naturally reduces the yield of photon-producing excimers by a factor $\tau_i^\prime/\tau_i$ for both the singlet and triplet populations~\cite{quenching_o2,quenching_n2}. The effective quenched triplet lifetime $\tau_t^\prime$, provided as an input parameter to the simulation, is based on average values observed in data.

Since LAr scintillation time structure is convolved with the four components of TPB fluorescence, with lifetimes ranging from $\approx$~\SI{1}{ns} to as high as $\SI{3.55}{\mu s}$~\cite{segreto}, a conversion is needed between the measured late-light lifetime from waveforms to the ``true'' $\tau_t^\prime$. To accomplish this, a Monte Carlo (MC) study was performed in which photon arrival-time distributions were generated, incorporating LAr physics and TPB reemission effects (as described in the following subsection), with different levels of quenching applied.  A linear relationship between $\tau_t^\prime$ and the measured lifetime, $\tau_{\text{meas}}$, was determined by fitting averaged waveforms to an exponential in the range of 0.4-2~$\mu$s: 
\begin{equation}\label{eq:efftau}
\tau_t^\prime  \left[ \text{ns}\right] = [0.965]\tau_{\text{meas}} - [41.7 \text{ ns}].
%a &= 0.965 \nonumber \\
%b &= 41.7\text{ ns} \nonumber
\end{equation}
This $\tau_t^\prime$ is used to calculate $\tau_q$, thus enabling the calculation of the effective singlet decay time, $\tau_s^\prime = (1/\tau_q + 1/\tau_s)^{-1}$, and the overall yield reduction factors necessary for proper simulation of $N_\gamma$ quenching. 

Because the characteristic quenching timescale in LAr is usually long compared to that of the singlet-state emission, this prompt scintillation component is affected very little by the quenching.  For typical measured late-light lifetimes in the data, $\tau_{\text{meas}} \approx \SI{1200}{ns}$ ($\tau_t^\prime\approx \SI{1120}{ns}$), we expect negligible quenching of the fast-decaying singlet population ($\tau_s^\prime / \tau_s \approx 1$) and a $14\%$ reduction in the slower-decaying triplet population ($\tau_t^\prime / \tau_t \approx 0.86$).

% ---------------------------------------------------------------
\subsection{\label{sec:opsim} Photon propagation}

\begin{figure}
\centering
\includegraphics[width=0.95\columnwidth]{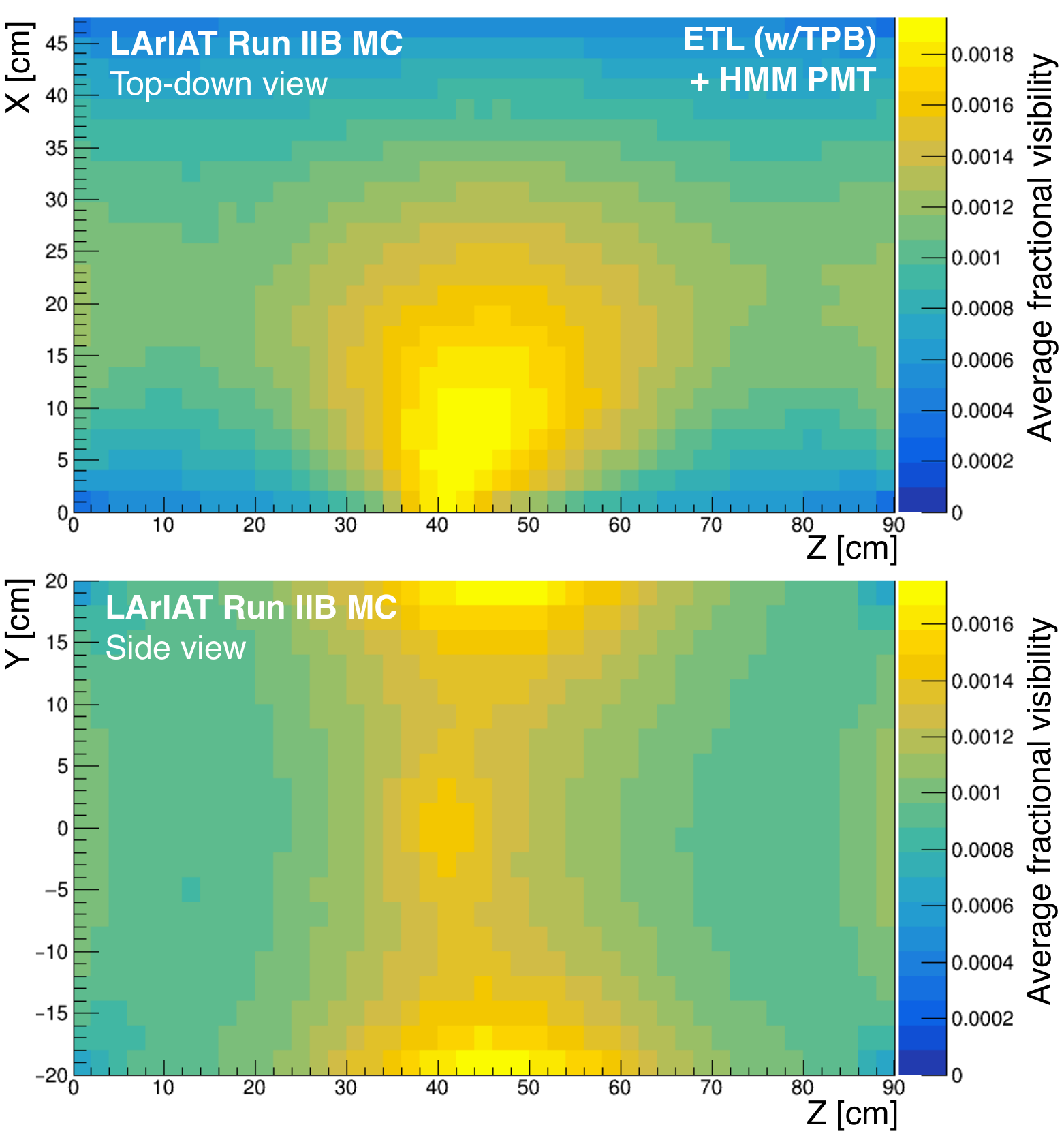}
\caption{The 2D projections of the visibility map, or the average fraction of scintillation photons reaching the PMT windows, for Run~II. 
%The distinct distribution of visibility is a consequence of the use of TPB-coated reflector foils on LArIAT's field cage walls.
}
\label{fig:vismaps}
\end{figure}

Detailed Geant-level simulation of photon propagation is computationally taxing given the hundreds of thousands of photons that need to be tracked in every event. Instead, a photon propagation MC simulation is used to generate 3D maps quantifying the photon detection probability and photon travel time for scintillation produced at all points throughout the active volume. When simulating physics events, the number of detected photons and the distribution of their arrival times for each energy deposition can be quickly determined by drawing from these maps.  

To create the photon visibility and timing maps, the active volume is subdivided into a number of 3D voxels of size $\sim$~2~$\times$~2~$\times$~2~cm$^3$. %in  2 24~(x) $\times$ 20~(y) $\times$ 45~(z) 3D voxels. 
Within each voxel, 300,000 VUV photons are emitted isotropically.  Photons are assigned energy $E_{\gamma} = 9.69\pm\SI{0.25}{\text{eV}}$, corresponding to a peak wavelength $\lambda = \SI{128}{nm}$, with a full width at half-max of \SI{8}{nm}~\cite{scintoflar}. The propagation velocity and Rayleigh scattering length are determined using parametrized functions of energy. 
The functional form for velocity was found from a linear fit in the local neighborhood of $E_{\gamma}$ to values of the energy-dependent photon group velocity, $v_g = c/[n(E) + dn/d(\ln{E})]$, where $n$ is the energy-dependent index of refraction for LAr.
The parametrization of the energy-dependent Rayleigh scattering length is a polynomial fit to recently reported values~\cite{grace}, with an average of 62~cm.

Reflection and reemission from TPB are assumed to be Lambertian with a 100\% VUV-to-visible conversion fraction and 95\% visible reflection efficiency~\cite{ardm}.  Reflections from the copper cathode are described according to the GLISUR model~\cite{geant4} using a reflectivity of 0\% for VUV light and 17\% for visible~\cite{cu-reflect}.  Visible photons are assigned a uniform velocity of 24~cm/ns and a Rayleigh scattering length of 320~m, corresponding to a wavelength of \SI{400}{nm}~\cite{grace}. The probability of transmission through each of the three wire planes is parametrized, using simple geometrical arguments, as a function of the angle of the incoming photon projected onto the plane normal to the wires.
No reflectivity is simulated at the glass windows of the PMTs, so any photon which reaches a PMT is counted.  

The fraction, per voxel, of the total simulated photons that reach each PMT as visible light is scaled by the PMT's photodetection efficiency~\cite{pmt_tests}, and saved into the visibility map as $f_{\vis}$. Metrics describing the time distribution of both visible and VUV photons -- the minimum, average, and standard deviation of arrival times -- are also saved to form timing maps.  Two-dimensional projections of $f_{\vis}$, for both PMTs combined, are shown in Fig.~\ref{fig:vismaps}.  Based on this simulation, the total fractional visibility to scintillation averaged throughout the active volume is $\langle f_{\vis} \rangle \approx 9 \times 10^{-4}$.

When simulating energy deposition by particles in the TPC, the $f_{\vis}$ corresponding to the voxel in which the deposition occurs is used to calculate the number of detected photoelectrons.  Stochastic fluctuations are simulated as well.  The mean and standard deviation of the photon propagation times corresponding to the voxel are retrieved from the map and used to generate an arrival time.  Therefore, each individual photon's final arrival time depends on the particle's time of creation, the quenched LAr scintillation lifetime $\tau_s^{\prime}$ or $\tau_t^{\prime}$, the retrieved propagation time, and the characteristic reemission time structure of TPB fluorescence.

%%%%%%%%%%%%%%%%%%%%%%%%%%%%%%%%%%%%%%%%%%%%%%%%%%%%%%%%%%%%%%%%%%%%%%%%%%%%%%%
\subsection{Optical smearing and trigger efficiency}

The PMT electronics response is not included in our simulation. To reproduce detector resolution, integrated photoelectron counts from MC are smeared to mimic fluctuations from the single photoelectron (SPE) resolution, 
$\sigma_{\text{pe}}/\text{pe}\sim0.3$, as well as from the average waveform RMS noise measured in data for each PMT.  

\begin{figure}\centering
\includegraphics[width=\columnwidth]{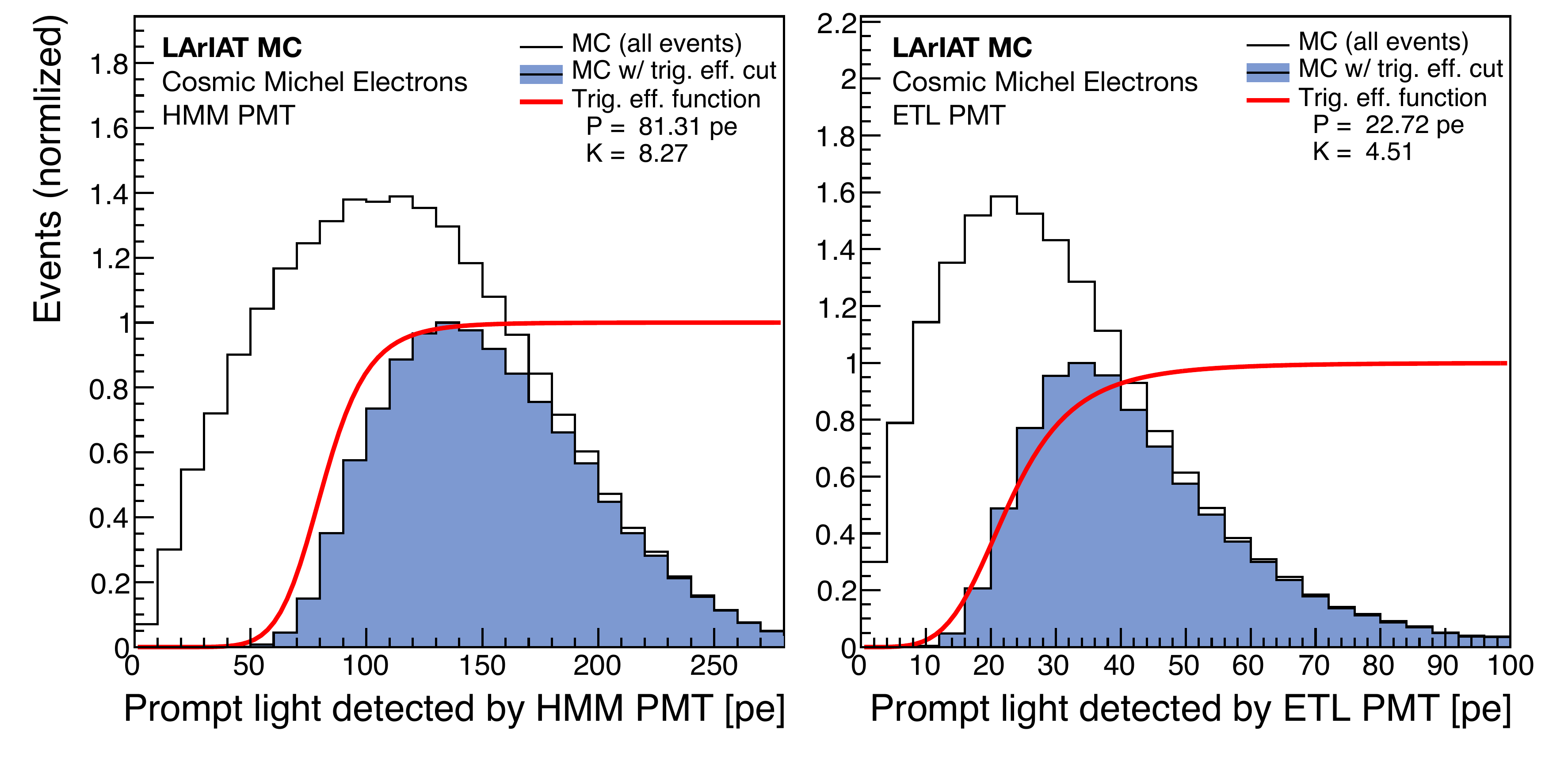}
\caption{The trigger efficiency function $f_{\text{trig}}$ (red) overlaid with its respective $S_\textrm{100}$ distribution for both PMTs.  In each case the function is scaled arbitrarily and is drawn only for illustrative purposes.  The unfilled distribution represents $S_\textrm{100}$ for all MC events, prior to the trigger cut.}
\label{fig:trigeff}
\end{figure}

Additional modifications to the reconstructed light in MC, derived from tuning to Michel electron scintillation spectra in data, are described below:

\vspace{-\topsep}
\begin{enumerate}[i)]
\setlength\itemsep{0pt}
\item \emph{Additional smearing factor} $\sigma^\prime$: This constant fractional smearing is applied only to $S_\textrm{total}$ (following muon late light corrections) and accounts for effects like intermittent noise oscillations which cannot be easily modeled.

\item \emph{Scale factor} $F_{\text{\text{scale}}}$: Considering the approximations used in the photon propagation simulation, combined with the $\pm2.5\%$ uncertainty in the reported PMT collection efficiencies~\cite{pmt_tests}, we apply a scale correction of a few percent to the simulated light distributions to match distributions in data.

\item \emph{Trigger efficiency parameters $P$ and $K$:} The Michel trigger creates a bias toward more luminous, higher-energy electrons.  To replicate this in MC, a trigger efficiency function $f_{\text{trig}}$ is used to determine the event acceptance probability based on the prompt light $S_\textrm{100}$,
\begin{equation}
f_{\text{trig}}(S_\textrm{100}) = 1 - \left[ 1 + (S_\textrm{100}/P)^K \right]^{-1}.
\end{equation}
Parameter $P$ sets the approximate cutoff point while $K$ controls the sharpness of this boundary.  A visual overlay of the trigger efficiency functions with each (arbitrarily normalized) prompt light distribution from the MC is shown in Fig.~\ref{fig:trigeff}. 

\end{enumerate}

\begin{table}
\centering
\caption{Smearing, scaling, and trigger efficiency cut parameters applied to the MC samples.}
\label{table:smearing}
\setlength{\tabcolsep}{10pt}
\begin{tabular}{l  c  c  c }
\hline\hline
Parameter				& HMM PMT 	& ETL PMT \\
\hline
$\sigma^\prime$ [pe]				
& 0.002	
& 0.238  \\
$F_{\text{scale}}$				
& 1.021	
& 0.957 \\
$P$ [pe]	
& 81.31 
& 22.72  \\
$K$							
& 8.271			
& 4.506	  \\
\hline\hline
\end{tabular}
\end{table}

\begin{figure*}
\centering
\includegraphics[width=0.65\textwidth]{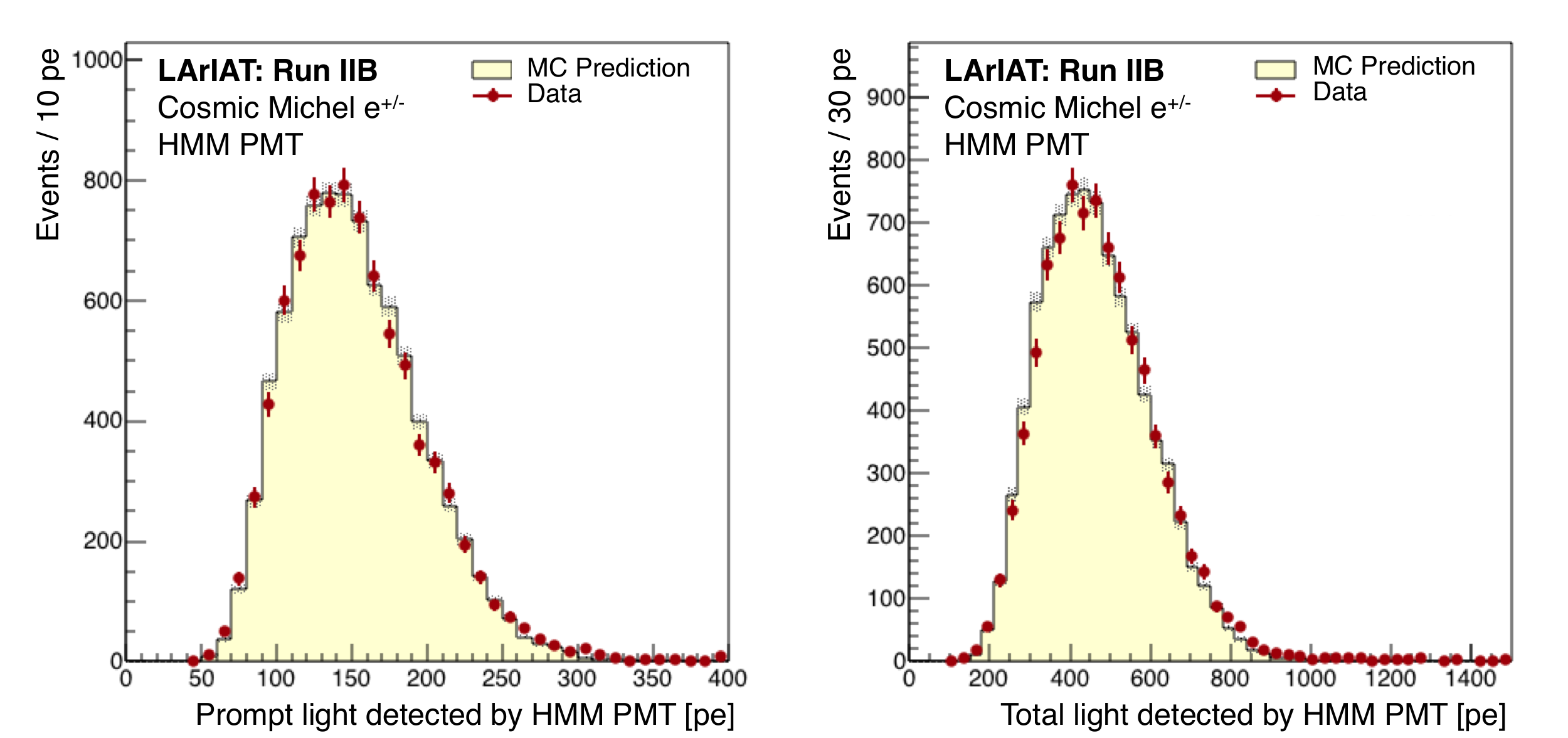}
\includegraphics[width=0.65\textwidth]{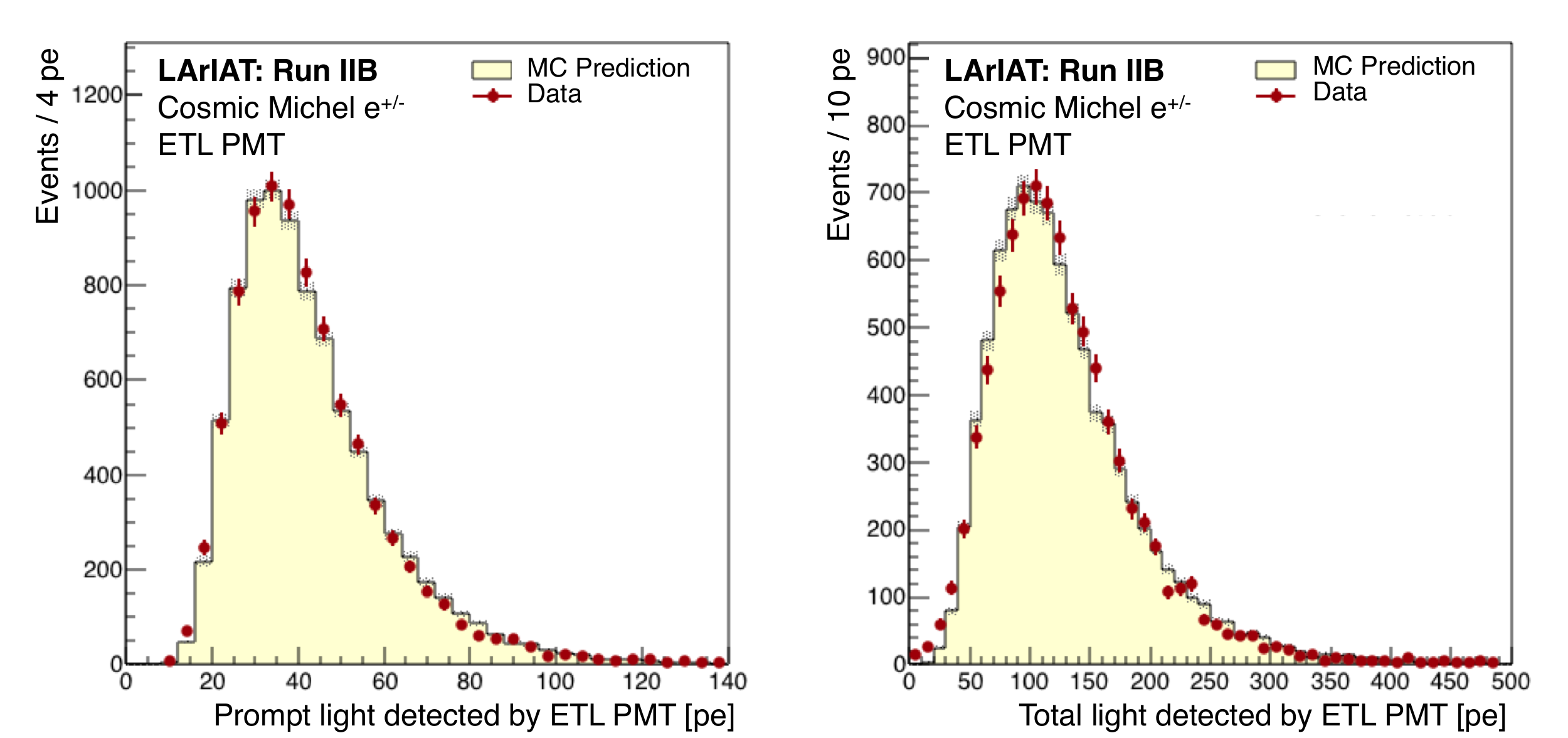}
\caption{Distributions of $S_\textrm{100}$ and $S_\textrm{total}$, in units of photoelectrons (pe), compared to MC after smearing, scaling, and trigger efficiency parameters are optimized.}\label{fig:run2_pe}
\end{figure*}

The four parameters ($\sigma^\prime$, $F_{\text{scale}}$, $P$, and $K$) are determined by fitting to the integrated light distributions in each PMT from Michel candidate events in data after applying a decay time cut of $\Delta T > \SI{1.8}{\mu s}$ and requiring the event to have a reconstructed 2D shower. 
The minimization package $\texttt{MINUIT}$ in ROOT~\cite{root} is first used to find the three-fold combination of $P$, $K$, and $F_{\text{scale}}$ that optimizes the match between $S_\textrm{100}$ distributions in data and MC. 
With the values of $P$, $K$, and $F_{\text{scale}}$ then fixed, a value of $\sigma^\prime$ is chosen to optimize the match for $S_\textrm{total}$ distributions.
The optimized parameters are summarized in Table~\ref{table:smearing}, and the resulting distributions after optimization are overlaid with data in Fig.~\ref{fig:run2_pe}. It should be noted that the ETL PMT requires a much larger additional smearing ($\sigma^\prime$) to match data compared to that of the HMM PMT. This is likely due to the lower light yield in the ETL PMT, as well as the higher level of sporadic electronic noise affecting the ETL PMT's waveforms.

%#################################################
% Section 4: Calorimetry results
%#################################################
\section{\label{sec:calo}Calorimetry Results}

%Here we describe the calorimetric reconstruction of Michel electrons.  First, the event selection and determination of energy from light and charge are outlined, including the corrections needed on both quantities to compensate for attenuation and reconstruction effects.  Then the reconstructed energy spectra for electron ionization tracks as well as for showers using the formulaic constructions of energy are presented. Based on these results, the light yield (LY) of the LArIAT TPC is estimated.  Next, we prepare a maximum-likelihood fitter tailored specifically for the Michel electron sample in LArIAT's Run~II and present the resulting energy spectrum for comparison.  Monte Carlo (MC) resolution studies are carried out to gauge the improvement in Michel electron energy resolution using the three different reconstruction methods.

Presented here is the calorimetric reconstruction of Michel electrons.  
First, the event selection cuts are outlined. Procedures for determining energy from both charge and light are then described, including corrections needed on both quantities to compensate for attenuation during propagation and drift. Reconstructed energy spectra are presented.  Studies are performed using MC to gauge the differences in the Michel electron energy resolution using the different reconstruction methods.

Finally, a sample of electrons is simulated to determine LArIAT's energy resolution for isolated electron showers, which more closely approximate the event topology expected from low-energy $\nu_e$-CC interactions.

%%%%%%%%%%%%%%%%%%%%%%%%%%%%%%%%%%%%%%%%%%%%%%%%%%%%%%%%%%%%%%%%%%%%%%%%%%%%%%%%%%%%
\subsection{Event selection}

Selection cuts are made to ensure only well-reconstructed events contribute to the final physics samples.  First, events containing a single identified stopping muon-candidate track are selected. Following this, a series of selection cuts are made based on optical reconstruction and on the results of 2D and 3D shower reconstruction. These selection cuts are outlined below, with specific cut values and event reduction statistics presented in Table~\ref{table:evtcuts}.

\vspace{-\topsep}
\subsubsection{Michel optical criteria}
\vspace{-\topsep}
This set of cuts selects good-quality events that are consistent with the expected Michel optical topology. First a cut is made excluding events with high RMS noise on the PMTs. Exactly two hits are required on both PMTs with decay times that match to within \SI{15}{ns}.  Cuts on the full-width-half-max of optical pulses are made to exclude accidental noise hits. %No saturation cut is needed since no events were found that saturate the digitizer readout scale. 
A cut on minimum and maximum allowable $\Delta T$ is made to exclude events that are not consistent with the coincidence gate width and gate delay used in the trigger.

\vspace{-\topsep}
\subsubsection{2D shower criteria} 
\vspace{-\topsep}
The muon and electron clusters on the collection plane are required to have a minimum number of wire hits ($N^{\mu} \ge 8$ hits, $N^\text{el} \ge 4$ hits). Events with a muon track of average linearity less than 0.7 are excluded.  There must be at least five hits of high linearity available to fit the muon's terminal direction.  The reconstructed 2D decay angle must fall within $15^{\circ} < \theta_\textrm{2D} < 165^{\circ}$, to exclude events where the last few hits of the muon cluster are likely to be misidentified as belonging to the electron.

\vspace{-\topsep}
\subsubsection{3D shower criteria} 
\vspace{-\topsep}
Michel electron showers must be reconstructed in both planes.  Since we do not intend to extract calorimetric measurements from the induction plane, we do not impose the same level of quality cuts on the induction plane shower as we do on the collection plane.  At least three 3D space points must be reconstructed, with at least 15\% of hits from the smaller cluster matched in time. The charge-weighted 3D centroid of the shower must be located within a fiducial volume defined by a \SI{4}{cm} margin at the edges of the TPC.

\vspace{-\topsep}
\subsubsection{Decay time criteria} 
\vspace{-\topsep}
A final decay-time cut of $\Delta T>\SI{1.8}{\mu s}$ is imposed in order to ensure sufficient separation between the muon and electron pulses such that, after the muon's late light is fit and subtracted from the electron pulse, the effect of residual contamination is negligible.

\begin{table}
\centering
\caption{Event reduction table for the cosmic Michel electron data.  The fractional change in the number of events for each cut is shown in parentheses.}
\label{table:evtcuts}

\begin{tabular}{ p{0.6\columnwidth}  r  c }
\hline\hline

Cut 
& \multicolumn{2}{c}{Number of Events}\\
\hline 

Stopping $\mu$-candidate track    	
& \multicolumn{2}{c}{179054} 	\\[5pt]

\emph{Optical criteria:}	&& \\
RMS noise $<$ 1.8 ADC					
& 174341 &(-2.6\%)\\
2 optical hits				
& 71605 &(-59\%)\\
$\Delta T$ match between PMTs 		
& 69653 &(-2.7\%)	\\
Hit widths $>$ 10 ADC				
& 68709 &(-1.4\%)  \\
%2nd pulse saturation			
%& 68,709 &(-0.0\%) 	&-- & -- \\
\SI{350}{ns} $< \Delta T <$ \SI{7200}{ns} 	
& 64741 &(-5.8\%)  \\[5pt]

\emph{2D shower criteria:}	&& \\
Cluster boundary found			
& 43316 &(-33\%) 	 \\
$N^{\mu} \geq 8, N^\textrm{el} \geq 4$			
& 33479 &(-23\%)	\\
$\mu$ linearity $>$ 0.7 	
& 31478 &(-6.0\%) 	 \\
$\mu$ direction fit points $\geq$ 5		
& 29974 &(-4.8\%) \\
$15^{\circ} < \theta_\text{2D} < 165^{\circ}$	
& 25932 &(-14\%)  \\[5pt]

\emph{3D shower criteria:}					&&  \\
Shower on both planes 	
& 17512 &(-33\%) 	\\
$N_\textrm{3D} \ge 3$			
& 13772 &(-21\%) 	 \\
Frac. 3D hits $>$ 0.15 		
& 13456 &(-2.3\%) 	  \\
Centroid fiducial cut			 
& 12004 &(-11\%)  \\[5pt]

\emph{Decay time criteria:} && \\
$\Delta T > \SI{1.8}{\mu s}$ 				
& 5361 &(-55\%) \\[5pt]
\hline\hline

\end{tabular}

\end{table}

%%%%%%%%%%%%%%%%%%%%%%%%%%%%%%%%%%%%%%%%%%%%%%%%%%%%%%%%%%%%%%%%%%%%%%%%
\subsection{Energy reconstruction methods}

Energies of Michel electrons are reconstructed using the three methods outlined below:
\vspace{-\topsep}
\begin{enumerate}[i)]
\setlength\itemsep{0pt}
% -------------------------------------------------
\item \emph{$E_Q$: ``charge-only'' (Q-only) method.} This is the traditional approach, used in large neutrino detector LArTPCs, which relies only on the reconstructed charge, $Q$:
\begin{align}
E_Q &= (Q \times \langle R \rangle^{-1}) \times \Wion \label{eq:EQ}
\end{align}
The average recombination factor, $\langle R \rangle~=~N_e/\Ni = 0.69$, is chosen using the Modified Box model assuming an average Michel electron stopping power of 2.3 MeV/cm (an assumption also made by MicroBooNE~\cite{microboone_michels}).

% -------------------------------------------------
\item \emph{$E_{QL}$: ``charge-plus-light'' (Q+L) method.} In this approach, the sum of the collected charge and light together is used to determine the energy deposited. To incorporate $L$, we first require that showers were constructed successfully on both planes.  The charge-weighted average visibility of the 3D space points, $f_{\textrm{vis}}$, is calculated. The total scintillation light, $L$, is then calculated from $S_\textrm{total}$.  To correct for quenching, the late component of the scintillation light is scaled upward.  We then calculate the charge-plus-light energy as
\begin{equation}\label{eq:EQL}
E_{QL} = (Q + L) \times \Wph.
\end{equation}

% -------------------------------------------------
\item $E_{QL}^{\text{likelihood}}$: \emph{maximum-likelihood method.} 
Here we employ a likelihood-based hypothesis testing framework, tailored specifically for the Michel electron sample in LArIAT, which returns the most probable deposited energy that would lead to a particular measured combination of $Q$ and $L$.  This method requires knowledge of the detector's response, averaged over all spatial positions, as a function of the deposited energy. We achieve this through the construction of parametrized probability distribution functions (PDFs), $f_Q$ and $f_L$, which approximate the probability of measuring any $Q$ or $L$ given some amount of energy deposited by an electron shower within the TPC. The advantage of the likelihood method is that it naturally incorporates the performance of the detector in its estimation of energy, given that noise levels and light detection efficiencies are reflected in the MC samples which are used to generate the PDFs.  Details on the maximum-likelihood technique and the construction of PDFs can be found in Appendix~\ref{sec:developingfitter}.

\end{enumerate}

%%%%%%%%%%%%%%%%%%%%%%%%%%%%%%%%%%%%%%%%%%%%%%%%%%%%%%%%%%%%%%%%%%%%%%%%%
\subsection{Reconstructing Q and L}

\begin{figure*}
\includegraphics[width=0.75\textwidth]{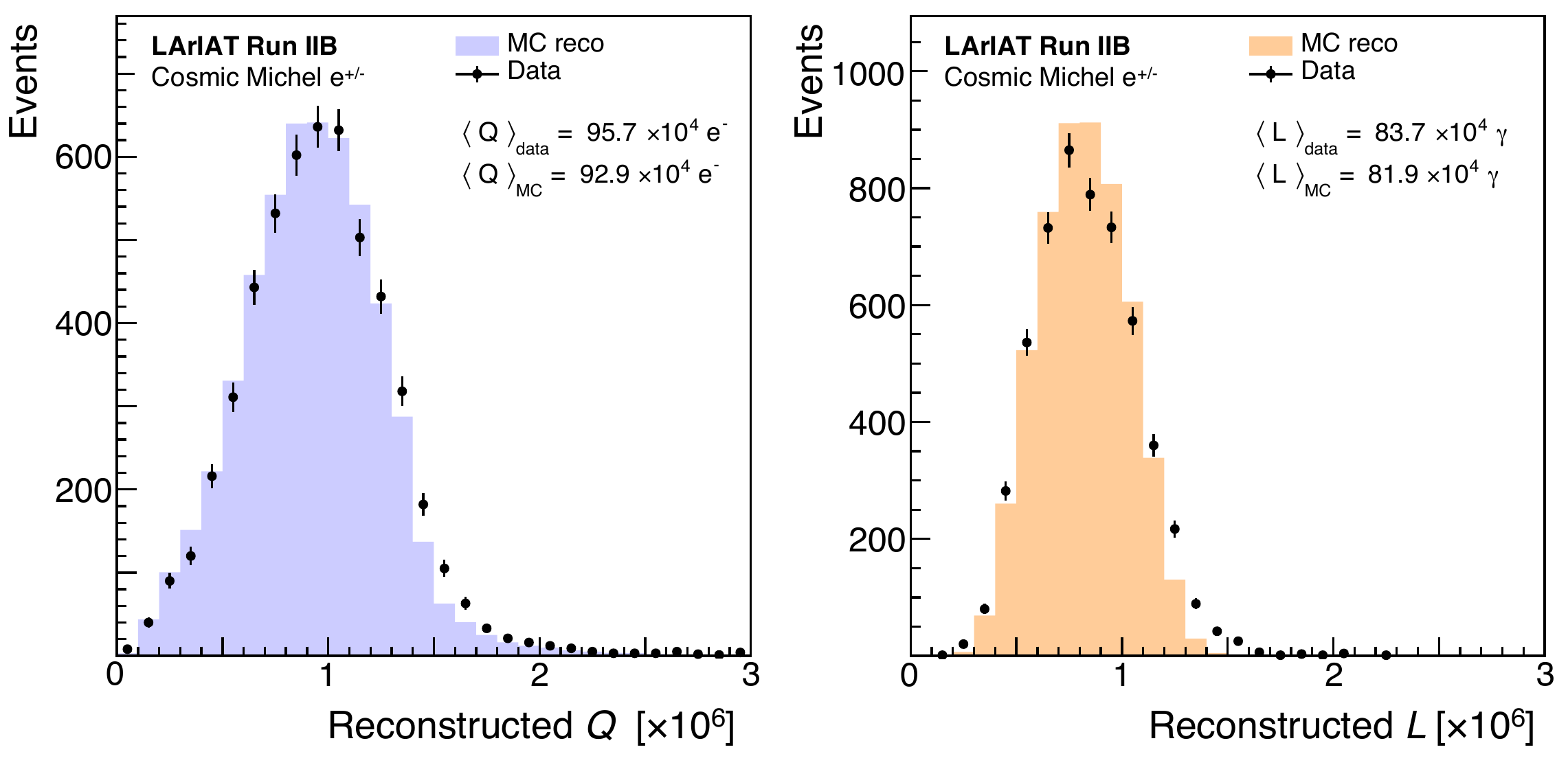}
\caption{Reconstructed distributions of free ionization electrons $Q$ (left) and scintillation photons $L$ (right) for the Michel electron data sample.}
\label{fig:QL}
\end{figure*}

The $Q$ for a topology containing $N$ reconstructed wire hits is simply the summation of all individual charge deposits ($q_i$), each corrected for its drift attenuation,
\begin{equation}\label{eq:Q}
Q = C^{\text{cal}}_{e^-} \times \sum_i^N \left[ q_i^{\text{ADC}} \times e^{t_i/\tau_e} %\exp(t_i/\tau_e) 
\right],
\end{equation}
where $q_i^{\text{ADC}}$ is the integrated charge of each individual hit in ADC counts, and $C_{e^-}^{\text{cal}}$ is the ADC-to-electron calibration constant. This calibration constant is a property of the readout electronics and is measured by fitting the $dE/dx$ as a function of residual range along well-reconstructed stopping muon tracks in our sample~\cite{lariat_detpaper}.

Likewise, $L$ is calculated by summing the light from both PMTs after making necessary corrections to account for quenching and losses during propagation:
\begin{equation}\label{eq:L}
L = \sum_j^{2} \left[ 
S_\textrm{total}^{\text{ADC}} 
\times C_{\text{pe}}^{\text{cal}}   
\times f_{\text{vis}}^{-1}  
\times C_{\text{quench}} 
\right]_j.
\end{equation}
The term $S_\textrm{total}^\text{ADC}$ is the total integrated charge of the candidate pulse on PMT $j$ in ADC units. The calibration constant $C_{\text{pe}}^{\text{cal}}$ is the inverse of the PMT's single photoelectron (SPE) response.   
The visibility and quenching corrections in Eq.~\ref{eq:L} are analogous to the electron attenuation lifetime corrections applied to get $Q$ in Eq.~\ref{eq:Q}, in that the collected light is scaled up to account for the losses from propagation and impurity quenching.  

The average photon visibility $f_{\text{vis}}$ is calculated as the charge-weighted average of the visibilities at reconstructed 3D space points. For PMT $j$,
\[
f_{\text{vis}} = \sum_i^{N_\textrm{3D}} \left[ 
q_i \times f_{j}(\vec{p}_i) %x_i,y_i,z_i)
\right] \times \left( \sum_i^{N_\textrm{3D}} q_i  \right)^{-1},
\]
where $f_j(\vec{p}_i)$ is the fraction of scintillation photons that would reach PMT~$j$ if emanating from a space point at position $\vec{p}_i$, according to the 3D visibility map.

Since the quenching of prompt light from singlet excimer decays is negligible (see Sec.~\ref{sec:sim-ql}), the correction factor $C_\text{quench}$ is constructed such that only the late component of the total light is scaled up. We define prompt light simply as the light arriving prior to some time $T_{\text{pr}}$ (nominally \SI{100}{ns} for this analysis). The quenching correction factor is thus redefined as a function of the prompt fraction, $f_{\textrm{pr}} = S_\textrm{100}/S_\textrm{total}$, as well as the triplet excimer lifetime $\tau_t$ and the effective quenched late lifetime $\tau_t^\prime$ found from fits to data,
\begin{equation}\label{eq:cquench}
C_{\text{quench}} = f_{\text{pr}} +  (1-f_{\text{pr}}) C_{\text{quench}}^{\text{late}},
\end{equation}
The term $C_{\text{quench}}^{\text{late}}$ is derived mathematically as the ratio of the expected total integral of triplet light following $T_{\text{pr}}$ to the integral of light after quenching:
\begin{align}
\begin{split}
C_{\text{quench}}^{\text{late}} &= \int_{T_{\text{pr}}}^{\infty} \frac{e^{-t/\tau_t}}{e^{-t/\tau_t^\prime}} \,dt = \int_{T_{\text{pr}}}^{\infty} e^{t(1/\tau_t^\prime - 1/\tau_t)} \,dt \\
&= \left(\frac{\tau_t}{\tau_t^\prime}\right) \exp\Big[T_{\text{pr}}\left(\frac{1}{\tau_t^\prime}-\frac{1}{\tau_t} \right) \Big].
\end{split}
\end{align}

 The resulting distributions of both $Q$ and $L$ for our sample are presented in Fig.~\ref{fig:QL}.

%%%%%%%%%%%%%%%%%%%%%%%%%%%%%%%%%%%%%%%%%%%%%%%%%%%%%%%%%%%%%%%%%%%%%%%
\subsection{Michel electron energy spectrum in LArIAT}

Here we present reconstructed energy spectra for Michel electrons in LArIAT using charge and light.  

\begin{figure}\centering
\includegraphics[width=0.8\columnwidth]{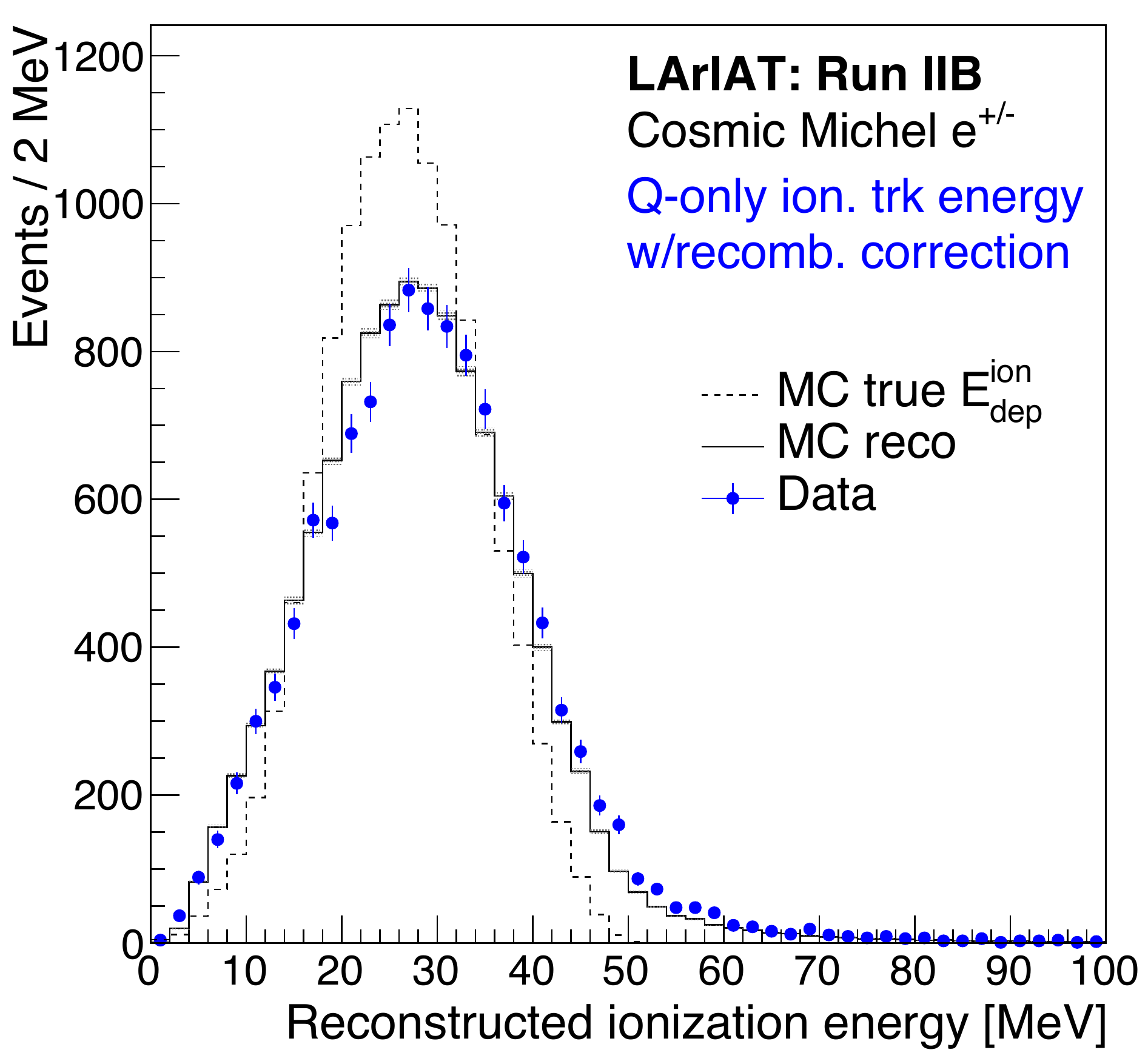}
\caption{Reconstructed energy $E_Q^{\text{ion}}$ deposited by hits identified as the direct electron-induced ionization.}
\label{fig:EQtrk}
\end{figure}

Figure~\ref{fig:EQtrk} shows the charge-based energy from ionization generated directly by the decay electron itself, referred to as $E_Q^{\text{ion}}$, using Eq.~\ref{eq:EQ}. This energy is reconstructed using only hits included in the original electron cluster, excluding displaced photon-induced energy depositions. %The  %falling within the 2D shower acceptance cone.

\begin{figure}

\subfigure[][\hspace{1ex}`Q-only' energy spectrum]{
	\includegraphics[width=0.82\columnwidth]{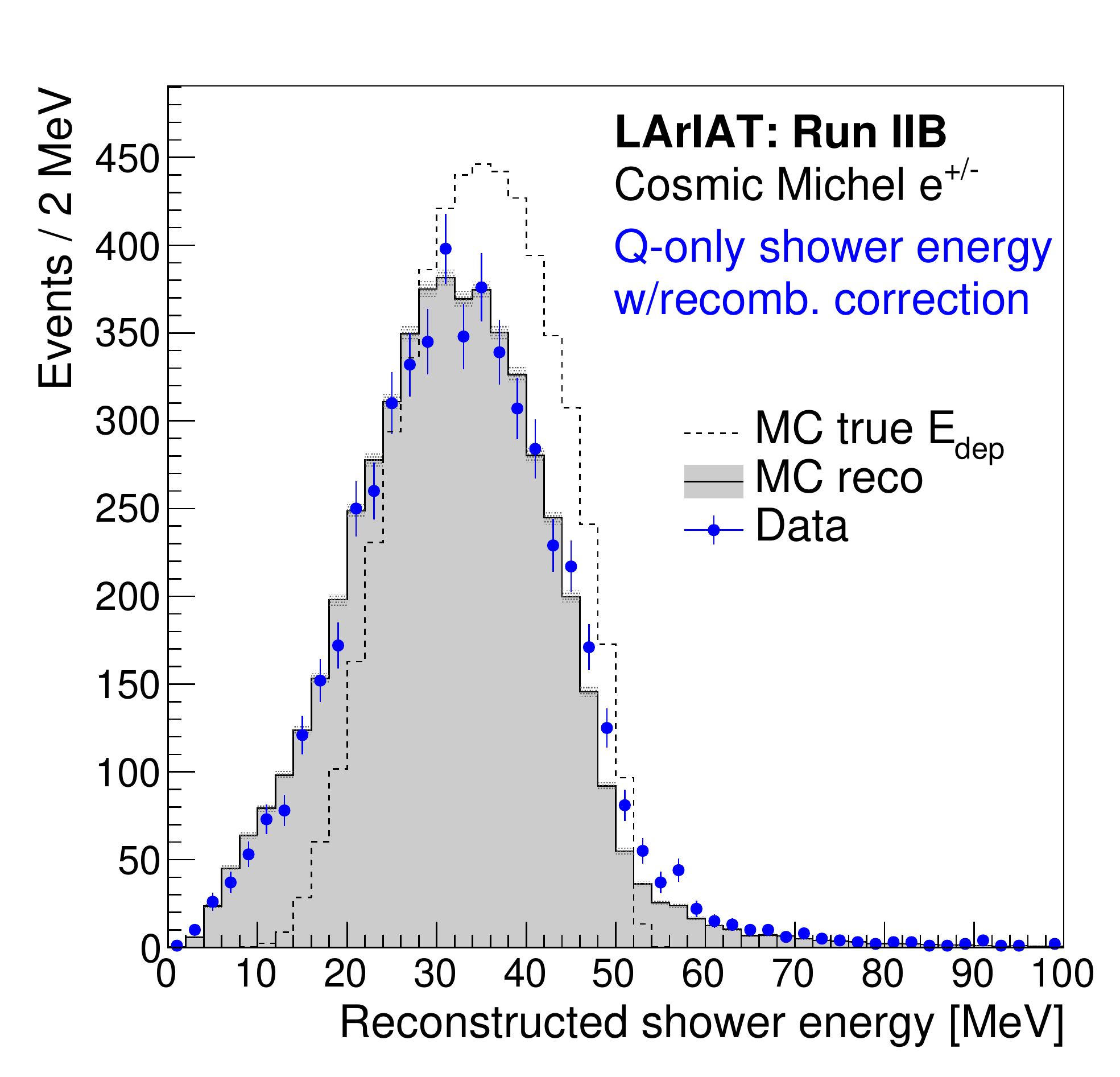}
	\label{fig:energyplots_EQ}
}

\vspace{-1\baselineskip}

\subfigure[][\hspace{1ex}`Q+L' energy spectrum]{
	\includegraphics[width=0.82\columnwidth]{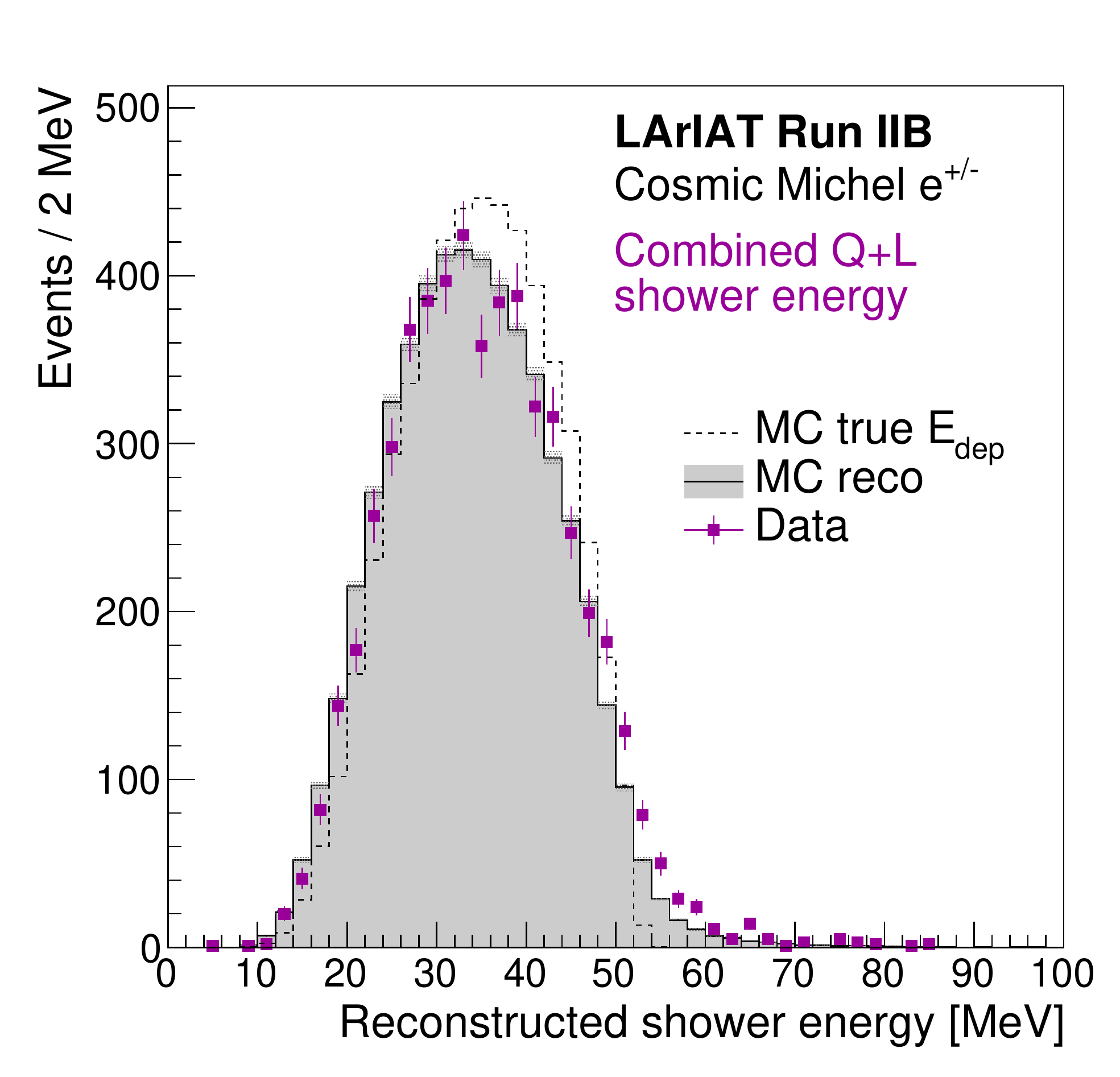}
	\label{fig:energyplots_EQL}

}

\vspace{-1\baselineskip}

\subfigure[][\hspace{1ex}`Q+L' maximum-likelihood energy spectrum]{
	\includegraphics[width=0.82\columnwidth]{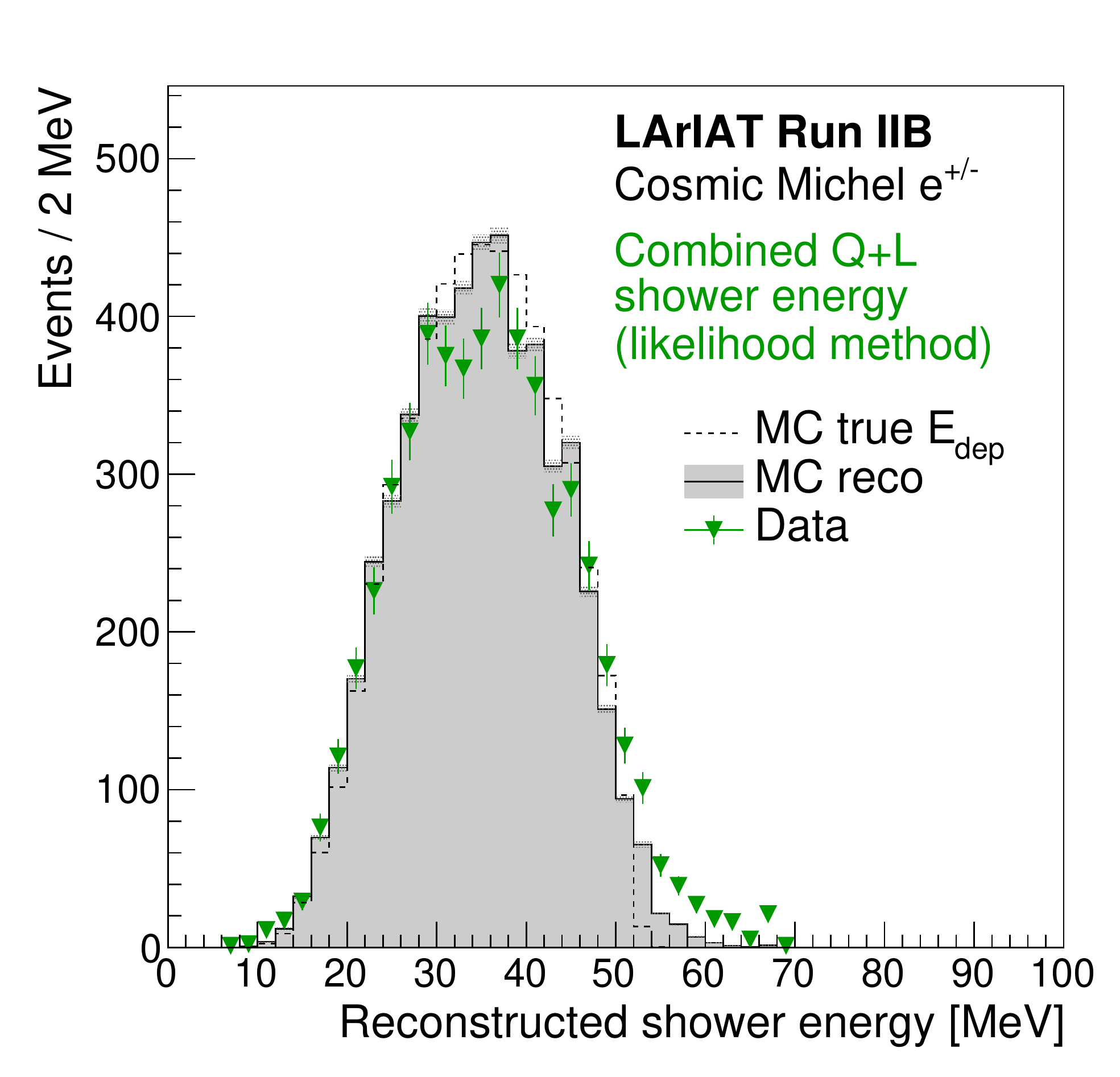}
		\label{fig:energyplots_EQL_LogL}

}

\caption{A comparison of the Michel electron shower energy spectrum in LArIAT reconstructed using the two simple formulaic constructions, $E_Q$  (\emph{a}) and $E_{QL}$ (\emph{b}), and the likelihood fitting technique, $E_{QL}^{\text{likelihood}}$ (\emph{c}).}
\label{fig:energyplots}
\end{figure}

The reconstructed energy spectra of the full electron shower using Eqs.~\ref{eq:EQ} and~\ref{eq:EQL}, including both the direct electron ionization as well as any photon-induced hits that were clustered into the shower on the collection plane, are presented in Figs.~\ref{fig:energyplots_EQ} ($E_{Q}$) and ~\ref{fig:energyplots_EQL} ($E_{QL}$). Only events that satisfy the 3D-shower selection criteria and have $\Delta T > \SI{1.8}{\mu s}$ are included. A tail extending to lower energies in $E_Q$ is mitigated by the addition of light in $E_{QL}$. This underscores the power of $E_{QL}$ -- information that would be lost using only charge, either due to higher-than-assumed recombination, incomplete clustering, or hit thresholding, is recovered to some extent through the inclusion of optical data.

Fig.~\ref{fig:energyplots_EQL_LogL} shows the resulting spectrum of Michel electron shower energy using the likelihood method, $E_{QL}^{\text{likelihood}}$.  The likelihood method performs similarly to $E_{QL}$, though its match to the true deposited energy is better.

It is worth noting the measured values of $Q$ and $L$, and consequently the  energy derived from them, slightly exceed MC predictions by approximately 3\%. The precise cause of this ``high-energy tail'' is not known, though it falls within the 5\% systematic uncertainty in the energy scale itself due to the contribution from $\Wph$ (=~$19.5\pm\SI{1.0}{eV}$~\cite{w_ph,doke}) described in Sec.~\ref{sec:intro}.

%%%%%%%%%%%%%%%%%%%%%%%%%%%%%%%%%%%%%%%%%%%%%%%%%%%%%%%%%%%%%%%%%%%%%%
\subsection{Monte Carlo energy resolution}

With the simulations validated by data, MC can now be used to compare the energy reconstruction performance of the three calorimetric techniques for low-energy electrons in LArIAT.  We begin by studying Michel electrons and quantifying the resolution on measuring the total energy they deposit in the LArIAT volume.

The Michel electron reconstruction, however, is vulnerable to inaccurate muon-electron boundary determination, muon-electron charge overlap, incomplete shower clustering, and optical contamination from the late light of the muon. Considering these complications, the sample serves as a poor representation of the low-energy electron showers that would be induced by supernova or solar neutrinos in a deep underground LArTPC. To better study this simpler topology, we also reconstruct a simulated sample of lone electrons positioned randomly in the LArIAT active volume.   

Resolution is a metric for quantifying precision -- but in the analysis that follows, its exact definition is slightly fluid.  When characterizing resolution, a histogram of the energy variance, $\delta E$, is filled on an event-by-event basis:
\begin{equation}\label{eq:variance}
\delta E = \frac{E^{\text{reco}} - E^{\text{true}}}{E^{\text{true}}}.
\end{equation}
In an idealized detector, $\delta E$ forms a perfect Gaussian and we take the resolution to be its fitted width or standard deviation (RMS). However, when the variance takes on a non-Gaussian shape, as will be seen for the Michel electron sample, the resolution becomes ill-defined and the definition must be modified to fit the situation.

\subsubsection{Resolution of Michel electrons in LArIAT}

Multiple energy variance histograms are constructed from the MC sample for Michel electron events with different true deposited energy, $E_\text{dep}^\text{true}$. Each histogram includes events within $\pm~\SI{0.5}{MeV}$ of particular evenly spaced values of $E_\text{dep}^\text{true}$ starting at \SI{10}{MeV}.  

\begin{figure}
\includegraphics[width=\columnwidth]{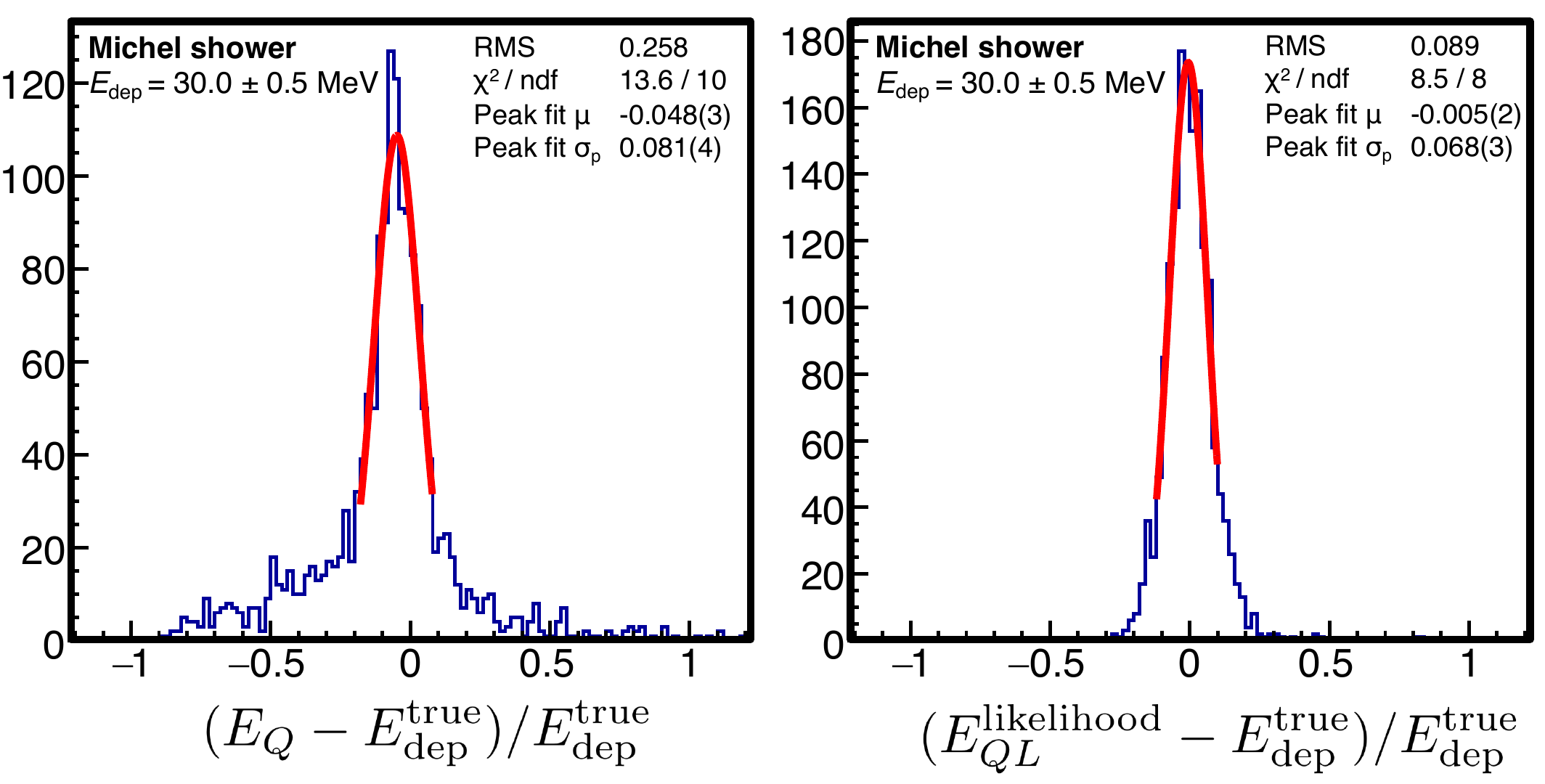}
\caption{The charge-based energy resolution histogram for Michel electron events with $E_{\text{dep}} = \SI{30}{MeV}$ with a Gaussian function fitted to the peak (red). Note that the width of the peak from the fit (7.9\%) is much smaller than the overall RMS (25.8\%).}
\label{fig:michel_resExample}
\end{figure}

Figure~\ref{fig:michel_resExample} shows an example of the energy variance distribution for Michel electrons with $E_\text{dep}^\text{true}$ around \SI{30}{MeV}.   The distribution is clearly non-Gaussian, consisting of a central ``peak'' region of relatively well-reconstructed events which sits on top of a more diffuse distribution from events where some fraction of charge is missed or added by the reconstruction.
In characterizing the resolution, we take into account both the width of the peak as well as the RMS of the entire distribution. To fit the peak in each energy variance histogram, a fit region is defined which extends from the peak bin 
to points on either side where the distribution drops to 1/3 of the maximum height.

\begin{figure}
\includegraphics[width=0.9\columnwidth]{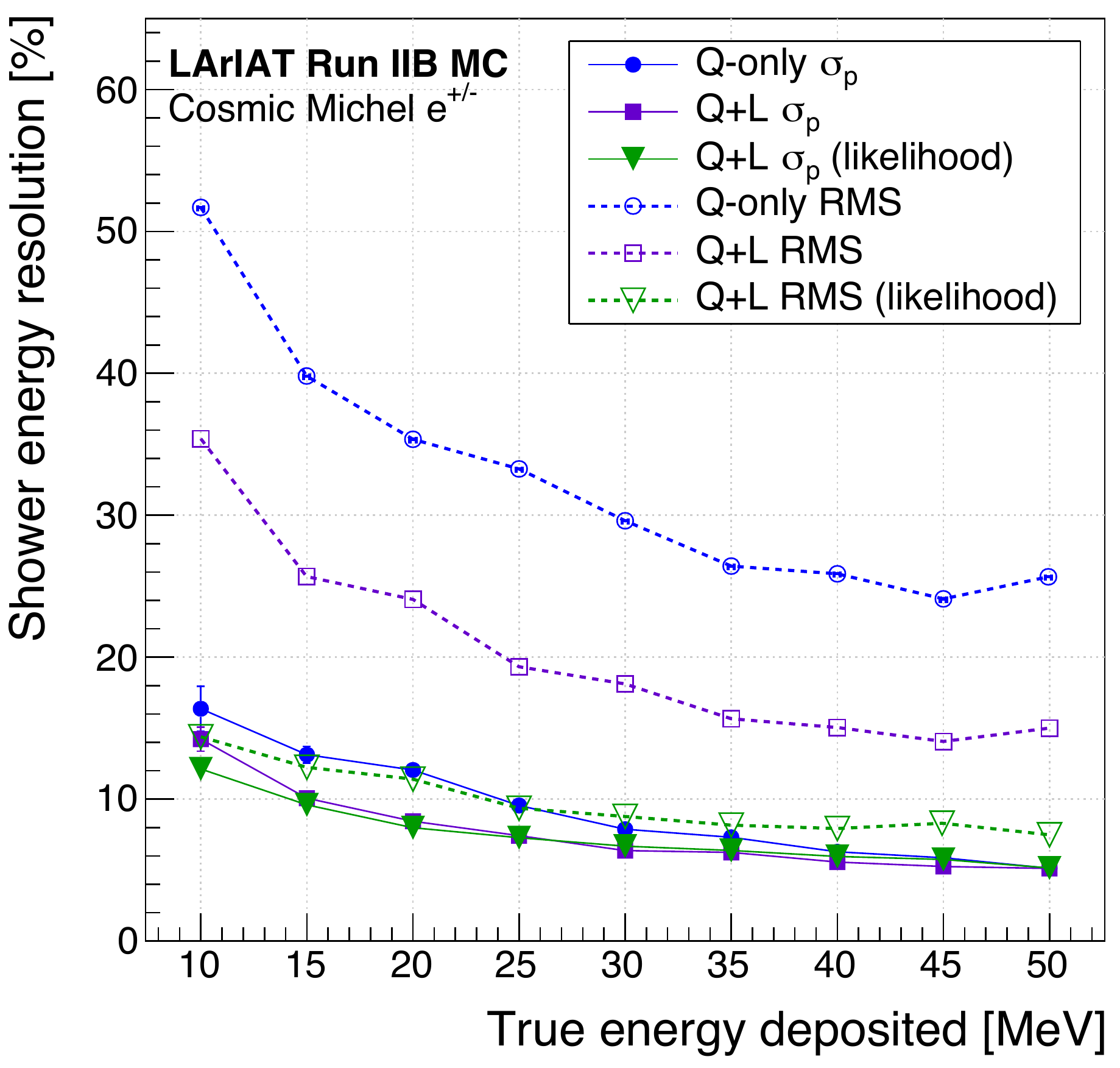}
\caption{The energy resolution of Michel electron events in LArIAT for energies reconstructed from charge (blue) as well as from both charge and light (violet and green).  The dotted lines show RMS resolution of the energy variance histograms, while the solid lines trace the resolution as defined by the width of a Gaussian fit to the central peak.}
\label{fig:michel_res}
\end{figure}

In Fig.~\ref{fig:michel_res}, the peak width and distribution RMS are plotted against the deposited energy for each of the three different calorimetric methods. A drastic improvement is seen in the RMS when using $E_{QL}$ and $E_{QL}^{\text{likelihood}}$ compared to the traditional charge-based $E_Q$, thanks to information being recovered by optical data in events where a significant portion of the deposited charge is not reconstructed.  The resolution of the peak is comparable for the $Q$+$L$ and likelihood techniques, and both slightly outperform the $Q$-only technique.

% ..........................................................................
\subsubsection{Resolution of isolated electrons in LArIAT} \label{sec:iso_electrons}

For the simulated isolated electron sample, we require each event be fully contained in that neither the electron nor any of its daughters in the EM shower escape the TPC.  This better mimics the case of a large neutrino detector and allows us to use the initial energy of the electron ($E^\text{true}_e$) as our target energy, instead of the visible deposited energy as we did for the Michel electrons. For isolated showers at fixed energy, both $Q$ and $L$ distributions are well described by simple Gaussian functions, and the procedure outlined in Appendix~\ref{sec:developingfitter} for constructing parametrized PDFs of $Q$ and $L$ is repeated.

The energy resolution for all three calorimetry methods are now compared using  this sample of lone electron showers.  To account for any deviations from Gaussian shapes in these distributions, we incorporate both the width of a fitted Gaussian ($\sigma_{\text{fit}}$) as well as the distribution's RMS ($\sigma_{\text{RMS}}$) in our definition of energy resolution. We assign a relative weight ($w$) to $\sigma_\text{RMS}$ based on the overall goodness-of-fit:
\begin{equation}\label{eq:e_res}
\sigma = \frac{\sigma_{\text{fit}} + w\sigma_{\text{RMS}}}{1+w}
\end{equation}
where $w = \sqrt{\chi^2_\nu}$.  With this definition, the resolution from variance distributions that stray from a Gaussian shape ($\chi^2_\nu > 1$)
will tend to more heavily weight the RMS rather than the Gaussian fit. For well-behaved distributions ($\chi^2_\nu \approx 1$), the fit and RMS will be weighted equally and the assigned resolution will be approximately the average of the two. 

\begin{figure}
\centering
\includegraphics[width=0.90\columnwidth]{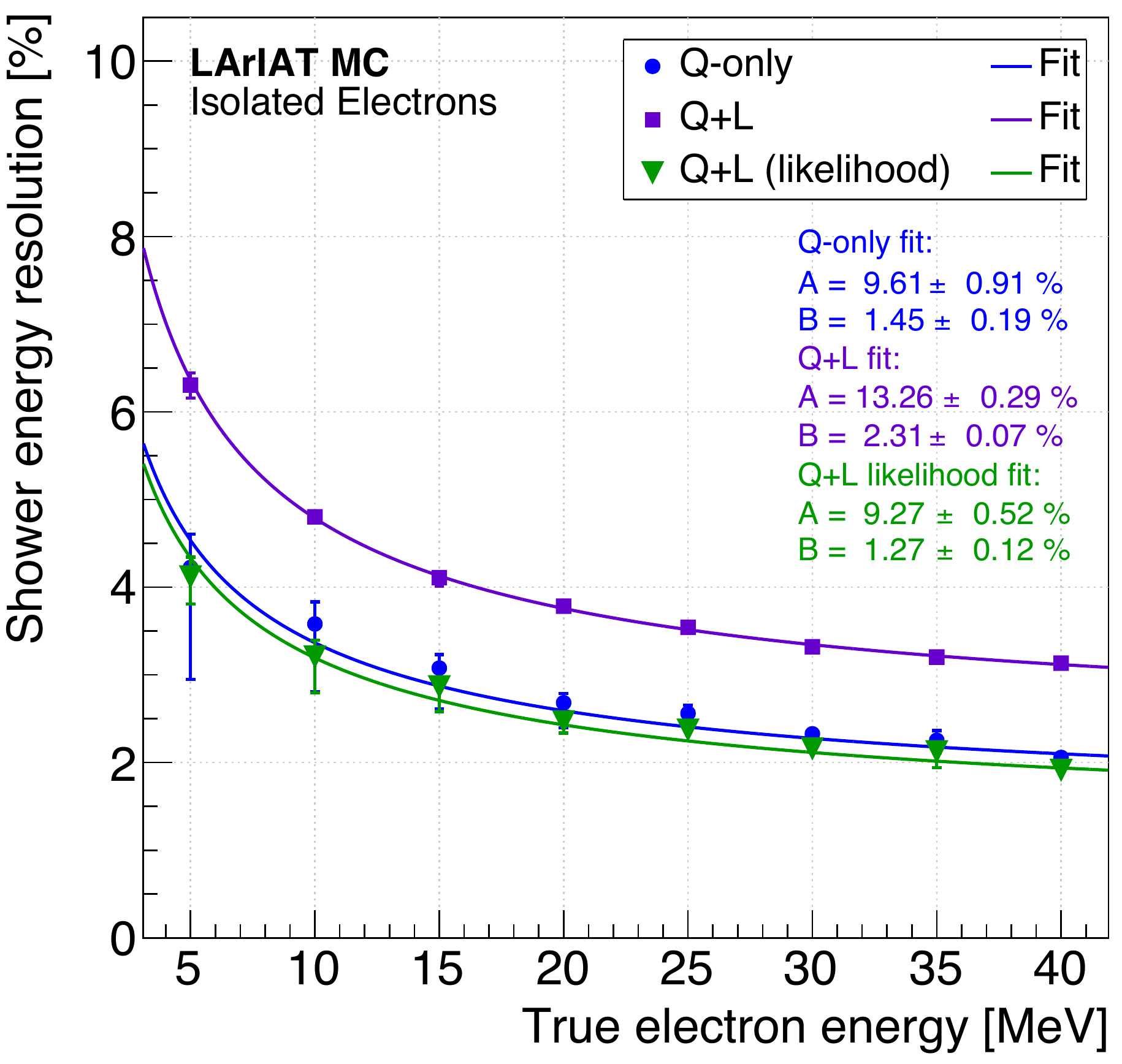}
\caption{The energy resolution, as defined by Eq.~\ref{eq:e_res}, of isolated electron showers in the LArIAT TPC under Run~IIB operating conditions.}
\label{fig:e_res_nom}
\end{figure}

The resulting energy resolution curves for isolated electrons in LArIAT using the three calorimetric methods are plotted in Fig.~\ref{fig:e_res_nom}. Each set of points is fit to the function,\begin{equation}
\sigma(E) = \frac{A\text{ [\%]}}{\sqrt{E\text{ [MeV]}}} \oplus B\text{ [\%]},
\end{equation} where the first term ($A$) is meant to model any noise dependence while the second term accounts for reconstruction-specific effects related to hit fitting and thresholding that limit the achievable resolution~\cite{icarus_recomb}. Since the resolution in the reconstructed charge now exceeds that of light due to the simpler charge topology of the isolated electrons, $E_{Q}$ outperforms $E_{QL}$. However, as expected, the best resolution is still achieved through the maximum-likelihood combination of $Q$ and $L$. 

For low-energy electrons from 5-50 MeV, we find that LArIAT achieves an energy resolution characterized as:
%\begin{linenomath*}
\begin{align}
\text{Charge-only:}& \quad \sigma= \frac{9.6\%}{\sqrt{E\text{ [MeV]}}} \oplus 1.5\% \\
\text{Charge and light:}& \quad \sigma = \frac{9.3\%}{\sqrt{E\text{ [MeV]}}} \oplus 1.3\%
\end{align}
%\end{linenomath*}

%%%%%%%%%%%%%%%%%%%%%%%%%%%%%%%%%%%%%%%%%%%%%%%%%%%%%%%%%%%%%%%%%%%%%%
\section{\label{sec:calpotential}Calorimetric Potential of Future LArTPCs}

Here we expand the MC resolution studies for isolated low-energy electron showers to determine the relative impact of light-augmented calorimetry for larger LArTPCs with different wire signal-to-noise levels (S/N) and light yields.  

In Run~II of LArIAT, S/N~$\approx$~50 was observed on the raw collection plane wire signals, measured using the most probable pulse height from minimally-ionizing beam particle tracks. LArIAT's light collection system achieved LY~=~\SI{18}{pe}/MeV for this same run period~\cite{lariat_detpaper}. In comparison, DUNE's baseline design requires S/N~$>$~10 due to its longer wires, with expected LY~$\approx$~\SI{5}{pe}/MeV~\cite{dune}.  Due to its use of TPB-coated reflector foils, the Short-Baseline Near Detector (SBND)~\cite{sbn} at Fermilab is expected to achieve LY $>$~\SI{50}{pe}/MeV for all drift distances~\cite{sbnd_ly}. %However, results presented here are less applicable to SBND given its surface location and small size relative to DUNE. 
However, our electron MC sample more closely approximates the quieter conditions expected in an underground detector which, free from cosmic ray pileup, is more capable of studying low-energy astrophysical $\nu_e$.

\begin{figure}
\centering
\includegraphics[width=0.90\columnwidth]{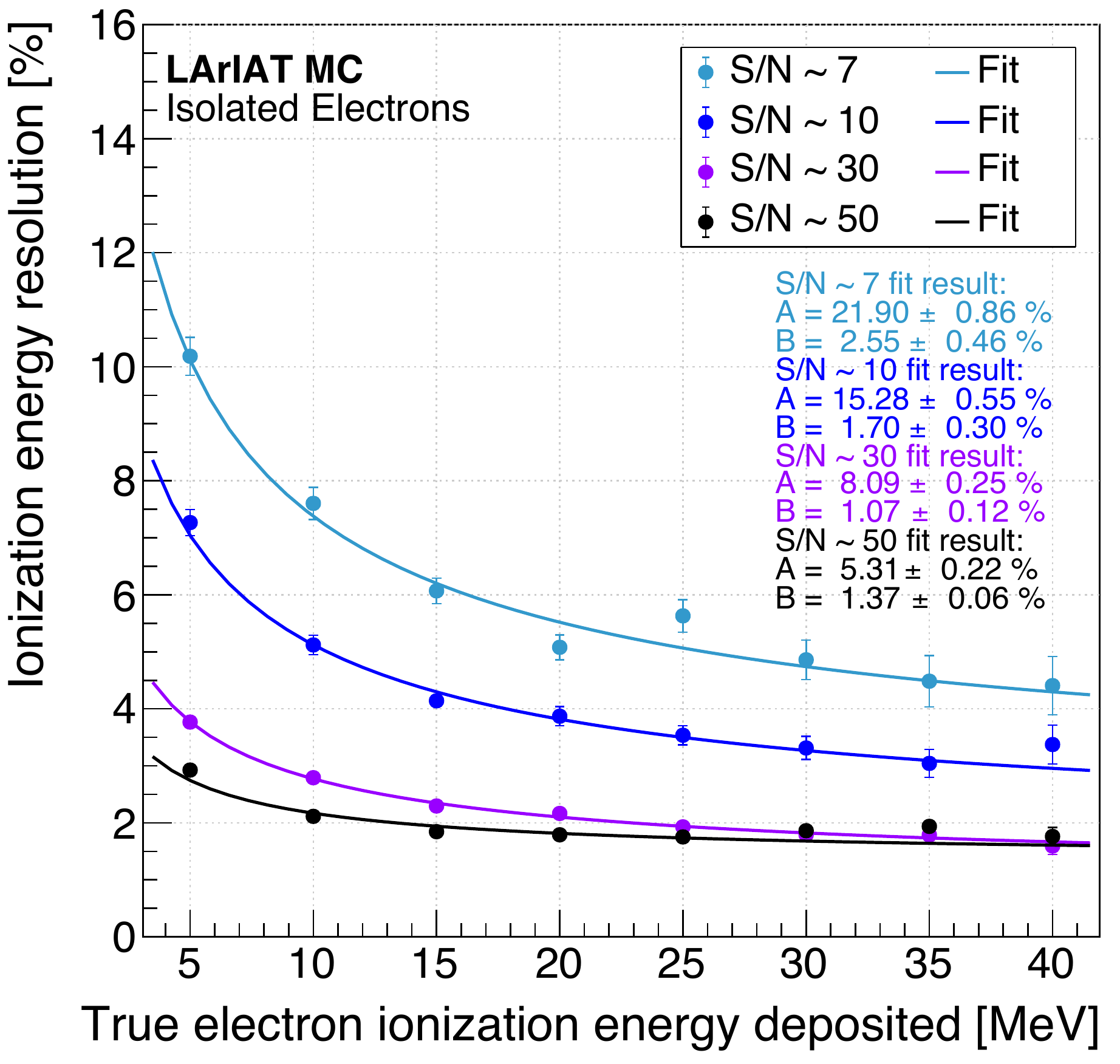}
\caption{Charge-based energy resolution of the primary electron (excluding shower photons) for the three simulated values of S/N on the wires.}
\label{fig:e_res_ion}
\end{figure}

Several isolated electron samples in LArIAT are simulated with the amplitude of the raw wire signal noise on the collection plane tuned to achieve S/N ranging from 50 (LArIAT conditions) to as low as 7. The simulated LY ranges over representative values from \SI{2}{pe}/MeV to \SI{100}{pe}/MeV. We neglect the LArIAT-specific optical smearing and instead apply a smearing that replicates the expected resolution for a system capable of a single photoelectron resolution of $\sigma_\text{pe}/\text{pe} = 0.1$, as has been achieved in many silicon photomultiplier (SiPM) devices.

We first examine the charge-based energy resolution of the primary electron ionization, $E_Q^\text{ion}$.  For each event, a proximity-based clustering procedure is repeated, starting from the wire hit most closely aligned in drift coordinates ($X$) with the true vertex of the simulated shower. Hits included within this cluster are categorized as electron-induced ionization.  The \emph{deposited} electron ionization energy is treated as the true energy in the resolution. We also limit the fit to the peak region of each energy variance histogram to minimize contributions from poorly-reconstructed events---i.e., where the clustering stops short of the electron endpoint, or photon deposits are accidentally clustered together with those of the electron. 

In Fig.~\ref{fig:e_res_ion} the $E_Q^{\text{ion}}$ resolution curves for the different S/N are plotted. The resolution behaves as expected with respect to wire noise, with the first term in the fit (which scales as $\sim$$1/\sqrt{E}$) increasing considerably with noise. The flat contribution term also increases slightly due to the increased hit-finding threshold required at higher noise levels. LArIAT, at S/N~$\approx$~50, achieves a resolution on $E_Q^{\text{ion}}$ of $5.3\% / \sqrt{E} \oplus 1.4\%$, which worsens to $15.3\% / \sqrt{E} \oplus 1.7\%$ when using the MC sample with wire noise adjusted to S/N $\approx$ 10.  These results are consistent with those found in a study by the ICARUS Collaboration of electrons from muon decays, which obtained a resolution for $E_Q^{\text{ion}}$ of $11\% / \sqrt{E} \oplus 2\%$ at S/N $\approx$ 14 using isolated electrons~\cite{icarus-michels,icarus-snr}.

\begin{figure}
\centering
\includegraphics[width=0.86\columnwidth]{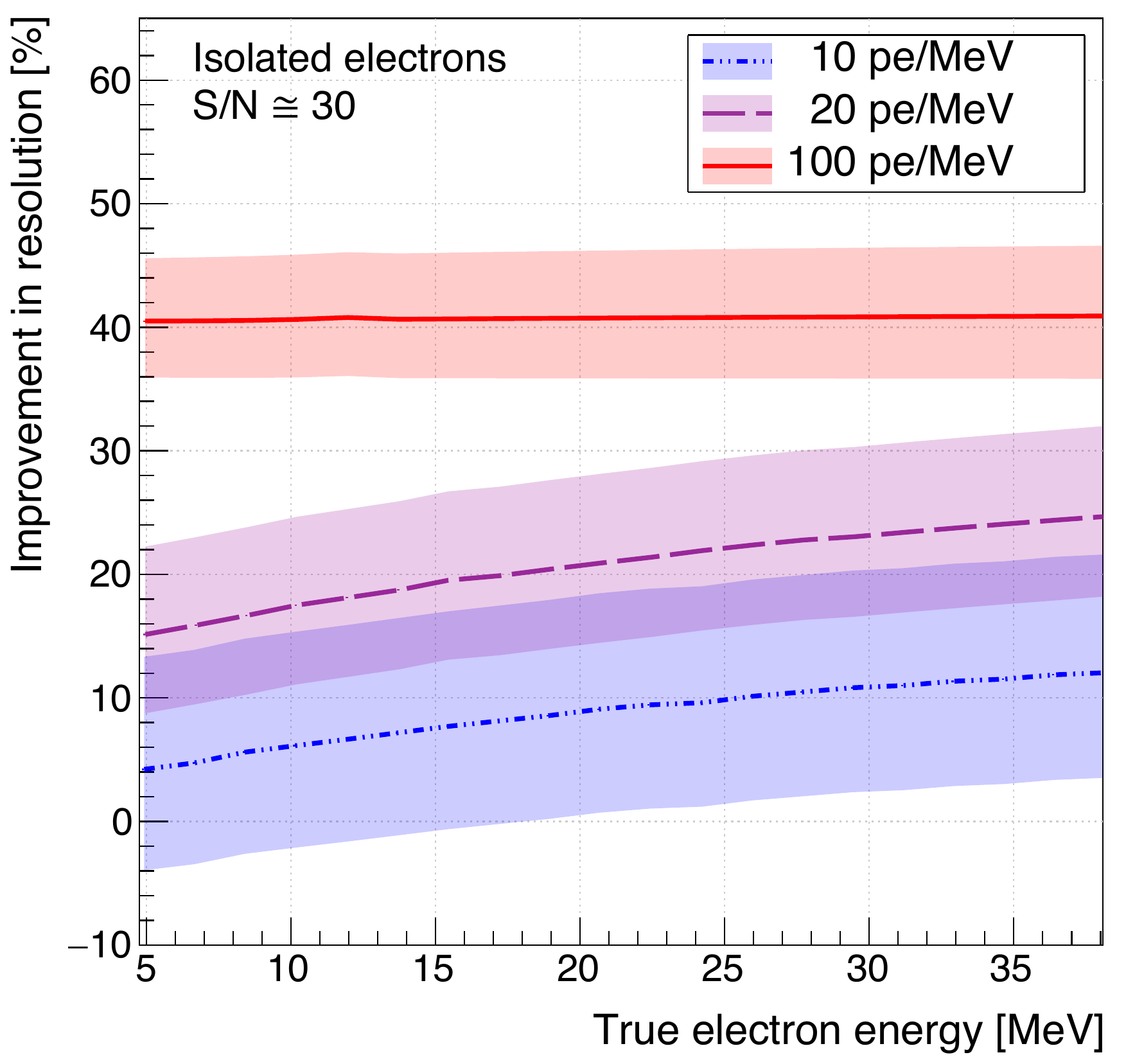}
\caption{The fractional improvement in resolution by combining charge and light, with wire S/N~$\approx$~30, for the four simulated light yields. The error bands reflect uncertainties on the fit parameters.} 
\label{fig:fracimprov}
\end{figure}

\begin{figure*}
\centering
\includegraphics[width=0.95\textwidth]{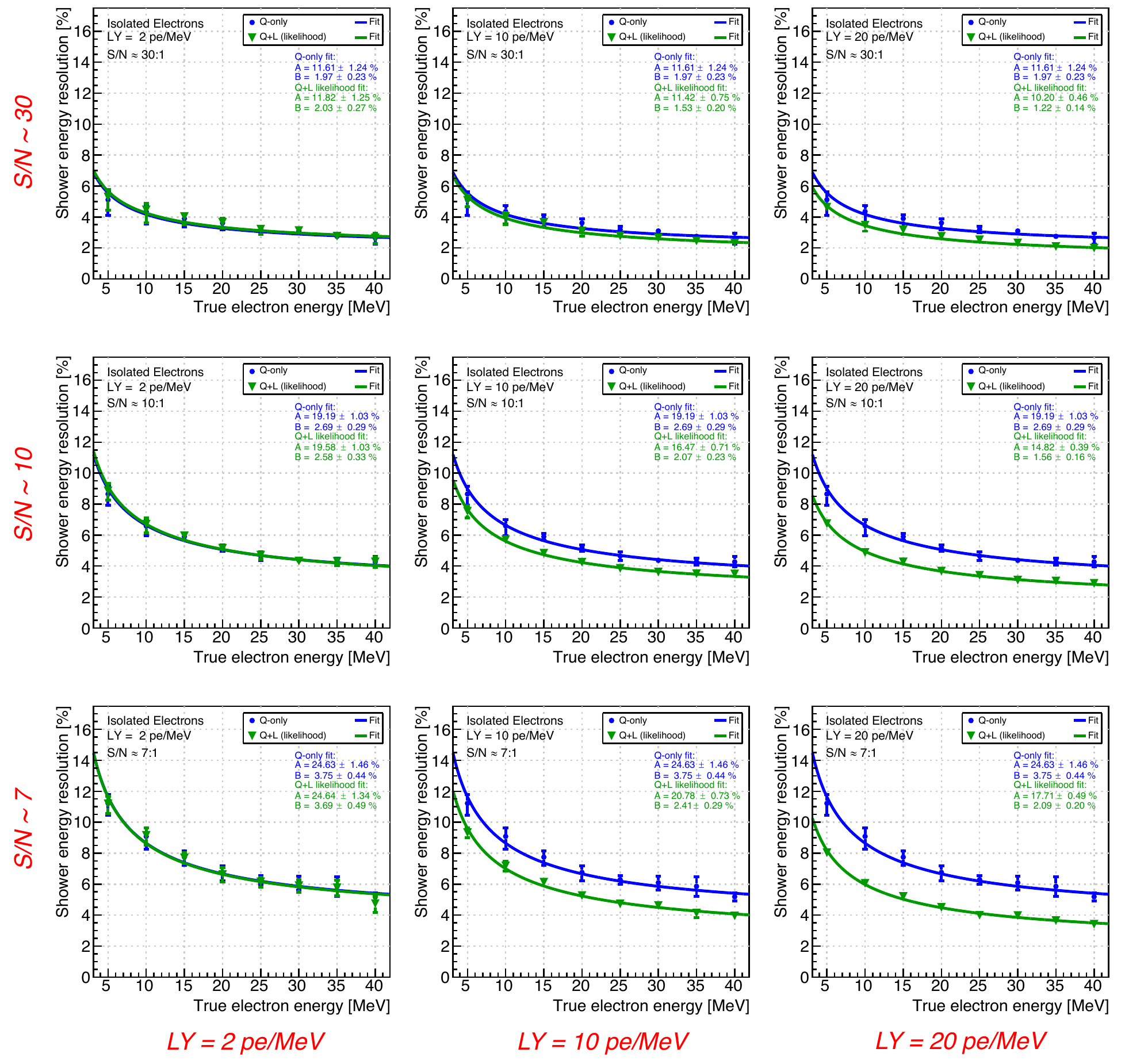}
\caption{Energy resolution for charge-only energy ($E_Q$) and the light-augmented likelihood energy ($E_{QL}^{\text{likelihood}}$) for different wire noise levels and light yields. Fits are performed to function $\sigma = A/\sqrt{E} \oplus B$.}
\label{fig:e_resArray}
\end{figure*}

Next, we examine the resolution for measuring the total electron energy by reconstructing the full EM shower.  Charge and light probability distributions for the likelihood fitter are again parametrized, this time as functions of the true electron energy $E_e$.  The reconstruction of $Q$ and $L$ are assumed to be independent of one another, so $f_L$ is parametrized separately for each simulated LY while $f_Q$ is parametrized separately for each S/N.  In Fig.~\ref{fig:fracimprov} the relative improvement in resolution of the light-augmented likelihood energy ($E_{QL}^{\text{likelihood}}$) compared to the charge-based energy ($E_Q$) is plotted as a function of true electron energy for each simulated LY, assuming a fixed collection plane S/N~$\approx$~30. Figure~\ref{fig:e_resArray} presents the array of fitted energy resolution curves, for both $E_Q$ and $E_{QL}^{\text{likelihood}}$, for each simulated scenario.   At LY = \SI{2}{pe}/MeV, the addition of $L$ has virtually no impact on the energy resolution, while at LY = \SI{10}{pe}/MeV and above we find that the addition of $L$ begins to noticeably improve the energy resolution. For example, at S/N~$\approx$~30, the resolution in reconstructed energy (relative to the charge-only method) improves by 5-12\% at \SI{10}{pe}/MeV, by 15-25\% at \SI{20}{pe}/MeV, and by 40\% at \SI{100}{pe}/MeV. This improvement in resolution with increasing LY is more pronounced at smaller charge collection S/N.

%#################################################
% Section 5: Conclusion
%#################################################
\section{\label{sec:conclusion}Conclusions}

LArIAT has demonstrated light-augmented calorimetry for low-energy electrons in a LArTPC using a sample of Michel electrons from cosmic muons.  A light-based trigger was implemented to obtain this sample, and an automated reconstruction was carried out to determine the charge deposited and light produced by these electron-induced EM showers. A total of 25,932 good-quality Michel electron showers were successfully reconstructed in 2D using the collection plane wires for the Run~IIB dataset. By incorporating information from the induction plane wires, 12,004 of those events were reconstructed in 3D as well.

For complicated multiparticle events in LArTPCs, the addition of scintillation light greatly improves the ability to measure the total visible energy.  Through light, information is recovered that would otherwise be lost or distorted from reconstruction effects like the muon-electron charge overlap that affects our Michel electron sample. This improvement is reflected in the energy spectra as well as in both the RMS and the peak-fitted energy resolutions from Sec.~\ref{sec:calo}.  These results imply that similarly-complicated events, where the charge-energy ($E_Q$) would suffer from clustering or hit-finding inefficiencies, would benefit from the collection of scintillation light.

Even with a relatively simple likelihood fitter, described in Appendix~\ref{sec:developingfitter}, which models $Q$ and $L$ as functions of the total deposited energy, the precision of calorimetric measurements is improved.
A more detailed model that incorporates additional parameters, like shower location and direction,  would perform even better.  In principle, if the physics in a simulation is sufficiently detailed---i.e., incorporating the nonradiative quenching processes that manifest at high ionization densities as described in Sec.~\ref{sec:intro}---then likelihood modeling is expected to improve energy resolution for a wide variety of particle species.

Our simulation of the LArIAT detector, having been validated by comparisons to data, is used to estimate the calorimetric performance of larger underground LArTPCs like DUNE in reconstructing low-energy electrons (see  Sec.~\ref{sec:calpotential}). We find that at a minimum baseline expectation of wire S/N~$\approx$~10 and LY~$\approx$~\SI{1}{pe/MeV}, the collected light has negligible (if any) impact on energy resolution. However, at LY =~\SI{10}{pe/MeV}, $\approx$~15\% improvement in resolution can be achieved for these events, with $\approx$~25\% improvement possible at \SI{20}{pe/MeV}. At S/N~$\approx$~30, we find more modest energy resolution improvements of $\approx$~10\% and $\approx$~20\% at \SI{10}{pe/MeV} and \SI{20}{pe/MeV}, respectively. Accurate energy resolution for electrons down to 5~MeV will directly aid in the reconstruction of $\nu_e$ from a potential core-collapse supernova event.

It should be mentioned that our projections of detector performance in Sec.~\ref{sec:calpotential} are built on assumptions of a relatively uniform LY, which LArIAT achieves through use of TPB-coated foils surrounding the active volume. A less uniform LY will require greater diligence in properly modeling the variability in photon visibility across the TPC.  In addition, we assume a SiPM-like SPE resolution ($\simeq$ 0.1~pe) and an optical reconstruction that incorporates photon-counting and more selective integration techniques instead of direct \emph{brute-force} integration of optical pulses (out to 7~$\mu$s) as was done in LArIAT.

We hope these results inspire further discussion on the role light can play in LArTPC neutrino detectors. A more holistic treatment of reconstruction, exploiting both the ionization and scintillation produced in charged-current neutrino interactions, has the potential to extend the physics reach of these detectors.

%#################################################
% Acknowledgements
%#################################################
\acknowledgements

This document was prepared by the LArIAT collaboration using the resources of Fermilab, a U.S. Department of Energy, Office of Science, HEP User Facility. Fermilab is managed by Fermi Research Alliance, LLC (FRA), acting under Contract No. DE-AC02-07CH11359. We extend a special thank you to the coordinators and technicians of the Fermilab Test Beam Facility, without whom this work would not have been possible. This work was directly supported by the National Science Foundation (NSF) through Grant No. PHY-1555090. We also gratefully acknowledge additional support from the NSF; Brazil CNPq Grant No. 233511/2014-8; Coordena\c{c}\~ao de Aperfei\c{c}oamento de Pessoal de N\'ivel Superior - Brazil (CAPES) - Finance Code 001; S\~ao Paulo Research Foundation - FAPESP (BR) grant number 16/22738-0; the Science and Technology Facilities Council (STFC), part of the United Kingdom Research and Innovation; The Royal Society (United Kingdom); the Polish National Science Centre Grant No. Dec-2013/09/N/ST2/02793; and the JSPS grant-in-aid (Grant No. 25105008), Japan.

%#################################################
% Appendices for supplemental plots and fits
%#################################################
\begin{appendix}

\section{SAMPLE VALIDATION USING MUON DECAY SPECTRUM}
\label{sec:mudecaytimes}

The reconstructed muon decay time spectrum is used to validate the sample and estimate its purity. Muons that come to a rest behave differently depending on their charge sign.  As a negative muon ($\mu^-$) slows down and approaches rest, it becomes bound to a nucleus due to the attractive Coulomb potential and immediately cascades to the 1S atomic orbital where it can then undergo nuclear capture via the interaction $\mu p \rightarrow \nu_{\mu} n$, .  The decay of bound muons thus competes with the capture process, resulting in an effective $\mu^{-}$ lifetime,
\begin{equation}
\tau_{\mu^-} = \left( \frac{1}{\tau_\text{c}} + \frac{Q}{\tau_{\text{free}}} \right)^{-1}
\label{eq:tauminus}
\end{equation}
where $\tau_{\text{free}}$ is the free muon lifetime of 2197~ns, $\tau_\text{c}$ is the capture lifetime, and Q ( =~0.988 for Ar~\cite{nucl_capture_rate}) is the Huff factor, a minor corrective term to account for the reduction in decay rate for bound $\mu^-$~\cite{measday}.

\begin{figure}
\centering
\includegraphics[width=0.95\columnwidth]{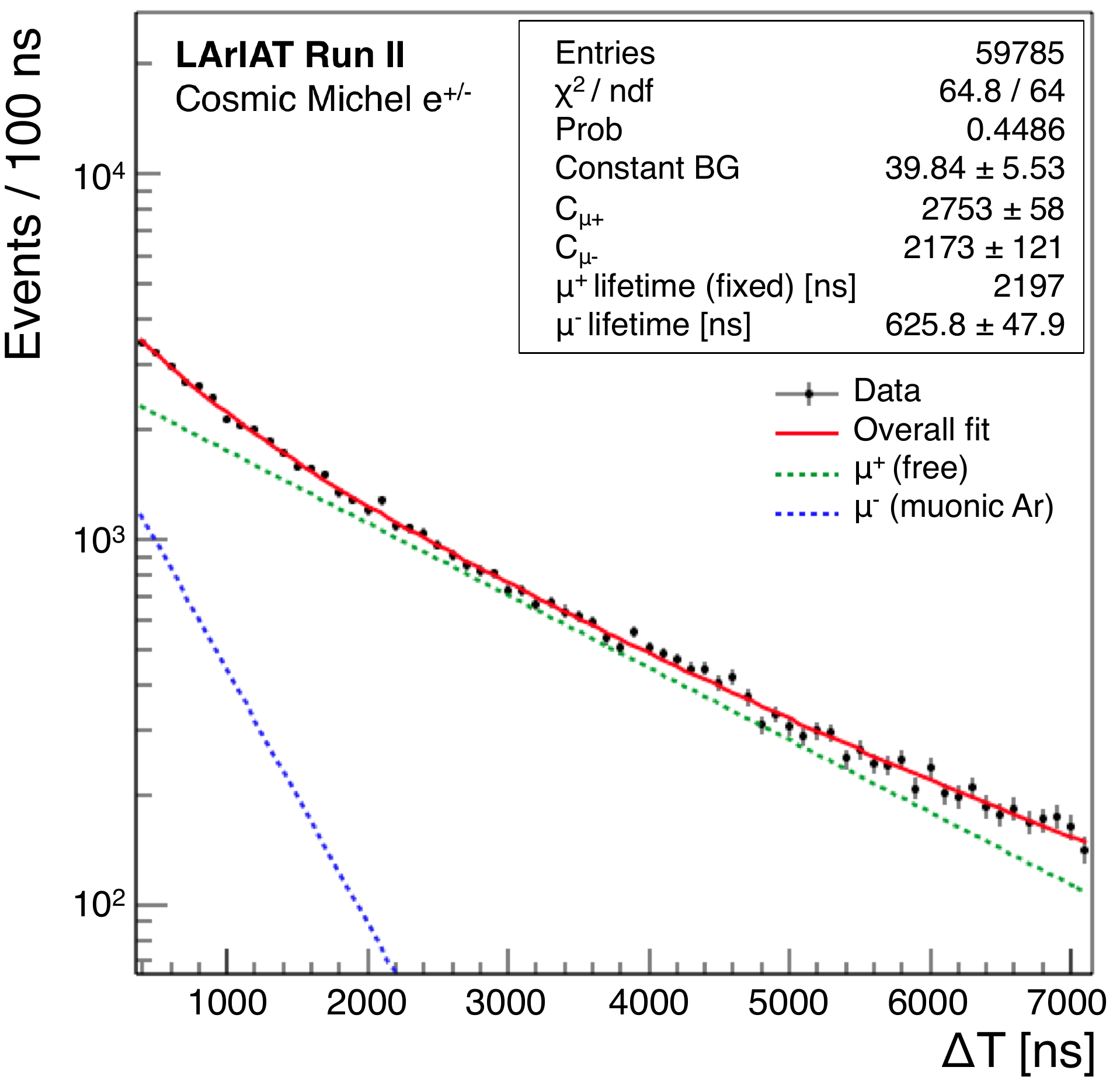}
\caption{Muon decay time spectrum for LArIAT's Run~IIB Michel electron dataset.}
\label{fig:result_decaytime}
\end{figure}

The muon decay time spectrum for the LArIAT Run~II Michel electron data is shown in Fig.~\ref{fig:result_decaytime}. Since there is no intention to use any calorimetric information from the wires here, the only requirement is that a stopping 3D track was identified with an optical topology consistent with a delayed Michel electron decay.  Cuts are made on pulse width and amplitude to exclude likely noise hits.  These cuts are unlikely to affect the measured decay spectrum since decay time and Michel electron energy are largely independent.  The $\Delta T$ distribution is fit with two exponentials,
\begin{equation}\label{eq:dtfit}
f(t) = \left[ C_{\mu^+} \times e^{-t/\tau_{\mu^+}} \right] + \left[ C_{\mu^-} \times e^{-t/\tau_{\mu^-}} \right] + B,
\end{equation}
where $B$ is a constant background term, $\tau_{\mu^+}$ is the positive muon lifetime of 2197 ns~\cite{muonlifetime}, and $\tau_{\mu^-}$ is the effective lifetime of negative muons decaying in orbit from muonic Ar.  From the fit, we find
$\tau_{\mu^-} = 626\pm48$ ns,
which suggests a muon capture lifetime of  
\[
\tau_\text{c} = 871 \pm \SI{93}{ns}, 
\] 
in agreement with the theoretical expectation and previous measurements~\cite{measday,nucl_capture_rate}.   
The corresponding capture probability derived from our data is 
\[
P_\text{c} = \frac{k_c}{k_{\text{total}}} = \left( \frac{\tau_\text{c}}{\tau_{\mu^-}} \right)^{-1} = 71.8\pm2.2 \%.
\]

The flat background term, $B = 40 \pm 6$~events/bin, comprises $\approx$2740 events integrated across the relevant decay times. We therefore estimate the contamination from non-Michel events to be  $<5$\%.  Additional event quality cuts, described in Sec~\ref{sec:calo}, are expected to improve upon this purity in the final selected sample used for calorimetric studies.

We can use the relative normalization of the $\mu^+$ and $\mu^-$ populations to estimate the cosmic muon charge-ratio in our sample,
\begin{align}
\frac{\mu^+}{\mu^-} 
= \frac{N_{\mu^+}}{N_{\mu^-}} 
= \frac{(C_{\mu^+})(\tau_{\mu^+})}{(C_{\mu^-}) (\tau_{\mu^-}) (1-P_\text{c})^{-1}} 
\end{align}
where $C_{\mu^+}$ and $C_{\mu^-}$ are the normalization factors for the two exponential decay components in the fit. The result,
\[
\frac{\mu^+}{\mu^-} = 1.27 \pm 0.16,
\]
is consistent with a previous measurement of $\mu^+/\mu^-~=~1.25$ from CMS~\cite{cms-mucharge}, which used a sample of muons in the momentum range of 5~GeV/c to 1~TeV/c. Our sample probes the lowest-energy tail of the cosmic muon momentum distribution since the muons must be low enough in energy to stop in the \mbox{40-cm-tall} LAr active volume of the LArIAT TPC.  In fact, simulations indicate the stopping muons in our selected sample have an average initial momentum of $170\pm 40$~MeV/c.

The measurements presented in this section serve only to validate the cosmic stopping muon sample. Systematic errors have not been evaluated in detail. Studies are needed to estimate the impact of optical hit-finding efficiencies on measurements based on the muon decay time spectrum.

%%%%%%%%%%%%%%%%%%%%%%%%%%%%%%%%%%%%%%%%%%%%%%%%%%%%%%%%%%%%%%%%%%%%%%%%
\section{Q+L MAXIMUM-LIKELIHOOD FITTER}
\label{sec:developingfitter}

Here we describe the assembly of a more sophisticated tool for reconstructing Michel electron energy from charge and light which goes beyond the simple prescriptive formulations presented in Eqs~\ref{eq:EQ} and~\ref{eq:EQL}. We make use of the maximum-likelihood hypothesis technique which finds the \emph{most likely} energy that would produce each measured combination of $Q$ and $L$ given the detector's expected performance in reconstructing these two quantities.

The likelihood of reconstructing a Michel electron event with measured $Q$ and $L$, given a true deposited energy $E$, is given by
\begin{equation}\label{eq:likelihood}
\mathcal{L}(Q,L;E) = f_Q(Q;E) \times f_L(L;E),
\end{equation} 
where $f$ denotes the probability distribution function (PDF) for the measured charge or light.  For each event we seek to find the $E$ that maximizes $\mathcal{L}$, so we perform a minimization over the negative logarithm:
\begin{align}\label{eq:loglikelihood}
F(Q,L;E) &= -\log\mathcal{L}(Q,L;E) \\ 
&= -\log f_Q(Q;E) -\log f_L(L;E).
\end{align}

\begin{figure}[t]
\centering
\includegraphics[width=0.95\columnwidth]{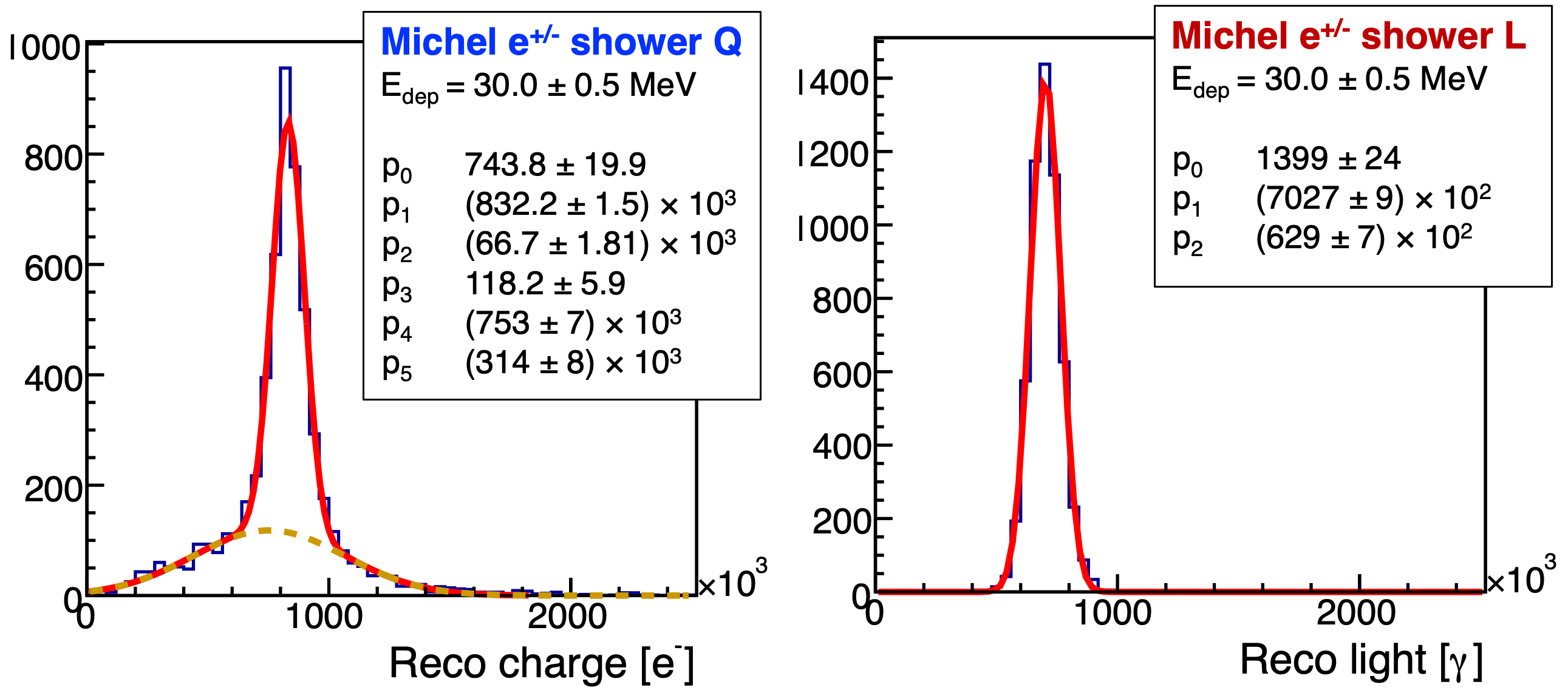}
\caption{Examples from MC of reconstructed $Q$ (left) and $L$ (right) distributions for simulated Michel electrons depositing \SI{30}{MeV} in the LArIAT TPC. The red lines are fits to the distribution. For $Q$ the fit is to a double-Gaussian function, while $L$ is fit to a single Gaussian at all energies. The dotted orange line in $Q$ illustrates the ``background'' (BG) Gaussian in the double-Gaussian fit.}
\label{fig:michel_QLfits_example}
\end{figure}

To determine the energy-dependent PDFs, we first use the Monte Carlo sample (with trigger efficiency cuts turned off) to assemble histogrammed distributions of reconstructed charge and light at different values of true deposited shower energy. These ``slices'' in energy are made at regular intervals of 5~MeV and are relatively narrow ($\pm$\SI{0.5}{MeV}) to minimize smearing of the distributions due to contributions from events of widely differing energies.  We then find that each $L$ distribution can be fit to a single Gaussian:
\[
f^*_L(L) = N \times \exp\left[ -\frac{(L-\mu)^2}{2\sigma^2} \right].
\] 
However, to fit $Q$, we require two Gaussians -- one modeling the central ``peak'' and another that models the more diffuse ``background'' population of events:
\[
f^*_Q(Q) = \left[ N_p \times e^{ -\frac{(Q-\mu_p)^2}{2\sigma_p^2} } \right]
+ \left[ N_{BG} \times e^{ -\frac{(Q-\mu_{BG})^2}{2\sigma_{BG}^2} } \right].
\]
This distinct non-Gaussian distribution in charge is a result of reconstruction effects specific to the Michel electron sample such as charge overlap between the muon and electron and incomplete clustering of the Michel shower. Events with $Q$ values that populate the central peak of these distributions are presumed to be well-reconstructed.  Examples of  fitted distributions at 30~MeV for both $Q$ and $L$ are shown in Fig.\ref{fig:michel_QLfits_example}.

\begin{figure*}
\centering

\includegraphics[width=0.8\textwidth]{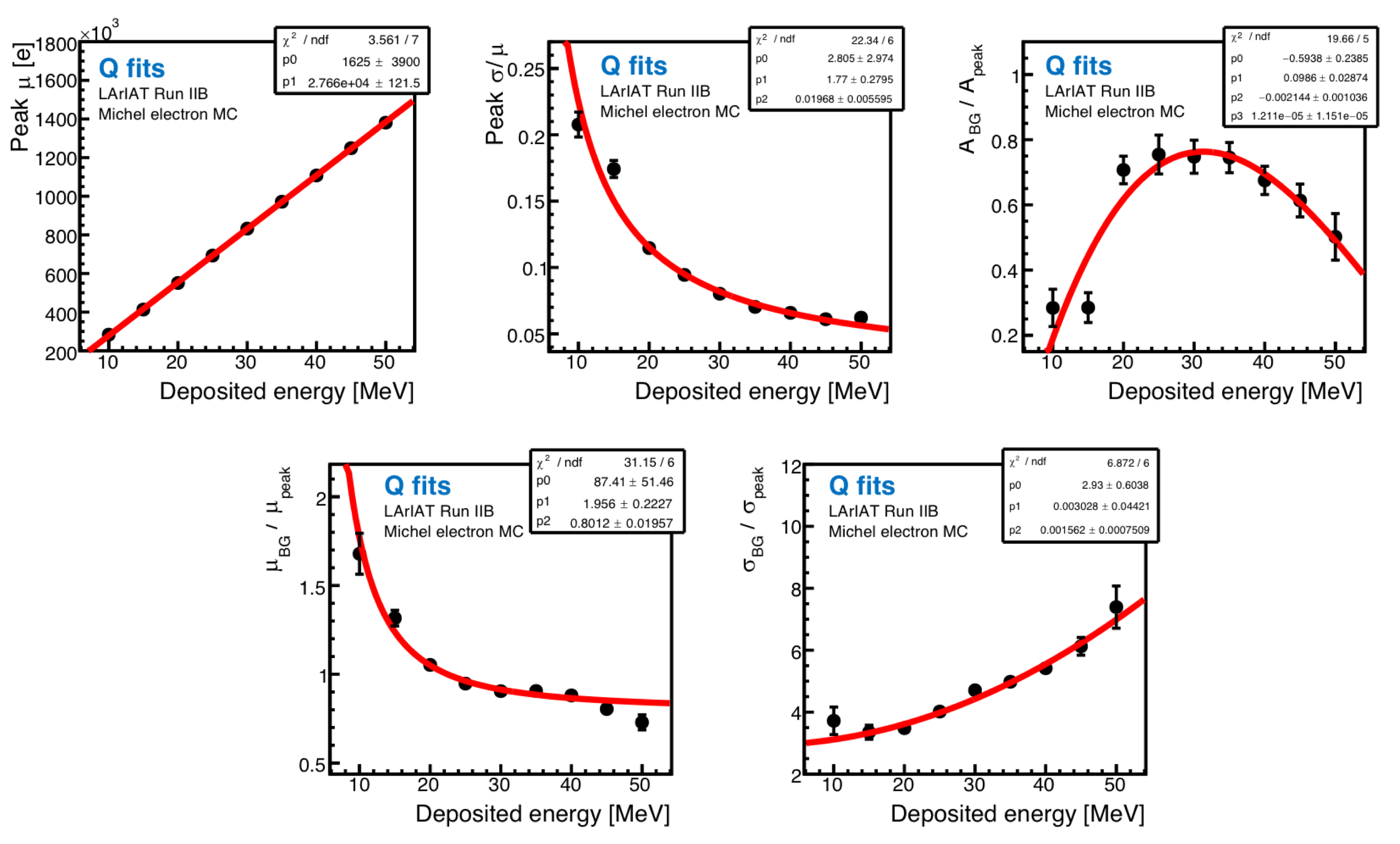}\\
\caption{Parameters describing the charge PDF $f_Q$ modeled as functions of energy deposited in the TPC by the Michel electron shower using Eqs.~\ref{eq:Qparams1}-\ref{eq:Qparams2}.}
\label{fig:michel_params_Q}
\vspace{5mm}
\includegraphics[width=0.57\textwidth]{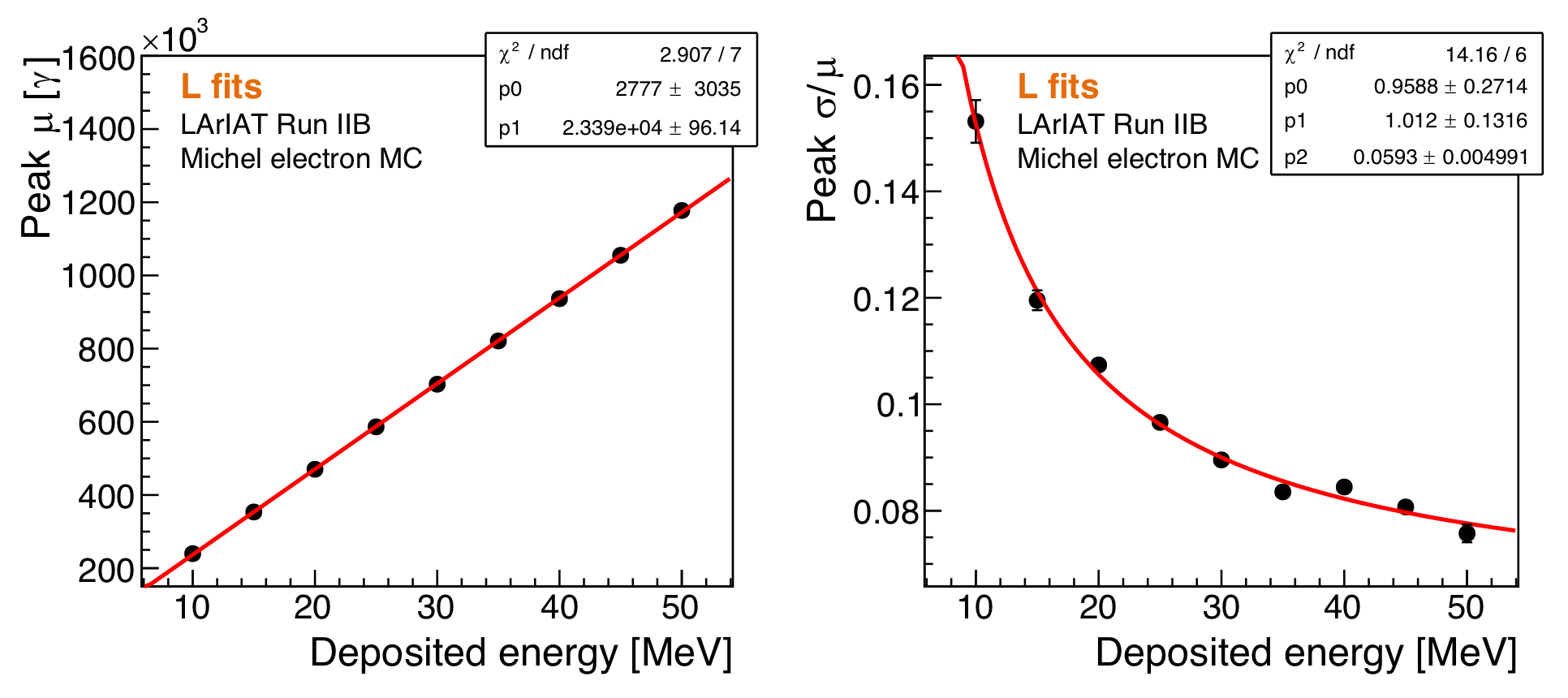}
\caption{Parameters describing the scintillation light PDF $f_L$ modeled as functions of energy deposited in the TPC by the Michel electron shower using Eqs.~\ref{eq:Lparams1}-\ref{eq:Lparams2}.}
\label{fig:michel_params_L}
\end{figure*}
% Tingjun: stat boxes are too small

The functions $f_Q^*$ and $f_L^*$  when normalized to unity are equivalent to the PDFs $f_Q(Q;E)$ and $f_L(L;E)$ needed to compute likelihood $\mathcal{L}$. In order to extrapolate between the disparate energy bins used to make the nine fits and predict $f_Q$ and $f_L$ for all deposited energies, we model the parameters of each fit as arbitrarily-chosen functions of $E$ over the range of relevant deposited energies (5-50~MeV). 

For $f_Q^*$, neglecting the overall normalization (which will be fixed to one), we find that specific combinations of parameters can be modeled as follows:
\begin{align} 
[\mu_p]_Q &= p_0 + p_1E \label{eq:Qparams1} \\
[\sigma_p/\mu_p]_Q &= p_0/E^2 + p_1/E + p_2 \\
[A_{BG}/A_{p}]_Q &= p_0 + p_1E + p_2E^2 + p_3E^3 \\
[\mu_{BG}/\mu_p]_Q &= p_0/E^{p_1} + p_2 \\ 
[\sigma_{BG}/\sigma_p]_Q &= p_0 + p_1E + p_2E^2 \label{eq:Qparams2}
\end{align}

The term $A$ above refers to the integral of the Gaussian component, $A = \sqrt{2\pi} N \sigma$. Similarly for $f_L^*$, we are able to reproduce each fit completely with only two parametrizations:
\begin{align} 
[\mu]_L &= p_0 + p_1E \label{eq:Lparams1} \\
[\sigma/\mu]_L &= p_0 /E^{p_1} + p_2 \label{eq:Lparams2}
\end{align}

These fitted parametrizations are displayed in Figs.~\ref{fig:michel_params_Q} and \ref{fig:michel_params_L}. Using these, we are then able to construct the likelihood distributions of measured $Q$ and $L$ for events of any deposited energy. 
With $f_Q$ and $f_L$ defined as the unity-normalized functions $f_Q^*$ and $f_L^*$, we now have all the pieces necessary to construct the event likelihood in Eq.~\ref{eq:likelihood}. 

\end{appendix}

%#################################################
% References
%#################################################


\begin{thebibliography}{9}

%\def\splitJINST#120#2#3#4#5#6#7#8#9{\href{http://www.iop.org/EJ/abstract/1748-0221/#1/#5#6/#4#5#6#7#8#9}
%    {20#2#3 {\it JINST }{\bf #1} #4#5#6#7#8#9}}
%\newcommand\jinst[3]{\splitJINST#1#2#3}
%\newcommand\arxiv[2]{\href{http://arxiv.org/abs/#1}{arXiv:#1 [\texttt{#2}]}}

%\def\splitJINST#120#2#3#4#5#6#7#8#9{\href{http://iopscience.iop.org/article/10.1088/1748-0221/#1/#5#6/#4#5#6#7#8#9}
%    {20#2#3 {\it JINST }{\bf #1} #4#5#6#7#8#9}}
\def\splitJINST#120#2#3#4#5#6#7#8#9{\href{http://iopscience.iop.org/article/10.1088/1748-0221/#1/#5#6/#4#5#6#7#8#9}
    {{J. Instrum. }{\bf #1}, #4#5#6#7#8#9 (20#2#3)}}
\newcommand\jinst[3]{\splitJINST#1#2#3}
\newcommand\jphys[5]{\href{#1}{J. Phys. #2 {\bf #3} #4 (#5)}}
\newcommand\physrev[5]{\href{#1}{Phys. Rev. #2 {\bf #3} #4 (#5)}}
\newcommand\physrevlett[4]{\href{#1}{Phys. Rev. Lett. {\bf #2}, #3 (#4)}}
%\newcommand\arxiv[2]{\href{http://arxiv.org/abs/#1}{arXiv:#1 [#2]}}
\newcommand\arxiv[2]{\href{http://arxiv.org/abs/#1}{arXiv:#1}}
\newcommand{\etal}{\emph{et al}.}
%\newcommand{\etal}{et al.}


%\bibitem{microboone}
%MicroBooNE Collaboration, 
%\emph{The MicroBooNE Technical Design Report}, 
%\href{http://www-microboone.fnal.gov/publications/TDRCD3.pdf}{microboone.fnal.gov}.

\bibitem{sbn}
R. Acciarri \etal { } (MicroBooNE and LAr1-ND and ICARUS-WA104 Collaborations),
%\emph{A Proposal for a Three Detector Short-Baseline Neutrino Oscillation Program in the Fermilab Booster Neutrino Beam},
\arxiv{1503.01520}{physics.ins-det}.


\bibitem{dune}
DUNE Collaboration, 
%\emph{Long-Baseline Neutrino Facility (LBNF) and Deep Underground Neutrino Experiment (DUNE)} 
\arxiv{1512.06148}{physics.ins-det}.

\bibitem{dune_solar}
F. Capozzi, S. Weishi Li, G. Zhu, and J. F. Beacom,
%\emph{DUNE as the Next-Generation Solar Neutrino Experiment}, 
\href{https://journals.aps.org/prl/abstract/10.1103/PhysRevLett.123.131803}{Phys. Rev. Lett. {\bf 123}, 131803 (2019)}.
%\physrevlett{https://journals.aps.org/prl/abstract/10.1103/PhysRevLett.123.131803}{123}{131803}{2019}.
%\arxiv{1808.08232}{physics.hep-ph}.

\bibitem{dune_supernova}
A. Ankowski \etal, 
%\emph{Supernova Physics at DUNE} 
\arxiv{1608.07853}{physics.hep-ex}.

\bibitem{w_ion}
M. Miyajima, T. Takahashi, S. Konno, T. Hamada, S. Kubota, H. Shibamura, and T. Doke, 
%\emph{Average energy expended per ion pair in liquid argon}, 
%\href{https://journals.aps.org/pra/abstract/10.1103/PhysRevA.9.1438}{Phys. Rev. A 9, 1438}.
\physrev{https://journals.aps.org/pra/abstract/10.1103/PhysRevA.9.1438}{A}{9}{1438}{1974}; \href{https://journals.aps.org/pra/abstract/10.1103/PhysRevA.10.1452}{{\bf 10}, 1452(E) (1974)}.


\bibitem{excitons_in_lar}
S. Kubota, A. Nakamoto, T. Takahashi, S. Konno, T. Hamada, M. Miyajima, A. Hitachi, E. Shibamura, and T. Doke, 
%\emph{Evidence of the existence of exciton states in liquid argon and exciton-enhanced ionization from xenon doping}, 
\physrev{https://journals.aps.org/prb/abstract/10.1103/PhysRevB.13.1649}{B}{13}{1649}{1976}.

\bibitem{aprile_book}
Aprile, A. E. Bolotnikov, A. I. Bolozdynya, and T. Doke, 
\emph{Noble Gas Detectors}, 
(Wiley-VCH Verlag GmbH and Co. KGaA, New York, 2006.

\bibitem{hitachi_timedependence}
A. Hitachi \etal,
%\emph{Effect of ionization density on the time dependence of luminescence from liquid argon and xenon}, 
\physrev{https://journals.aps.org/prb/abstract/10.1103/PhysRevB.27.5279}{B}{27}{5279}{1983}.

\bibitem{scintoflar}
T. Heindl \etal, 
%\emph{The scintillation of liquid argon},
\href{https://doi.org/10.1209/0295-5075/91/62002}{Europhys. Lett. {\bf 91} 62002 (2010)}.

\bibitem{nuclear_recoil_scint}
D. M. Mei, Z. B. Yin, L. C. Stonehill, and A. Hime, 
%\emph{A Model of Nuclear Recoil Scintillation Efficiency in Noble Liquids}, 
\href{https://journals.aps.org/prb/abstract/10.1103/PhysRevB.27.5279}{Astropart. Phys. {\bf 30}, 12 (2008)}.
%Astropart. Phys. 30:12-17, 2008,
%\arxiv{0712.2470}{physics.nucl-ex}.

% Lindhard's theory
\bibitem{lindhard}
J. Lindhard, M. Scharff, and H. E. Schi{\o}tt, 
\href{http://gymarkiv.sdu.dk/MFM/kdvs/mfm\%2030-39/mfm-33-14.pdf}{Mat. Fys. Medd. Dan. Vid. Selsk. 33(14), 1 (1963)}.

% Hitatchi treatment (bi-excitonic quench)
\bibitem{biexcitonic_quenching}
%A. Hitachi, in Proceedings of the Fourth International Workshop, York, UK, 2002, edited by Neil J. C. Spooner and Vitaly Kudryavtsev (University of Sheffield, United Kingdom, 2003), P. 357; in 5th International Workshop on the Identification of Dark Matter, IDM2004, Edinburgh, Scotland, 2004 (unpublished).
A. Hitachi,
in \href{http://inspirehep.net/record/610256}{Proceedings of the 4th International Workshop on the Identification of Dark Matter (IDM 2002), Conference C02-09-02.4, p. 357-362}; 
in \href{http://inspirehep.net/record/672168}{Proceedings of the 5th International Workshop on the Identification of Dark Matter (IDM 2004), Conference C04-09-06.3, p. 396-401}.

% Penning quenching:
\bibitem{penning_quenching}
I. D. Clark, A. J. Masson, and R. P. Wayne,
\href{https://www.tandfonline.com/doi/abs/10.1080/00268977200100981}{Molecular Physics {\bf 23}, 995 (1972)}.

\bibitem{w_ph}
T. Doke, K. Masuda, and E. Shibamura, 
%\emph{Estimation of absolute photon yields in liquid argon and xenon for relativistic (1 MeV) electrons}, 
\href{https://www.sciencedirect.com/science/article/pii/016890029090011T}{Nucl. Instrum. Methods Phys. Res., Sect. A {\bf 291}, 617 (1990)}.

\bibitem{doke}
T. Doke, A. Hitachi, J. Kikuchi, K. Masuda, H. Okada, and E. Shibamura, 
%\emph{Absolute Scintillation Yields in Liquid Argon and Xenon For Various Particles}, 
\href{https://iopscience.iop.org/article/10.1143/JJAP.41.1538/meta}{Jpn. J. Appl. Phys. {\bf 41}, 1538 (2002)}.

\bibitem{birks_law}
J. B. Birks, 
%\emph{Scintillations from Organic Crystals: Specific Fluorescence and Relative Response to Different Radiations}, 
\href{https://iopscience.iop.org/article/10.1088/0370-1298/64/10/303/meta}{Proc. Phys. Soc. A \textbf{64}, 874 (1951)}.

\bibitem{thomas_imel}
J. Thomas, D.A. Imel, 
%\emph{Recombination of electron-ion pairs in liquid argon and liquid xenon}, 
%\href{https://journals.aps.org/pra/abstract/10.1103/PhysRevA.36.614}{Phys. Rev. \textbf{A} 36 (1987), 614}.
\physrev{https://journals.aps.org/pra/abstract/10.1103/PhysRevA.36.614}{A}{36}{614}{1987}.

\bibitem{icarus_recomb}
S. Amoruso \etal,
%\emph{Study of electron recombination in liquid argon with the ICARUS TPC}, 
\href{https://www.sciencedirect.com/science/article/pii/S0168900204000506}{Nucl. Instrum. Methods Phys. Res., Sect. A \textbf{523}, 275 (2004)}.

\bibitem{argoneut_recomb}
R. Acciarri \etal,
%\emph{A study of electron recombination using highly ionizing particles in the ArgoNeuT Liquid Argon TPC}, 
\jinst{8}{2013}{P08005}.

\bibitem{sorel}
M. Sorel,
%\emph{Expected performance of an ideal liquid argon neutrino detector with enhanced sensitivity to scintillation light},
\jinst{9}{2014}{P10002}.

%\bibitem{lariat}
%F. Cavanna, M. Kordosky, J. Raaf, B. Rebel on behalf of the LArIAT Collaboration, 
%\emph{LArIAT: Liquid Argon In A Testbeam}, 
%\arxiv{1406.5560v3}{physics.ins-det}.

\bibitem{lariat_detpaper}
LArIAT Collaboration, 
%\emph{The Liquid Argon In A Testbeam (LArIAT) Experiment},
\href{http://inspirehep.net/record/1766977}{FERMILAB-PUB-19-460-ND},
\arxiv{1911.10379}{physics.ins-det}.

\bibitem{ftbf}
FTBF (Fermilab Test Beam Facility), \href{http://ftbf.fnal.gov}{http://ftbf.fnal.gov}.

\bibitem{argoneut}
C Anderson \etal (ArgoNeuT Collaboration),
%\emph{The ArgoNeuT Detector in the NuMI Low-Energy beam line at Fermilab}, 
\jinst{7}{2012}{P10019}.

\bibitem{pmt_tests}
R. Acciarri \etal,
%\emph{Demonstration and Comparison of Operation of Photomultiplier Tubes at Liquid Argon Temperature},
%\arxiv{1108.5584}{physics.ins-det}.
\jinst{7}{2012}{P01016}.

\bibitem{warp_fast_dig}
A. M. Szelc, N. Canci, F. Cavanna, A. Cortopassi, M. D’Incecco, G. Mini, F. Pietropaolo, A. Romboli, E. Segreto, and R. Acciarri,
%\emph{First Tests of a New Fast Waveform Digitizer for PMT Signal Read-out from Liquid Argon Dark Matter Detectors}, 
\href{https://www.sciencedirect.com/science/article/pii/S1875389212018123}{Phys. Procedia \textbf{37}, 1131 (2012)}.

\bibitem{nevis}
M. Bardon \etal,
%\emph{Measurement of the Momentum Spectrum of Positrons from Muon Decay},
%\href{http://journals.aps.org/prl/pdf/10.1103/PhysRevLett.14.449}{Phys. Rev. Lett. 14, 449 (1965)}.
\physrevlett{http://journals.aps.org/prl/pdf/10.1103/PhysRevLett.14.449}{14}{449}{1965}.

\bibitem{larsoft}
E.  Church,
%\emph{LArSoft: a software package  for  liquid  argon  time  proportional  drift  chambers}, 
\arxiv{1311.6774}{physics.ins-det}.

\bibitem{trajcluster}
B. Baller, 
%\emph{Liquid argon TPC signal formation, signal processing and reconstruction techniques}, 
\jinst{12}{2017}{P07010}.
%\arxiv{1703.04024}{physics.ins-det}.

\bibitem{pmtrack}
M. Antonello, B. Baibussinov, P. Benetti, E. Calligarich, N. Canci, S. Centro, A. Cesana, K. Cieslik, D.B. Cline, A.G. Cocco \etal,
\href{https://www.hindawi.com/journals/ahep/2013/260820/}{Adv. High Energy Phys. \textbf{2013}, Article ID 260820 (2013).}

\bibitem{microboone_michels}
R. Acciarri \etal,
%\emph{Michel Electron Reconstruction Using Cosmic-Ray Data from the MicroBooNE LArTPC}
%\arxiv{1704.02927}{physics.ins-det}.
\jinst{12}{2017}{P09014}.

\bibitem{segreto}
E. Segreto, 
%\emph{Evidence of delayed light emission of tetraphenyl-butadiene excited by liquid-argon scintillation light}, 
\physrev{https://journals.aps.org/prc/abstract/10.1103/PhysRevC.91.035503}{C}{91}{035503}{2015}.
%Phys. Rev. C 91, 035503 (2015), 
%\arxiv{1411.4524}{physics.ins-det}.

\bibitem{geant4}
S. Agostinelli \etal, 
%\emph{GEANT4 - a simulation toolkit}, 
\href{https://www.sciencedirect.com/science/article/pii/S0168900203013688}{Nucl. Instrum. Methods Phys. Res., Sect A \textbf{506}, 250 (2003)}.

\bibitem{hofmann}
M. Hofmann \etal, 
%\emph{Ion-beam excitation of liquid argon}, Eur. Phys. J. C (2013) 73:2618, 
\href{https://link.springer.com/article/10.1140\%2Fepjc\%2Fs10052-013-2618-0}{Eur. Phys. J. C \textbf{73}, 2618 (2013)}.
%\arxiv{1511.07721}{physics.ins-det}.

\bibitem{singlet-to-triplet}
T. Heindl,  Ph.D.  thesis,  Technische  Universit\"at  M\"unchen, 2011, \href{http://mediatum.ub.tum.de/node?id=1080148}{http://mediatum.ub.tum.de/node?id=1080148}.

\bibitem{quenching_o2}
WArP Collaboration, 
%\emph{Oxygen contamination in liquid Argon: combined effects on ionization electron charge and scintillation light}, 
\jinst{5}{2010}{P05003},
%\arxiv{0804.1222}{physics.nucl-exp}. 
%[\href{https://arxiv.org/abs/0804.1222}{nucl-exp/0804.1222}]

\bibitem{quenching_n2}
R. Acciarri \etal, 
%\emph{Effects of Nitrogen contamination in liquid Argon}, 
\jinst{5}{2010}{P06003}.

\bibitem{grace}
E. Grace, A. Butcher, J. Monroe, and J. A. Nikkel, 
%\emph{Index of refraction, Rayleigh scattering length, and Sellmeier coefficients in solid and liquid argon and xenon}, 
%\arxiv{1502.04213}{physics.ins-det}.
\href{http://inspirehep.net/record/1345082?ln=en}{Nucl. Instrum. Methods A \textbf{867}, 204 (2017)}.

\bibitem{ardm}
V. Boccone \etal,
%\emph{Development of wavelength shifter coated reflectors for the ArDM argon dark matter detector}
\jinst{4}{2009}{P06001},
%\arxiv{0904.0246}{physics.ins-det}.

%Measurements at Cracow University of Technology.
\bibitem{cu-reflect}
J. Jaglarz \etal, J. Alloys Compd. \textbf{371}, 125 (2004).

%\bibitem{lbne_wireplaneshadow}
%LBNE/DUNE, \href{https://lbne2-docdb.fnal.gov/}{\texttt{lbnd-DocDB:6940-v2}} (Wireplane shadowing parameterization from Ben Jones).

\bibitem{root}
R. Brun and F. Rademakers, 
%\emph{ROOT - An Object Oriented Data Analysis Framework}, 
\href{http://inspirehep.net/record/458148?ln=en}{Nucl. Instrum. Methods Phys. Res., Sect. A \textbf{389}, 81 (1997)}; See also \href{root.cern.ch/}{http://root.cern.ch/}.

\bibitem{icarus-michels}
S. Amoruso \etal{ }(ICARUS Collaboration), 
%\emph{Measurement of the $\mu$ decay spectrum with the ICARUS liquid Argon TPC}, 
\href{https://link.springer.com/article/10.1140/epjc/s2004-01597-7}{Eur. Phys. J. C \textbf{33}, 233 (2004)}.
%\arxiv{0311040}{physics.hep-ex}.
%\href{http://arxiv.org/abs/hep-ex/0311040}{arXiv:0311040 [physics.hep-ex]}.

\bibitem{icarus-snr}
L. Bagby, B. Baibussinov, V. Bellini, M. Bonesini, A. Braggiotti, L. Castellani, S. Centro, T. Cervi, A.G. Cocco, and F. Fabris, 
%\emph{New read-out electronics for ICARUS-T600 liquid argon TPC. Description, simulation and tests of the new front-end and ADC system}, 
\jinst{13}{2018}{P12007}.
%\arxiv{1805.03931}{

\bibitem{sbnd_ly} D. Garcia-Gamez, The SBND High Efficiency Light Collection System, \emph{LIght Detection In Noble Elements (LIDINE)}, (SLAC, CA, 2017).

\bibitem{nucl_capture_rate}
T. Suzuki and D. F. Measday,
%\emph{Total nuclear capture rates for negative muons}, 
\physrev{https://journals.aps.org/prc/abstract/10.1103/PhysRevC.35.2212}{C}{35}{2212}{1987}.
%\href{https://journals.aps.org/prc/abstract/10.1103/PhysRevC.35.2212}{Phys. Rev. C \textbf{35}, 2212}.
%\prc{35}{1987}{2212}.
 	
\bibitem{measday}
D.F. Measday, 
%\emph{The nuclear physics of muon capture}, 
\href{http://www.sciencedirect.com/science/article/pii/S0370157301000126}{Phys. Rep. \textbf{354}, 234 (2001)}.

\bibitem{muonlifetime}
D.M. Webber \etal,
%\emph{Measurement of the Positive Muon Lifetime and Determination of the Fermi Constant to Part-per-Million Precision}, 
\physrevlett{https://journals.aps.org/prl/abstract/10.1103/PhysRevLett.106.041803}{106}{041803}{2011}.
%Phys. Rev. Lett. 106:041803 (2011),
%\arxiv{1010.0991}{physics.hep-ex}.
% [\href{https://arxiv.org/abs/1010.0991}{\texttt{hep-ex/1010.0991}}].

%\bibitem{jinr}
%A. Klinskih, V. Egorov, M. Shirchenko, D. Zinatulina, 
%\emph{Muon Capture in Ar. The Muon Lifetime and Yields of Cl Isotopes}, 
%\href{http://link.springer.com/article/10.3103\%2FS106287380806004X}{Bulletin of the Russian Academy of Sciences: Physics, 2008, Vol 72, No. 6, pp. 735-736}.

\bibitem{cms-mucharge}
CMS Collaboration, 
%\emph{Measurement of the charge ratio of atmospheric muons}, 
\href{https://www.sciencedirect.com/science/article/pii/S0370269310008725?via\%3Dihub}{Phs. Lett. B \textbf{692}, 83 (2010)}.
%\arxiv{1005.5332}{physics.hep-ex}.
%[\href{https://arxiv.org/abs/1005.5332}{hep-ex/1005.5332}].


\end{thebibliography}
\end{document}